\documentclass[twocolumn,showpacs,preprintnumbers,aps,floatfix,superscriptaddress]{revtex4-1}
\usepackage{graphicx}
\usepackage{amsmath,amssymb}
\usepackage{hyperref}
\usepackage{dcolumn}
\usepackage{bm}
\usepackage{color}

\begin{document}

\title{
Highly dispersive magnons with spin-gap like features
in the 
frustrated ferromagnetic $S=1/2$ chain compound Ca$_2$Y$_2$Cu$_5$O$_{10}$ 
detected by inelastic neutron scattering
}

\author{M.\ Matsuda}

\affiliation{Neutron Scattering Division, Oak Ridge National 
Laboratory, Oak Ridge, Tennessee 37831, USA}


\author{J.\ Ma}

\altaffiliation[Present address: ]{Department of Physics and Astronomy, 
Shanghai Jiao Tong University, Shanghai 200240, P. R. China}
\affiliation{Neutron Scattering Division, Oak Ridge National Laboratory, 
Oak Ridge, Tennessee 37831, USA}

\author{V. O.\ Garlea}
\affiliation{Neutron Scattering Division, Oak Ridge National 
Laboratory, Oak Ridge, Tennessee 37831, USA}

\author{T.\ Ito}
\affiliation{National Institute of Advanced Industrial Science and 
Technology (AIST), Tsukuba, Ibaraki 305-8562, Japan}
\author{H.\ Yamaguchi}
\affiliation{National Institute of Advanced Industrial Science and 
Technology (AIST), Tsukuba, Ibaraki 305-8562, Japan}
\author{K.\ Oka}
\affiliation{National Institute of Advanced Industrial Science and 
Technology (AIST), Tsukuba, Ibaraki 305-8562, Japan}

\author{S.-L.\ Drechsler}
\affiliation{Institute of Theoretical Solid State Physics, IFW Dresden, 
Helmholtzstra{\ss}e 20, D-01069 Dresden, Germany}
\author{R.\ Yadav}
\affiliation{Institute of Theoretical Solid State Physics, IFW Dresden,
 Helmholtzstra{\ss}e 20, D-01069 Dresden, Germany}
\author{L.\ Hozoi}
\affiliation{Institute of Theoretical Solid State Physics, IFW Dresden, 
Helmholtzstra{\ss}e 20, D-01069 Dresden, Germany}
\author{H.\ Rosner}
\affiliation{Max-Planck-Institute of Chemical Physics, N\"othnitzer Str.\ 40, 
D-01187 Dresden, Germany}
\author{R.\ Schumann}
\affiliation{Department of Phys., Inst.\ of Theor.\ Physics, 
TU Dresden, D-1062 Dresden, Zellescher Weg 17, Germany }
\author{R.O.\ Kuzian}
\affiliation{Institute for Problems of Materials Science NASU, 
Krzhizhanovskogo 3, 03180 Kiev, Ukraine}
\affiliation{Donostia International Physics Center (DIPC), Paseo Manuel de Lardizabal
4, San Sebastian/Donostia, 20018 Basque Country, Spain}
\author{S.\ Nishimoto}
\affiliation{Department of Phys., Inst.\ of Theor.\ Physics, 
TU Dresden,~Mommsenstra{\ss}e,~D-01069 Dresden, and \\
Inst.\ for Theoretical Solid State Physics, 
IFW-Dresden, Helmholtzstra{\ss}e 20, D-01069 Dresden, Germany}

\date{\today}

\begin{abstract}
\noindent
We report inelastic neutron scattering experiments in Ca$_2$Y$_2$Cu$_5$O$_{10}$ and map out the full one magnon dispersion which extends up to a record value of 53 meV for frustrated ferromagnetic (FM) edge-sharing CuO$_2$ chain (FFESC) cuprates. A homogeneous spin-1/2 chain model with a FM nearest-neighbor (NN), an antiferromagnetic (AFM) next-nearest-neighbor (NNN) inchain, and two diagonal AFM interchain couplings (ICs) analyzed within linear spin-wave theory (LSWT) reproduces well the observed strong dispersion along the chains and a weak one perpendicularly. The ratio $\alpha=|J_{a2}/J_{a1}|$ of the FM NN and the AFM NNN couplings is found as $\sim$0.23, close to the critical point $\alpha_{c}=1/4 $ which separates ferromagnetically and antiferromagnetically correlated spiral magnetic ground states in single chains, whereas $\alpha_{c}>0.25$ for coupled chains is considerably upshifted even for relatively weak IC. Although the measured dispersion can be described by homogeneous LSWT, the scattering intensity appears to be considerably reduced at $\sim$11.5 and $\sim$28 meV. The gap-like feature at 11.5~meV is attributed to magnon-phonon coupling whereas based on density matrix renormalization group simulations of the dynamical structure factor the gap at 28 meV is considered to stem partly from quantum effects due to the AFM IC. Another contribution to that gap is ascribed to the intrinsic superstructure from the distorting incommensurate pattern of CaY cationic chains adjacent to the CuO$_2$ ones. It gives rise to non-equivalent CuO$_4$ units and Cu-O-Cu bond angles $\Phi$ and a resulting distribution of all exchange integrals. The $J$'s fitted by homogeneous LSWT are regarded as average values. The record value of the FM NN integral $J_1=24$~meV among FFESC cuprates can be explained by a {\it non-universal } $\Phi (\neq 90^{\circ}$) and Cu-O bond length dependent {\it anisotropic} mean direct FM Cu-O exchange $\bar{K}_{pd}$$\sim$120 meV, similar to a value of 105 meV for Li$_2$CuO$_2$, in accord with larger values for La$_2$CuO$_4$ and CuGeO$_3$ ($\sim$110~meV) reported by Braden {\it et al.} [Phys.\ Rev.\ B {\bf 54}, 1105 (1996)] phenomenologically. Enhanced $K_{pd}$ values are also needed to compensate a significant AFM $J_{dd} \geq $~6~meV from the
$dd$ channel, generic for FFESC cuprates but ignored so far. 
\end{abstract}


\maketitle
\section {introduction}
One-dimensional (1D) antiferromagnetic (AFM) 
spin ($S$) 1/2
systems
have been studied intensively, since
they
 exhibit exotic quantum effects. 
 The spinon 
 is a typical 
 feature 
 generic for the 
 AFM Heisenberg chain.
 In contrast, 
 1D ferromagnetic (FM) systems do not show pronounced 
 quantum effects since the FM state is an eigenstate of the spin
 Hamiltonian. However, frustrating couplings, such 
 as a next-nearest neighbor (NNN) AFM $J_2$ and/or 
 AFM interchain couplings (ICs), can cause a more 
 interesting ground state \cite{bursil}.
 In particular, they may induce gaps
 of different 
 nature for excited states,
 strongly dependent on the sign of the nearest-neighbor (NN) exchange $J_1$: 
 well-known for
 AFM $J_1$ for 0.241 $\leq \alpha \stackrel{<}{\sim} 0.7$
 in the context of the spin-Peierls problem 
 \cite{Braden1996}
 and recently found for FM $J_1$ at $\alpha > \alpha_{\rm \tiny c}(=1/4)$ 
 due to quantum fluctuations 
 \cite{Agrapidis2017}, where the frustration  $\alpha$ reads
 \begin{equation}
  \alpha= J_2/\mid J_1\mid \ . 
 \label{frustratio}
 \end{equation}
In Ca$_2$Y$_2$Cu$_5$O$_{10}$ (CYCO)  and any other
 edge-sharing chain cuprates
 (see Sec.\ II and Fig.\ 1)
 described 
 by the 
 $S=1/2$
 $J_1$-$J_2$ model, $J_2>0$ 
 always holds 
 due to
the Cu-O-O-Cu superexchange. 
 Then $\alpha^{-1}$ measures the coupling 
 of
 two interpenetrating 
 ferromagnetically
interacting
 AFM Heisenberg chains,
 where the $J_1$-$J_2$ chain is regarded
 as a topologically equivalent
 zigzag chain with 
 different NN couplings. For
FM $J_1$, 
the ground state changes from a FM to 
an AFM spin-liquid with
noncollinear spiral fluctuations
for $\alpha > \alpha_c$ \cite{bursil}.
There are only few 
materials 
with long
edge-sharing CuO$_2$ chains and relatively large $J_1$- 
and $J_2$-values 
near such a critical point. 
We
mention
three of them: 
(i) Li$_2$CuO$_2$ (LICO),
with
FM inchain 
order below its N\'eel 
temperature $T_{\rm \tiny N} \approx 9$~K;
(ii) Li$_2$ZrCuO$_4$, with a spiral ordering 
with $\alpha \approx 0.33$
(at $T < 7$~K), predicted in Ref.\ \onlinecite{drechsler}
and confirmed in Refs.\ \onlinecite{Sirker2010, Tarui2008}
(see also 
\cite{Suppl});
(iii) 
further candidates 
near quantum criticality,
where the 
insight gained 
for CYCO
might be 
helpful to elucidate
their exchange interactions
and unusual magnetic states. 
Among them are 
La$_6$Ca$_8$Cu$_{24}$O$_{41}$ and derivatives which
contain besides two-leg spin ladders (TLL) similar frustrated FM
edge-sharing CuO$_2$ chains (FFESC).\\
\begin{figure}[b]
\includegraphics[width=7.2cm]{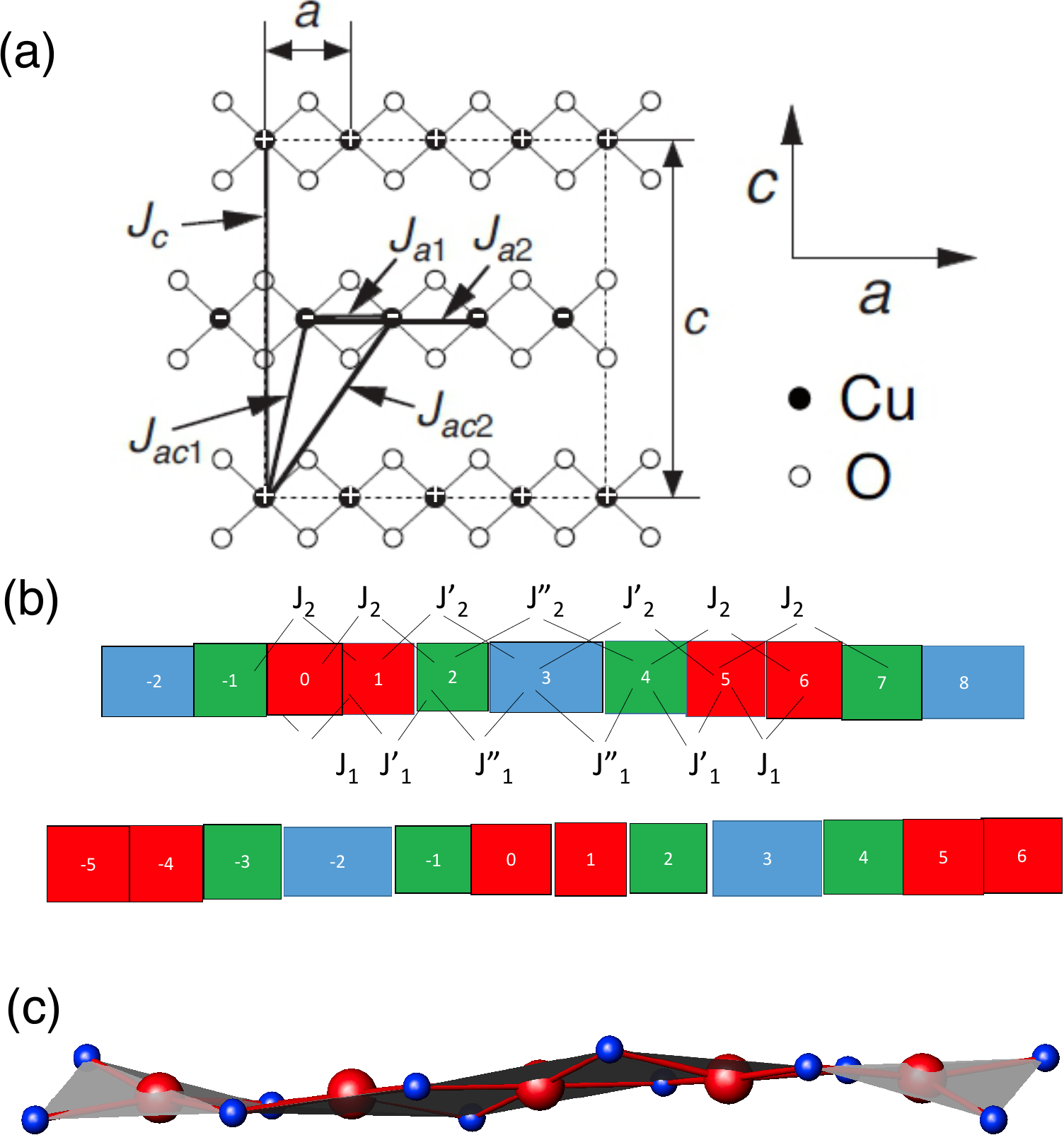}
\caption{(a) Schematic view of the 
CuO$_2$ spin chains 
in the $ac$ plane of Ca$_2R_2$Cu$_5$O$_{10}$ ($R$=Y and Nd) for an averaged 
idealized
structure. 
Inchain couplings $J_{a1}$ and $J_{a2}$ 
as well as the two 
diagonal NN and NNN ICs 
$J_{ac1}$, $J_{ac2}$, and $J_c$ are shown.
The spin 
order in the $ac$ plane is also depicted.
Spins along +$b$ ($-b$) directions are shown with $"+"$ ($"-"$), respectively.
(b) A non-ideal
Cu$_5$O$_{10}$ chain due to its misfit 
with the adjacent 
cationic Ca$_2R_2$ ($R$=Nd and Y) chains adopting 
symmetric distortions for simplicity. 
The 3 non-equivalent CuO$_4$ plaquettes of this case
are depicted by red, green, and 
blue rectangles. Here a chain has 3 different
boundaries (red-red, red-green, and red-blue pairs of bridging O)
and three different Cu-O-Cu
bond angles, giving rise to 3 different AFM contributions to each
NN coupling (see\ Sec.\ V).
At least 3 different
NN and NNN couplings denoted by $J_1$, $J_1$', and $J_1$" 
as well as $J_2$, $J_2$', and $J_2$"
[instead of 2 single $J_{a1}$ and $J_{a2}$ shown in (a)]
appear. The general asymmetric 
chain has
5 non-equivalent plaquettes and
a couple of 5 NN and NNN inchain couplings, respectively.
(c) A 
distorted single 
chain according to 
the model by Thar {\it et al.}
\cite{Thar2006} (view along the $c$ axis). 
Red (blue)
spheres 
denote
Cu (O) ions.}
\label{structure}
\end{figure}
 A recent inelastic neutron scattering (INS)
 study for LICO \cite{lorenz,Lorenz2011,Kuzian2018} revealed 
 a relatively large
$J_1$=$-$19.7 meV,
$\alpha=
0.332$, 
and a weak but nevertheless decisive
AFM IC of 0.78 meV. 
Although $\alpha > \alpha_{\rm \tiny c}$, 
a FM arrangement is realized in the 
chains due to specific AFM
ICs. 
Within a refined linear spin-wave
analysis employing the full magnon dispersion 
up to 53~meV and performing
measurements along those
scattering directions
where the small IC can be separated from the large inchain 
one, we will show that 
 $J_1$ of CYCO well
exceeds the largest
$J_1$-values 
reported so far for
LICO \cite{lorenz}
and Li$_2$ZrCuO$_4$ \cite{drechsler,Sirker2010}
among the FFESC family. 

As shown in detail 
below, CYCO with a FM stacking 
of 2D N\'eel planes along the $b$ axis is the ``2D-analogon'' of LICO.
CYCO has the highest $T_N$=$29.5$~K \cite{matsuda99,fong,Kargl2006}
and together
with 
LICO 
the largest ordered magnetic moments
among all FFESC.
The critical
point for the system of antiferromagnetically coupled mutually shifted NN chains 
by half a Cu-Cu distance in the chain direction (Fig.\ 1) 
is {\it upshifted} to $\alpha$$>$1/4, 
reflecting a stabilization of the
FM inchain ordering.
On the other hand, the spin-wave dispersion for 
LICO shows 
a flat minimum at the magnetic zone center, which reflects also incommensurate 
correlations along the chains. 
Here weak AFM 
O-mediated 
ICs between adjacent chains are relevant too
\cite{lorenz,Lorenz2011,Kuzian2018}.  

Since the large FM $J_1$ is the origin for 
a large magnon dispersion, its microscopic origin is of
interest for the cuprate physics in general to be addressed
in the framework of multiband Cu-O $pd$ models.
Then
$J_1$
depends first of all on
the direct FM exchange $K_{pd}$
between two holes on NN Cu and O sites and on Hund's exchange 
$J_H$
of two holes in two different O 2$p$ orbitals
on the same
O site which bridges twice two NN Cu sites.  Although being key
quantities, neither is precisely known.
In particular, the $K_{pd}=50~$~meV 
suggested in Ref.\ \onlinecite{mizuno} for LICO and CYCO 
and
{\it all} other FFESC 
differs 
by more than 200\%
from the empirical value 
for 
CuGeO$_3$ \cite{Braden1996} and even by 400\%
for the
corner-sharing
La$_2$CuO$_4$, where 
$K_{pd}$ and its weaker NN O-O counter part, $K_{pp}$, are 
known
from advanced many-body calculations
\cite{Hybertsen1990,Hybertsen1992}. 
We report on complementary quantum chemical (QC) and density functional theory (DFT) computations
for LICO, 
derive also a significant $K_{pd}$-value 
and stress
its 
key role for the 
FM $J_1$ 
and 
the magnon dispersion
\cite{lorenz,Lorenz2011}.
We 
resolve
a long-standing puzzle
for seemingly AFM 
or small FM
Curie-Weiss temperatures ($\Theta_{\rm \tiny CW}$) \cite{fong,Kargl2006D}
at odds with the 
 inchain FM alignment
of magnetic moments in the Ne{\'e}l-state
using a 
high-temperature 
expansion (HTE) for the spin
susceptibility $\chi(T)$ (Sec.\ V A).
In the SM \cite{Suppl} we present a rich collection of 
cuprates
with edge-sharing elements, provide 
further support for sizable
FM NN couplings, and explain 
special reasons for weak or even AM values.

\section {Previous and new results in CYCO}\label{SecHistory}
CYCO consists of edge-sharing CuO$_2$ chains \cite{davis}, 
each [CuO$_2$]$^{-2}$ unit carries a 
spin 1/2 \cite{remarkOmoment}.
The chains in the $ac$ plane are shown 
in Fig.\ \ref{structure}(a).
CYCO exhibits an AFM order below $T_{\rm \tiny N}$=29.5~K. 
The spins align ferromagnetically
along the $a$ axis (chain direction) and $b$ axis and 
antiferromagnetically
along the $c$ 
axis \cite{matsuda99,fong}. The ordered magnetic moment is 0.9$\mu\rm_B$. 
The magnetic structure in the $ac$ plane is shown in Fig.\ 1(a).
CYCO shows commensurate and incommensurate orders of Ca and Y, which gives 
rise to a supercell with 5$\times a$ and 4.11$\times c$ in the simplest 
approximation~\cite{davis}.
In this supercell, there are 4 Ca/Y and 5 Cu positions along the chain. 
The alternating Ca$^{2+}$/Y$^{3+}$ chains cause sizable 
shifts of the O ions
(Figs.\ 1b and 1c) at variance to chains with equivalent O 
sites \cite{Thar2006}.
However, it is 
unclear how much this superstructure (SS), affecting
mostly the O 
sites, does modify
the long-range magnetic order. The 
inchain spin arrangement is FM
and no magnetic SS
has been found so far. Hence,
the static magnetic order 
seems to be hardly affected  by the chain distortions, 
probably due to the few O spins. 
 
The spin-Hamiltonian for CYCO may be written as
\begin{equation}
\hat{H}=\hat{H}_{ch}+\hat{H}_{IC}+\hat{H}_{A},
\label{eq:spinH}
\end{equation}
where $\hat{H}_{ch}$($\hat{H}_{IC}$) describes in(inter)-chain isotropic
Heisenberg interactions of the form
\begin{equation}
\hat{H}_{ch}+\hat{H}_{IC}
=\frac{1}{2}\sum_{\mathbf{R,r}}
J_{\mathbf{r}}\hat{\mathbf{S}}_{\mathbf{R}}\hat{\mathbf{S}}_{\mathbf{R+r}},
\label{eq:Heis}
\end{equation}
$J_{\mathbf{r}}$ being exchange interaction between a pair of copper
spins $\hat{\mathbf{S}}_{\mathbf{R}}$ and $\hat{\mathbf{S}}_{\mathbf{R+r}}$
in the same (for $\hat{H}_{ch}$) or in different (for $\hat{H}_{IC}$) chains;
$\hat{H}_{A}$ denotes uniaxial pseudodipolar 
\begin{equation}
\noindent
\hspace{-0.3cm}\mbox{anisotropic interactions:} \quad 
\hat{H}_{A}=
\frac{1}{2}\sum_{\mathbf{R,r}}D_{\mathbf{r}}\hat{S}_{\mathbf{R}}^{z}\hat{S}_{\mathbf{R+r}}^{z}.
\label{eq:HA}
\end{equation}
In previous works \cite{Matsuda2001,matsuda052,matsuda05,kuzian},
the INS results were interpreted
assuming that all Cu sites are equivalent, i.e.\
ignoring a lattice
modulation.
The magnon
dispersion curves
were fitted by 
linear spin-wave theory (LSWT):
\begin{align}
  \omega_{q} &=\sqrt{A_{q}^{2}-B_{q}^{2}},\label{SWT_omega} \quad \mbox{with} \ \\
A_{q} & =  J_{a1}(\cos q_{a}-1)+J_{a2}(\cos2q_{a}-1)+J_{b}(\cos q_{b}-1) \nonumber\\
 & +  J_{c}(\cos q_{c}-1)+2J_{ab}(\cos\frac{q_{a}}{2}\cos\frac{q_{b}}{2}-1) \nonumber\\
 & +  2(J_{ac1}+J_{ac2})-D, \nonumber\\
B_{q} & =  2J_{ac1}\cos\frac{q_{a}}{2}\cos\frac{q_{c}}{2}
+2J_{ac2}\cos\frac{3q_{a}}{2}\cos\frac{q_{c}}{2}, \nonumber
\label{eq:AB}
\end{align}
where $J_{a1}$, $J_{a2}$ are FM NN and AFM NNN in-chain interactions,
$J_{ac1}$ and $J_{ac2}$ are NN and NNN ICs in $ac$ plane (Fig.
\ref{structure}), $J_{b}$ and $J_{c}$ are interactions along the
$b$ and $c$ directions, respectively; 
$\mathbf{q}=\left(q_{a},q_{b},q_{c}\right)=2\pi\left(h,k,l\right)$
is the magnon momentum. Only an averaged value $D$ of the anisotropy parameters
$D_{\mathbf{r}_{1}}$,$D_{\mathbf{r}_{2}}$
enters 
the dispersion 
\begin{equation}
D=\sum_{\mathbf{r}_{1}}D_{\mathbf{r}_{1}}-\sum_{\mathbf{r}_{2}}D_{\mathbf{r}_{2}},\label{eq:D}
\end{equation}
where the vectors $\mathbf{r}_{1}$($\mathbf{r}_{2}$) connect
sites of the same (different) AFM sublattice.

Let us recall that a FM state is an eigenstate of $\hat{H}_{ch}+\hat{H}_{A}$ and magnons 
are its \emph{exact} one-particle excitations. That is why the ordered moment in CYCO
is close to 1 $\mu _B$ and LSWT provides adequate values of exchange parameters
(cf.\ Sec.\ \ref{SecHTE}).
The only source of quantum fluctuations in CYCO is 
the relatively weak
$\hat{H}_{IC}$. 
It affects the magnon dynamics but does not change the overall
shape of the LSWT dispersion, Eq.\ (\ref{SWT_omega}) 
(see\ Sec.\ref{SecQuantumGap}). 

In Ref.\ \onlinecite{Matsuda2001}, the dispersion along 
high-symmetry
directions starting from
the zone center $\Gamma$
was measured
up to $W_{E}\sim$10 meV. The full dispersion curves were
available for the $b$ and $c$ directions $(0,k,0)$ and $(0,0,l)$,
thus the IC parameters $J_b=0.06$, $J_{ab}=−0.03$, $J_c=0$,
and $J_{s}=J_{ac1}+J_{ac2}\approx 2.24$~meV
were established. On the contrary, only a small part of the dispersion
was possible to be measured along the chain direction $a$. Inspection
of Eq. (\ref{SWT_omega}) shows that the dispersion along the
line $(h,0,0)$ is affected by the ICs.
The influence of the tiny
$J_{ab}$ can be ignored
but $J_{ac1,2}$ do substantially affect the
dispersion at small $\mathbf{q}$. Moreover, the dispersion depends
not only on the sum $J_{s}$, but also on the 
ratio $J_{ac1}/J_{ac2}$.
That is why the fit of those measurements
was ambiguous. Table \ref{Tab1} shows how the 
extracted inchain couplings became
more and more accurate 
by including 
data up to higher energies. 

New measurements along the lines $(h,0,1.25)$ and $(h,0,1.5)$
were performed in Ref.\ \onlinecite{kuzian}. The dispersion along the line
$(h,0,1.5)$ ($q_{c}=3\pi$) is
independent of $J_{ac1,2}$, and
its curvature near $h=0$ is determined
by $\alpha$. It reveals a substantial
value of $J_{a2}$ and allows
a new fit that
includes also broad excitation data up to 25 meV. Both $J_{a1}$ and
$J_{a2}$(4th row of Table \ref{Tab1}) were found to be much stronger
and consistent with theory.
As mentioned above,
for
$\alpha>$1/4, the ground state is an 
AFM
spiral state. 
In our previous study \cite{kuzian} we
found $\alpha\sim0.19$, 
below 
the critical value 
%
of a single chain.
In order to determine the overall profile of the 
magnon
dispersion 
and refine also
$\alpha$, we performed INS experiments using a time-of-flight 
chopper spectrometer. This way, we probed
the full dispersion 
that extends up to $\sim$53 meV and $\alpha$ was refined
as
$\sim$0.23, closer to $\alpha_c$.

As previously observed \cite{Matsuda2001}, the intensity of the
magnons appears 
to be reduced at $\sim$11.5 meV. In addition to this, we also found 
another gap-like behavior at $\sim$28 meV. We refine
the exchange parameters and discuss the origin of the
gap-like behavior in the 
magnon dispersion. The gap at $\sim$11.5 meV is related 
to the coupling with a weakly dispersive optical phonon.
The gap-like feature at $\sim$28 meV is ascribed to
quantum effects due to the 
AFM ICs \cite{Matsuda2001} and to the
SS 
mentioned above.

\section {Experimental method}
A CYCO single crystal 
was grown by the traveling solvent floating zone (TSFZ) method in air. 
The dimensions of the rod shaped crystal was $\sim$6$\Phi\times$25 mm$^{3}$. 
This crystal was already used in previous INS 
studies \cite{Matsuda2001,matsuda052,matsuda05,kuzian}. 
The present INS experiments were carried out on a hybrid neutron 
spectrometer HYSPEC \cite{HYSPEC} installed at the Spallation Neutron 
Source (SNS) and a triple-axis spectrometer HB-1 installed at the High 
Flux Isotope Reactor (HFIR) at Oak Ridge National Laboratory (ORNL). 
We utilized two incident energies of 27 and 60 meV on HYSPEC. 
Energy resolutions at the elastic position are $\sim$1.3 and $\sim$3.8~meV 
with $E\rm_i =27$ and 60~meV, respectively.
Neutrons with a final energy of 13.5~meV were used, together 
with a horizontal collimator sequence of $48'$--$80'$--S--$80'$--$240'$ on HB-1. 
The energy resolution at the elastic position 
amounts to $\sim$1.4 meV. Contamination from higher-order 
beams was effectively eliminated using pyrolytic graphite filters. The single crystal was 
oriented in the $(H,K,0)$ scattering plane and 
mounted in a 
closed-cycle $^4$He gas refrigerator on HYSPEC. On HB-1, the single crystal 
was oriented in the $(H,K,0)$ and $(H,0,L)$ scattering planes and 
mounted 
in a closed-cycle $^4$He gas refrigerator. The visualization of 
the HYSPEC data were performed using
the DAVE software \cite{DAVE}.
\section {Experimental results 
\& spin-waves} 
\subsection{Interchain couplings (ICs)}
As 
mentioned 
above, the previously studied
 dispersion along the
$(0, 0, l)$ direction  provides only the sum of ICs in the $ac$-plane 
$J_{s}=J_{ac1}+J_{ac2}\approx 2.24$~meV and the anisotropy parameter
$D=-0.45$~meV \cite{kuzian}. As pointed out in Ref.\ \onlinecite{kuzian}, 
the weak ICs can be fitted
more accurately from the dispersion 
relations at ($h$, $k$, $l$) with $h\ne$0 and any $k$ value, where the inchain 
couplings do not contribute. Hence, we probed the magnon
\begin{table}[t]
\caption{\label{Tab1} The inchain couplings $J_{a1}$ and $J_{a2}$ from INS data analyzed within
 LSWT, the maximum energy ($W_E$) below which the INS data were fitted,
  and $\alpha=|J_{a2}/J_{a1}|$. Values in the first row represents
   theoretical predictions from Ref.\ \onlinecite{mizuno}.}
\begin{tabular}{cccccc}
\hline
year & $J_{a1}$  & $J_{a2}$ & $W_E$  & $\alpha$ & Ref.\ \\
      & (meV) & (meV)& (meV) & & \\
\hline
1998 &  $-$2.2& 4.7 & --& 2.2& \cite{mizuno} \\
2001 &  $-$8 & 0.4& 10 & 0.05 & \cite{Matsuda2001} \\
         &  $-$6.9& 0.0 & 10 & 0 & \cite{Matsuda2001} \\
2012 & $-$19.6 & 3.7&25& 0.19 & \cite{kuzian} \\
2019 & $-$24& 5.5 &53  & 0.23 & present work\\
\hline
\hline
\end{tabular}
\end{table} 
\begin{figure}[b!]
\vspace{-0.0cm}
\includegraphics[width=7.0cm]{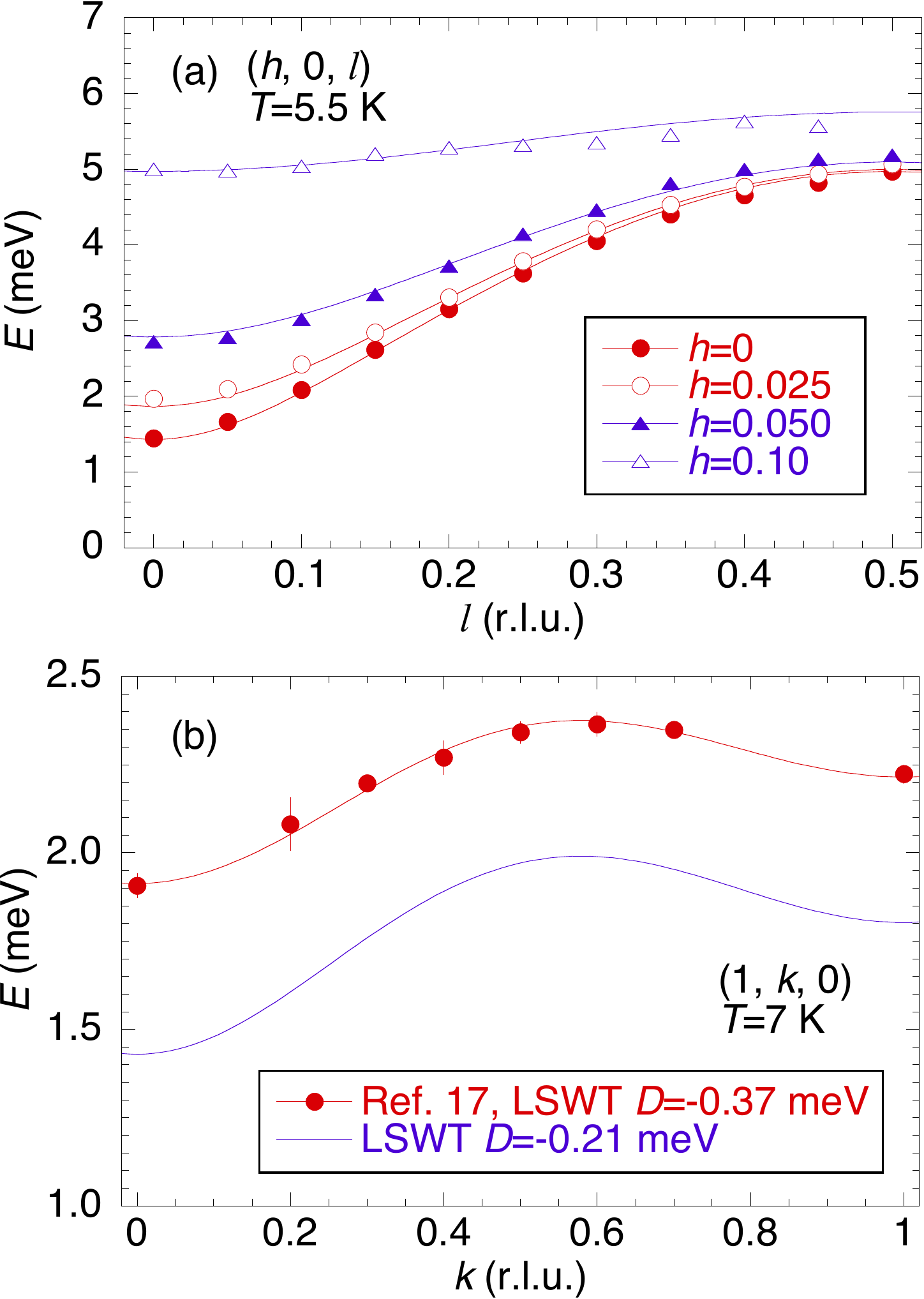}
\caption{(a) The weak magnon dispersion of 
CYCO perpendicular 
to the chain ($a$ axis) direction within the $ac$ plane [($h$, 0, $l$) with $h$=0, 0.025, 0.005, and 0.1], measured at $T$=5.5 K with $E\rm_f$=13.5 meV.
Solid curves: 
the dispersions 
calculated using
LSWT with the two skew ICs 
$J_{ac1}$=0.12, $J_{ac2}$=2.26, 
 and the anisotropy parameter $D$=$-$0.21~meV.
 The error bars are smaller than the size of the symbols.
 (b) The dispersion along the $b$ axis, which is well 
 reproduced with $J_b$=$-$0.0061, $J_{ab}$=$-$0.030, and $D_b$=$-$0.37 meV.
}
\label{dispersion_H}
\end{figure}
at ($h$, 0, $l$) with $h$=0, 0.025, 0.005, 
and 0.1 at $T$=5.5 K on HB-1. 
Its dispersion is
shown in Fig.\ \ref{dispersion_H}. 
The LSWT analysis 
yields a much larger NNN IC on the two adjacent
chains $J_{ac2}$=2.26 meV than the 
NN counter part
$J_{ac1}$=0.12 meV 
(cf.\ the ratio $\tilde{J}_1/\tilde{J}_2\sim$0.1 
in LICO
\cite{lorenz}).
The 
average anisotropic exchange parameter $D$ in Eq.\ (\ref{eq:D})
was found as $-$0.21~meV.

Analyzing the unusual intensity suppression near 30~meV with the aid of
density matrix renormalization group calculations (see\ Sec.\ \ref{SecQuantumGap}), 
we find that the  ICs would most effectively
suppress the intensity 
when $J_{ac1}\gg J_{ac2}$. 
This points to the imaginary part of the magnon 
self-energy $\Im \Sigma$, ignored in LSWT. It might cause
this different behavior while $\Re \Sigma$,
which governs 
the dispersion, is less sensitive to the ratio
$J_{ac1}/J_{ac2}$.

An almost flat dispersion along the $b$ axis 
is reproduced (Fig.\ref{dispersion_H}b) with the same coupling values
as in our previous
work \cite{Matsuda2001,matsuda05,kuzian}.
The difference of the anisotropy parameter $D_b$ from $D$ was 
explained in Ref.\ \onlinecite{Matsuda2001} by
a small deviation of  the spin-Hamiltonian's anisotropy 
from uniaxial symmetry; the deviation does not visibly split the spin wave branches
in the $ac$-plane and is not considered here. 
We mention the couplings as a useful reference 
for a realistic estimate of
the analogous
"face to face" interaction of CuO$_4$ plaquettes along the $a$ axis 
in LICO in view of its role in the FM alignment of magnetic moments 
along the chains under debate \cite{shu,Kuzian2018} and an order by disorder
scenario \cite{Xiang2007} vs.\ the AFM IC 
mechanism based on shifted adjacent chains 
in Refs.\ \cite{lorenz,Nishimoto2011,Nishimoto2015}.
To resolve this problem experiments around $\alpha_c$
would be helpful. In particular, an INS study under pressure and an 
analysis like ours
would be interesting in view of the 
pressure study of LICO \cite{You2009,Li2011}, where above 6 GPa a phase transition to
a monoclinic FM phase has been detected. Due to the larger IC coupling in CYCO higher
pressures might be necessary for a similar transition. Hence,
studies in La$_6$Ca$_8$Cu$_{24}$O$_{41}$ might be easier to perform
although any analysis of the dispersive magnon modes could be
difficult due to the presence of ladder spinons.
Anyhow, 
pressure 
is a 
promising tool
(see\ also the estimate of $J_1$ and $\alpha$ in Sec.\ E of SM \cite{Suppl}). 

\subsection{Inchain interactions}
The analysis given above rests on the 
assumption of {\it flat homogeneous} CuO$_2$ chains, practically unaffected by the 
incommensurate structure of the adjacent cationic CaY chains, as described in Sec. II. The opposite 
is depicted in Fig.~\ref{structure}(c) for the simplest case 
when a symmetric quasi-period 5 SS is induced in the cuprate
chain. A period 10 or 15 would give an even better approximation for
the incommensurate SS induced by the strong Coulomb interaction
between the differently charged cations and especially the closer O ions
of the CuO$_2$ chains. The latter case might be close to an inhomogeneous "lock-in" structure
containing "domains" of period 5 and period 10 units as well.
Then the two mechanisms of gap production proposed here would be cooperative,
resulting in a maximum experimentally observable effect.
Since in a general period-$m$ case 
the opening of $m-1$ gaps is allowed, in the present case
one is left with 
4 gaps for a period-5 model
while already 9 gaps for improved approximations of a period-10 (or even 14 gaps 
for a period-15)
are 
allowed. The simplest lock-in structure "5+10" 
has 13 gaps. 
The 
replacement of the incommensurate SS 
by a quasi-commensurate 
one containing an
{\it even} period component
is
essential because it allows the opening of a gap just at the wave vector 
of 1/4 where 
the gap near 28~meV has been 
found. The magnitude of all 
gaps depends on the distributions
of the 
local 
$\Phi$ and
of the distances between NN bridging O ions 
and also on the twisting
and/or other deformations away from the flat structure of ideal chains 
as in LICO.
Also the dispersion is slightly affected by the opening of gaps if they are included
in the fitting procedure (see\ Sec.\ V B and Sec.\ A in SM \cite{Suppl}).
Due to the largely
increasing number of corresponding couplings probably any distribution of gap 
amplitudes at the corresponding wave vectors generic for the adopted approximative commensurate
SS could be fitted. In this context we do not exclude the possibility 
of a lock-in transition of the incommensurate cationic chain into a real long periodic commensurate 
5$m$ periodic-chain, where $m$= 2, 3, 4 .... Thus, any real progress by convincing fits
should rest on a dialog examination of
various local
structural models compatible with the diffraction
patterns from neutron and x-ray scattering.
Improved detailed microstructure models in the real space
for inhomogeneous long periodic 
CuO$_2$ chains
have not yet been developed. The examination of
such alternative microstructure models with increasing complexity is extremely tedious. 
Therefore, it is
for a future study. Anyhow, we believe 
that the analysis of the gaps reported here
is very important to find effective models with a reasonable number of parameters.

The unexpected lacking of Zhang-Rice 
excitons
in a recent resonant inelastic x-ray scattering (RIXS) study
 \cite{Schmitt2017}, 
in sharp contrast to LICO
and
CuGeO$_3$ with 
"ideal" 
chains, is noteworthy. 
We suggest that the 
expected peak-like feature could not be resolved experimentally
due to a relatively broad distribution (more than 0.5~eV) of 
different "local" excitation energies 
(at 4.5~eV in LICO) 
caused by
the SS in CYCO.
Further consequences of the 
composite symmetry of CYCO, such as
suggested in Refs.\ \onlinecite{Thar2006,yamaguchi,Gotoh,Wizent2}
or within the 
approaches
proposed here, will be discussed elsewhere.

\begin{figure}[b]
\vspace{-0.5cm}
\includegraphics[width=8.0cm]{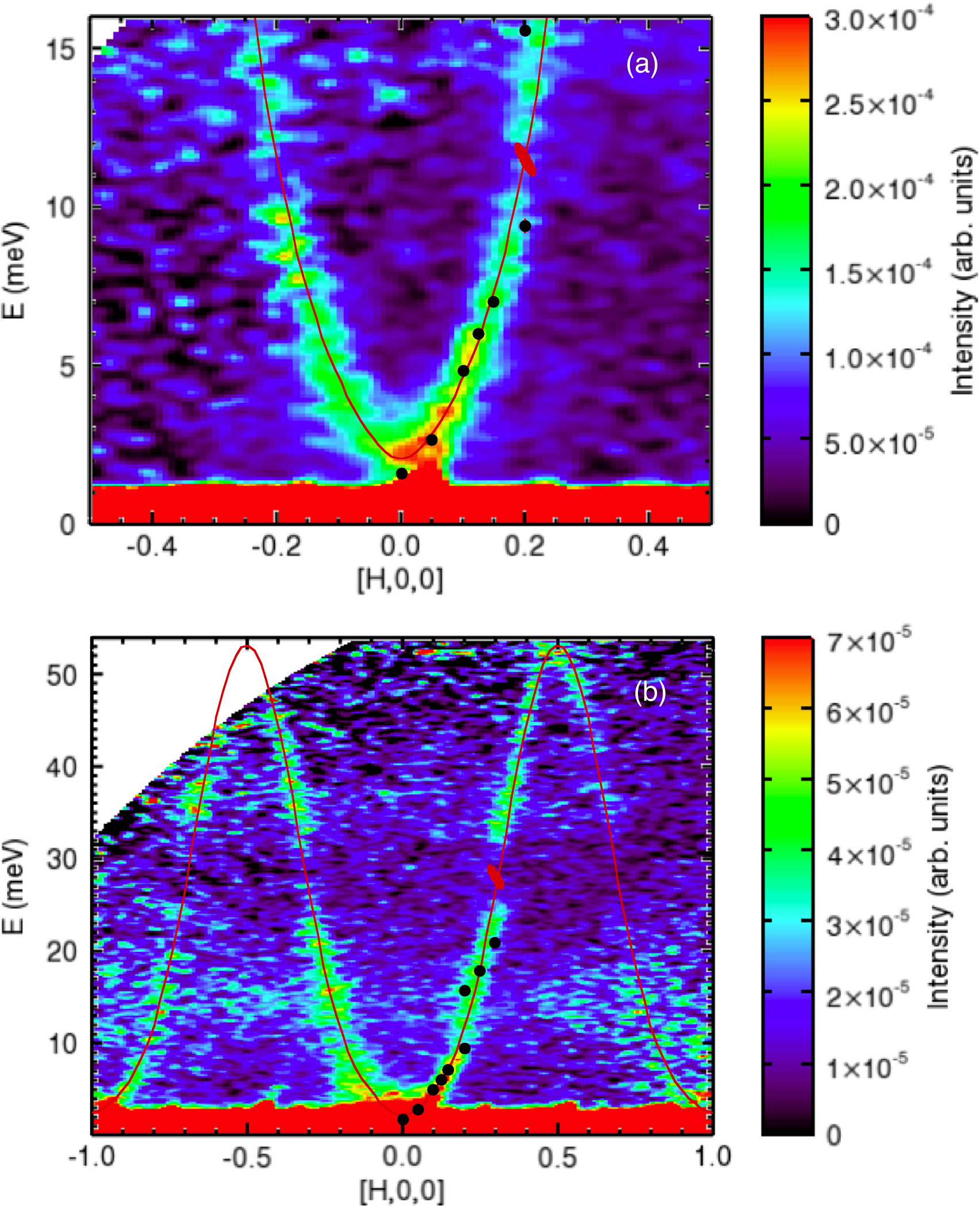}
\caption{Contour maps of the INS 
intensity $S(Q, E)$ for a CYCO
single 
crystal measured 
at 6~K with $E\rm_i$=27 meV (a) and 60 meV (b). 
Energy resolutions at the gap energies are estimated 
to be $\sim$0.7 meV at 11.5 meV with 27 meV $E\rm_i$ (a) and $\sim$2 meV 
at 28 meV with 60 meV $E\rm_i$ (b). The resolution volumes projected to the $E$-$Q$
space are shown around the gap energies with red ellipses.
Filled circles: data points reported in 
Refs.\ \cite{Matsuda2001,matsuda052}.
Solid curves: the dispersion relation calculated 
using LSWT with 
$J_1$=$-$24, $J_2$=5.5, 
$J_{ac1}$=0.12, $J_{ac2}$=2.26, and $D$=$-$0.21 meV.
}
\label{dispersion}
\end{figure}
\begin{figure}[b]
\includegraphics[width=8.0cm]{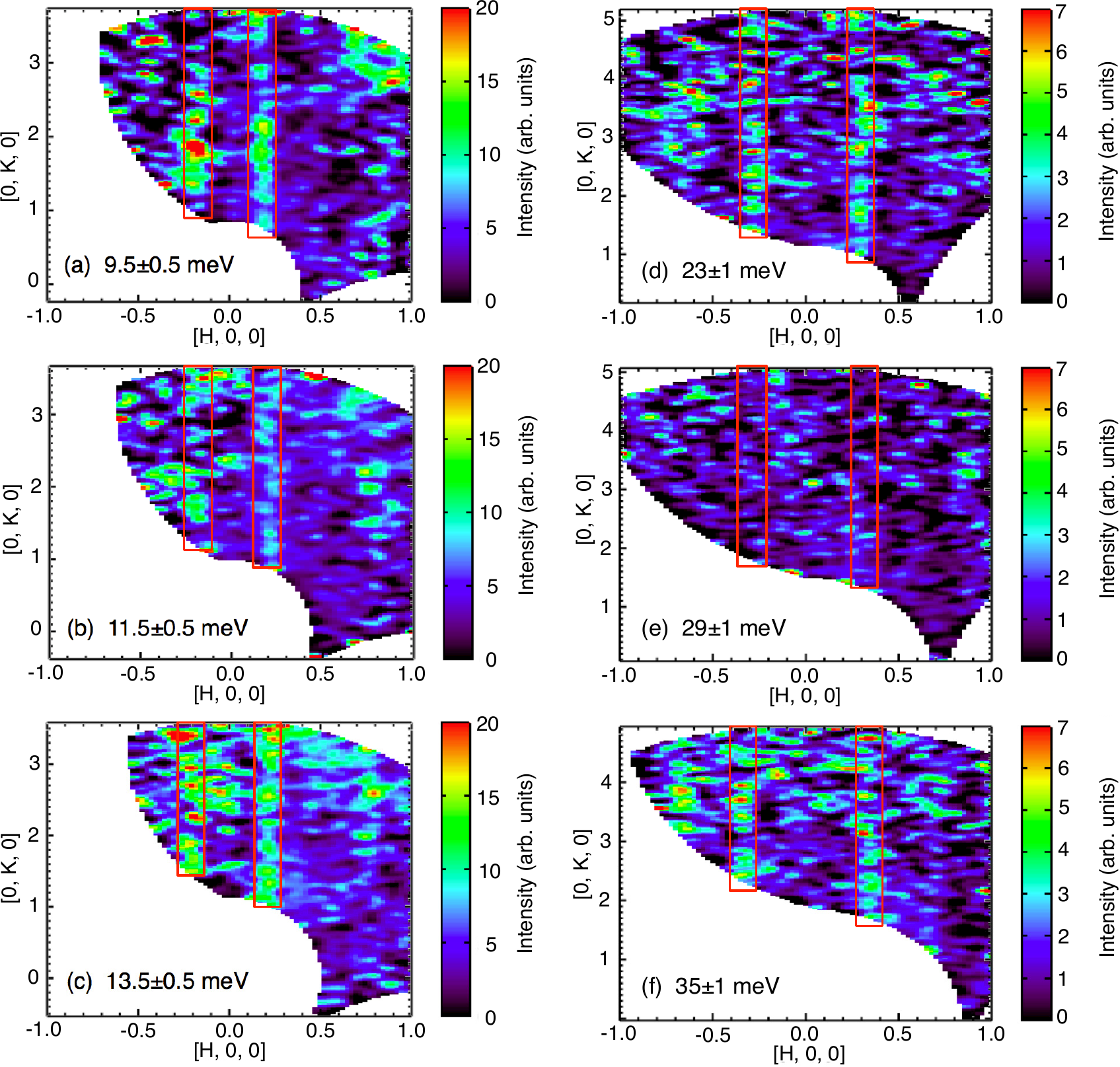}
\caption{Energy cuts of the INS intensity in 
the $(HK0)$ plane measured at 6 K. Spectra at 9.5 (a), 11.5 (b), 
and 13.5 meV (c), measured with $E\rm_i$=27 meV and integrated 
in the range of $-$0.1$\le L\le$0.1. Spectra at 23 (d), 29 (e), and 
35 meV (f), measured with $E\rm_i$=60 meV and integrated in 
the range of $-$0.2$\le L\le$0.2. Red rectangles: the expected regions 
for line-shaped magnetic excitations along $K$.}
\label{Ecut}
\end{figure}
Figure\ \ref{dispersion} shows the INS spectra 
$S(Q,E)$ from our CYCO single 
crystal measured at 6~K. 
Figure\ \ref{dispersion}(a) represents the low-energy excitations measured 
with $E\rm_i$=27 meV. The intensity is averaged over the range of 
1.8$\le K \le$ 3.2 and $-$0.1$\le L\le$0.1. The 
magnon dispersion along 
$K$ is almost flat and the band width is less than 0.2 meV \cite{Matsuda2001}.
Although the band width of the dispersion along $L$ is about 3 meV, the 
dispersion in the range $-$0.1$\le L\le$0.1 is less than 0.5 meV \cite{Matsuda2001}.
Therefore, the broadening due to the integration should be small. On the 
other hand, the scattering intensity with $E\rm_i$=60 meV was weak. In 
order to improve the statistics, the signal was integrated in a wide $Q$ range. 
In Fig.\ \ref{dispersion}(b), the intensity is averaged over $-$0.2$\le L\le$0.2, 
where the dispersion width is less than 1.5 meV, and entire $K$ range measured. 
The range of $K$ depends on the
excitation energy, e.g., 0.5$\le K \le$ 5.5 at 5 
meV and 3.0$\le K \le$ 4.7 at 51 meV. Therefore, the effective magnetic 
form factor gradually decreases with increasing energy and $H$ value, 
which reduces the averaged intensity. However, the overall dispersion 
curve can be generated with reasonably good statistics by this method. 
Figure\ \ref{dispersion}(b) clearly shows a single branch mode along $H$. 
The characteristic feature is that there are gap-like features 
at $\sim$11.5 meV and $\sim$28 meV, as shown in 
Fig.\ \ref{dispersion}(a) and (b), respectively.
\begin{figure}[t]
\includegraphics[width=6.5cm]{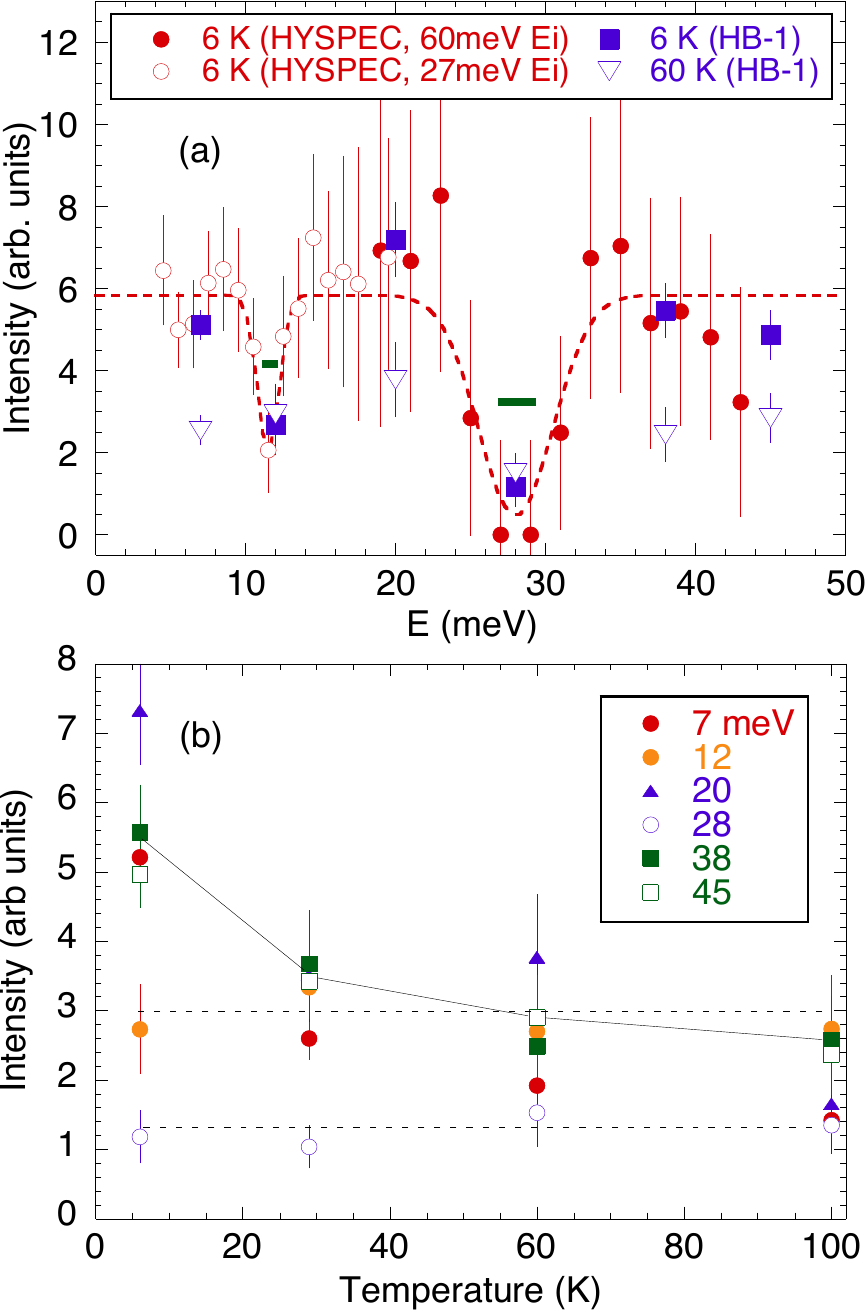}
\caption{Energy (a) and $T$ (b) dependences of the integrated 
intensities 
from constant energy cuts and scans. 
Solid and broken lines: 
guides to the eye. The thick horizontal bars 
near 11.5 and 28 meV in (a) are estimated instrumental resolution.
}
\label{summary}
\end{figure}
The observed magnon dispersion has been analyzed with the help of
LSWT (at $T=0$). For this purpose we 
have used
Eq.\ (5).
We fixed the ICs 
determined in Sec.\ III A ($J_{ac1}$=0.12, $J_{ac2}$=2.26, and 
$D$=$-$0.21 meV). The small $J_{c}$ was fixed at 0 meV \cite{Matsuda2001} 
for 
simplicity, since the dispersion shown in Ref.\ \onlinecite{Matsuda2001} yields 
tiny values of
$-$0.061 and 0.037~meV for the NN and NNN couplings, respectively. 
$J_{a1}$ and $J_{a2}$ were determined from the dispersion 
along $H$. $J_{a1}$ affects the magnon band width and 
$J_{a2}$ the dispersion shape in the low-energy region. 
We found that $J_{a1}$=$-$24 and $J_{a2}$=5.5 meV reproduce the overall 
dispersion, as shown in Figs.\ \ref{dispersion}(a) and (b). 
The resulting 
$\alpha$=0.23 is close to the critical $\alpha_{\rm \tiny c}$=1/4 
(see\ 
also Sec.\ E in SM \cite{Suppl} 
for the non-criticality of coupled chains).

Noteworthy, 
the increase by a factor of 3 of 
$|J_{a1}|$ 
found over the years 
(see Table 1). Thereby
the NNN inchain exchange
$J_{a2}\equiv J_2$ has been strongly raised
too, while 
$\alpha$ shows a more moderate increase. 
Possible disorder effects on the enhancement of $J_2$
in CYCO, Li$_2$ZrCuO$_4$, and LiCu$_2$O$_2$ are shown in Sec.\ I of SM \cite{Suppl}.
The 
unusually small $\alpha$-values in Table 1 
reflect 
the previous non-optimal fitting 
due to
the large number of involved 
couplings.
 This at first glance
surprising result is now well understood.
The increase of $|J_1|\equiv |J_{1a}|$
by 
$\sim$100~K as compared to 
that from a still non-optimal 
fit \cite{kuzian} is very instructive. 
Good fits can be 
achieved only by 
probing the full dispersion, $i.e.$ up to
energies
$E \geq 2 ]J_1|$,
if
the 3rd neighbor couplings
($J_3\equiv J_{3a}$) are reasonably small \cite{lorenz}.
 
\subsection{Gap-like features}
Figure \ref{Ecut} shows six energy cuts in the $(HK0)$ plane through 
$S(Q, E)$, measured with $E\rm_i$=27 meV (a)-(c) and $E\rm_i$=60 meV (d)-(f). 
Since the dispersion is almost flat along $K$, there are line-shaped 
dispersions along $K$, as indicated by red rectangles. Around the gap 
energies 11.5 and 29~meV, the intensity becomes weak throughout the 
whole $K$ range, indicating that the structure factor is modified 
considerably at these specific energies. In particular, the signal 
is very weak at 29~meV.

To show the intensity more 
quantitatively,
the scattering intensity 
was plotted as a function of the excitation energy, as shown in Fig.\ \ref{summary}(a).
The intensity was obtained by fitting the constant energy cut profile with 
a Gaussian function. In this plot, the integration ranges are 
 $-$0.1$\le L\le$0.1 and 1.5$\le K\le$2.5 for 27 meV $E_i$ data 
and $-$0.2$\le L\le$0.2 and 2.5$\le K\le$3.5 for 60 meV $E_i$ data. 
The intensities from the two sets of data are normalized using the 
data points around 20 meV. The correction of the inverse spin-wave 
velocity was made to convert from the $Q$ integrated intensity to the 
energy integrated $S(Q, E)$, plotted in Fig.\ \ref{summary}(a).
The gap-like behavior is distinct at 11.5 and 28 meV. The dip is broader
at 28 meV than at 11.5 meV, probably because of the combined effect of
broader energy resolution ($\sim$2 meV) and wider integration range 
along $L$, with dispersion width of $\sim$1.5 meV for the 60 meV $E_i$ 
data. The broader energy resolution with 60 meV $E_i$ also makes the
gap at 11.5 meV smeared out [Fig. \ref{dispersion}(b)], where 
the energy resolution is $\sim$2.5 meV.
Except the two dips, the intensity is almost constant throughout 
the whole energy range, which is expected for ferromagnets.
\begin{figure}[b]
\center{\includegraphics[width=7.5cm]{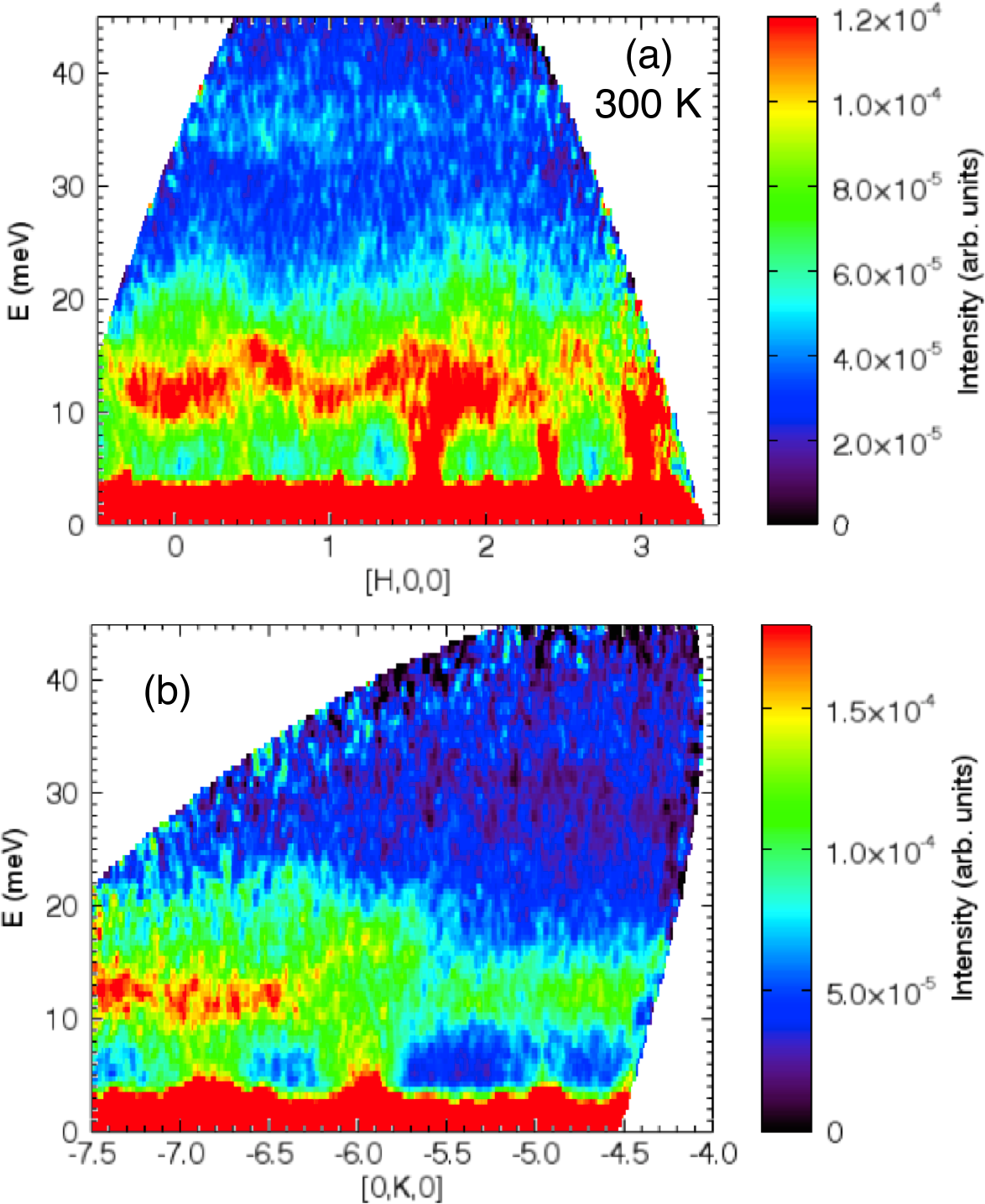}}
\caption{Contour maps of the phonon 
dispersions along $H$ (a) and $K$ directions (b) measured at 300 K 
with $E\rm_i$=60 meV. The intensities plotted in (a) and (b) are 
integrated in the range of $-$6.5$\le K\le$$-$4.5 and $-$0.5$\le H\le$0.5, respectively.
}
\label{phonon}
\end{figure}
One possibility to explain the gap behavior is the phonon-magnon coupling. 
Magnons can be interfered when a phonon mode is mixed. 
A magnon gap due to such an effect was actually reported 
in UO$_2$ \cite{cowley,caciuffo,caciuffo2011}, FeF$_2$ \cite{rainford,lovesey}, and 
La$_{1-x}$Ca$_x$MnO$_3$ \cite{hennion}. A gap behavior in the magnon 
dispersion was also reported for magnetite Fe$_3$O$_4$ below 
the Verwey transition temperature ($T$), where charge ordering is expected \cite{mcqueeney}. 
The acoustic magnon mode shows a gap at 43 meV and 
$q$=(0, 0, 1/2). The origin of the gap is still unknown, although both a
charge-density wave and magnetoelastic coupling are considered as possible causes.
We examined the phonon dispersions of CYCO carefully. 
A weakly modulating optical phonon mode along both $H$ and $K$ directions 
is observed around 11.5 meV, as shown in Fig.\ \ref{phonon}. This phonon 
can interfere with the magnon around 11.5 meV.
If there exists strong 
magnon-phonon 
coupling, a bending of the dispersion curve is usually observed as well 
as an excitation gap \cite{caciuffo2011}. No such bending was observed 
in the present measurements. As mentioned in Sec.\ III, the magnetic signal 
from the small magnetic moment ($S$=1/2) in CYCO is weak so that we need to integrate the signal 
in a wide range of $Q$ region to clearly show the dispersion curve. 
This integration is likely to make the bending unclear since the optical 
phonon mode around 12 meV is slightly dispersive.
In stark contrast, near 28 meV there is no phonon mode which would
intersect the magnon dispersion (Fig.\ \ref{phonon}).
Hence,
the latter gap 
cannot
be 
ascribed to phonon-magnon coupling.
As shown above,
the CuO$_2$ chains are
distorted due to the misfit with
the Ca-Y layer. Since the O distortions suggested in Ref.\ \onlinecite{Thar2006} are not so small, some changes of the dynamical structure factor and of the spin-wave dispersion might occur.
Our 
LSWT calculations suggest 
nevertheless a weak change in the dispersion starting from a homogeneous chain 
but more pronounced changes 
in the intensity leading even to the opening of quasi-gaps
have been found (see\ also Sec.\ A in SM \cite{Suppl} and below). Much more 
systematic studies of inhomogeneous 
models including
also Dzyaloshinskii-Moriya (DM) couplings [allowed in that case] 
and experimental 
refinement of the structural model are desired
to settle quantitatively 
this 
very complex problem. 
Similar studies for 
CYCO 
might be of interest too.
As first 
insights we show in Sec.\ V B
and in Sec.\ A of 
SM \cite{Suppl} the effect of various simple inhomogeneities.
\begin{figure}[b]
\includegraphics[width=7.9cm]{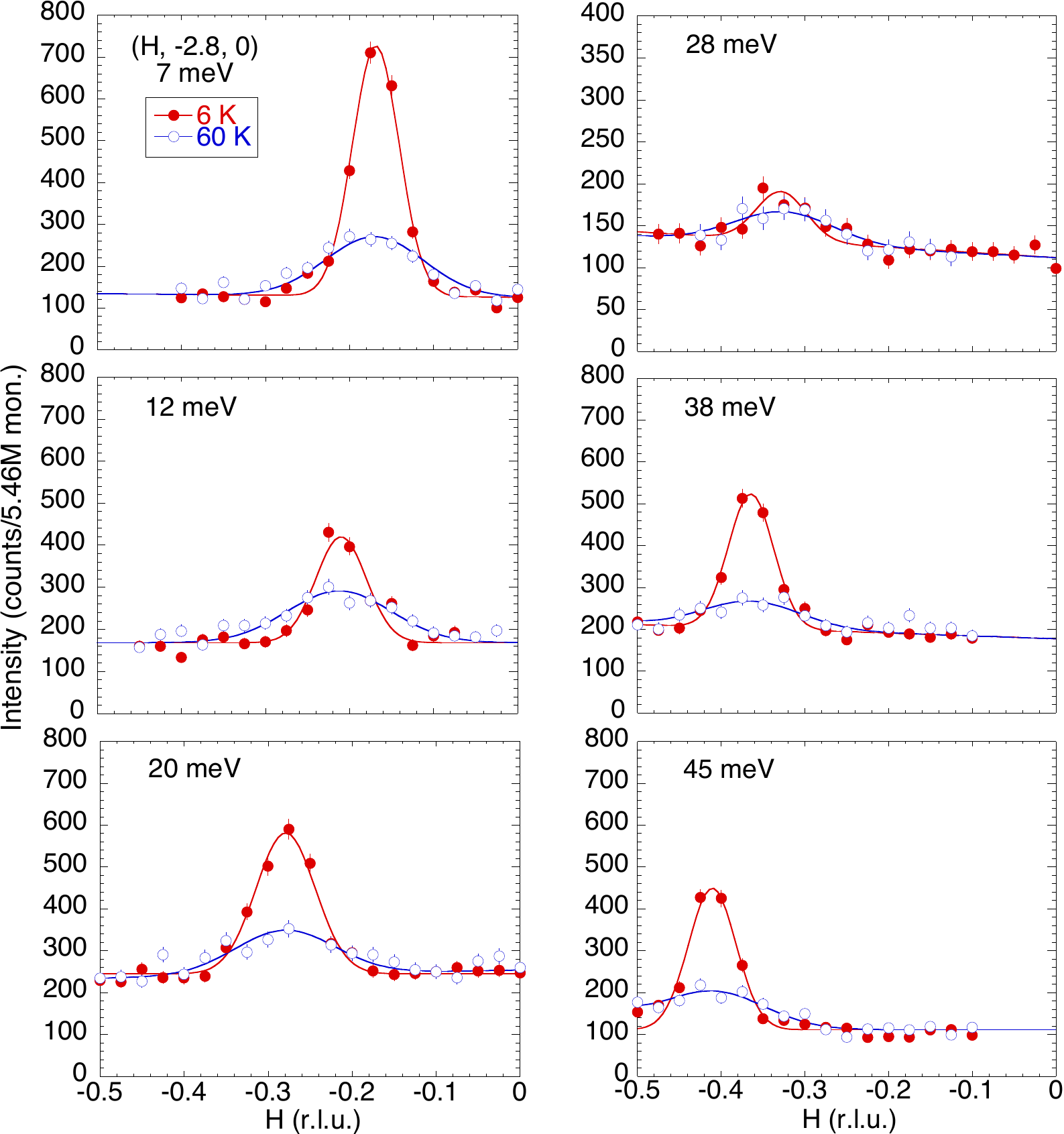}
\caption{Constant energy scans at ($H$, $-$2.8, 0) measured at $E$=7, 12, 
20, 28, 38, and 45 meV at $T$=6 and 60 K. 
Solid lines are the results 
of fits with a Gaussian function. r.l.u. represents 
reciprocal lattice units.
To emphasize the peak structure, the vertical scale of the 28 meV 
data is different from the others.
}
\label{HB1}
\end{figure}
Since $J_{a1}$ amounts to $-$24 meV (278 K), which is much larger than $T\rm_{N}$=29.5 K, 
a steep 
magnon dispersion is still expected above $T\rm_{N}$ along the chain. 
It is also expected that the dominant IC $J_{ac2}$(=2.26 meV) becomes less effective above $T\rm_{N}$. 
Therefore, the effect of the ICs can be elucidated by checking 
whether the gap-like behavior persists above $T\rm_{N}$. Figure\ \ref{HB1} 
displays constant-$E$ spectra at 7, 12, 20, 28, 38, and 45 meV, measured 
at 6 and 60 K on HB-1. The magnetic excitations persist even at 60~K, although 
the peak width becomes broader. The change of intensity depends on energy. 
The $T$-dependence of the integrated intensities is shown in 
Fig.\ \ref{summary}(b). Clearly, the intensities at 12 and 28 meV are 
$T$-independent. On the other hand, those at other energies decrease 
with increasing $T$. The integrated intensities at 6 and 60 K are 
shown as a function of energy in Fig.\ \ref{summary}(b). The intensities at 6 K 
are consistent with those measured on HYSPEC. Since the intensities at 12
and 28 meV are unchanged and those at other energies are reduced, the gap-like behavior 
becomes less distinct at 60~K. Hence,
the gap-like behavior 
at 28 meV is affected by the AFM ICs and
is in fact a quantum effect specific for FM chains, as suggested above.
The gap-like behavior at 11.5 meV may be due to 
magnon-phonon mixing, as mentioned above. 
Then, 
the
magnon-phonon coupling strength might be also
weakened at $T>T\rm_{\tiny N}$.

In view of the recently found
 strong renormalization of the charge transfer energy
$\Delta_{pd}$ in the RIXS spectra of LICO
by 
high-frequency O derived modes at 74~meV \cite{Johnston2016} and near 70~meV
for Ca$_{2+5x}$Y$_{2-5x}$Cu$_5$O$_{10}$ with $x$=0, 0.3, and 
0.33 \cite{Lee2013} (the latter 
two being 
hole ($h$)-doped derivatives of CYCO),
the present 
observation for another active 
phonon at much lower energy is 
interesting and deserves 
to be analyzed also in the 
general context of 
electron-phonon
coupling
in strongly correlated systems. Here, INS
brings a new low-energy scale not resolved in RIXS studies.
Since there is no low-energy gap near 11~meV seen in the INS data of 
Ref.\ \onlinecite{lorenz} for 
LICO,
we suggest that it might be an optical phonon
derived from the diatomic Ca/Y chain. Then a down shift of that phonon induced
gap-like feature might be expected for the sister compound 
Ca$_2$Nd$_2$Cu$_5$O$_{10}$, having a slightly reduced $T_N=24$~ K \cite{Wizent},
and even better with nonmagnetic isovalent substitutions of Y with Lu or Sc.
The insertion of magnetic rare earth ions 
provides
additional insight into the cuprate magnetism due to the 
interplay with the high-spin
rare earth subsystem
and the check of intrinsic quantum effects.
Substitution with Pr might also modify 
the magnetic structure of the chains through a decrease
of the O hole numbers $n_p$ due to the competing covalency with Pr 4$f$ electrons,
as it happens in PrBa$_2$Cu$_3$O$_{7-\delta} $\cite{Fehrenbacher}, with 
dramatic consequences for the corresponding $pd$
exchange integrals [Eqs.\ (5,6) in Ref. \cite{Fehrenbacher}]. In particular, a strong decrease
of $J_1$ might occur, if sizable O 2$p$-Pr $4f$ covalency is present. 
An 
ordering of the rare earth 
magnetic moments well above
few K (typical for dipole-dipole couplings) 
in the 
Pr-based quasi-2D cuprates
might explain details of the magnetic response in Ca$_2$Nd$_2$Cu$_5$O$_{10}$ \cite{Wizent}. A systematic
study of the whole rare earth series would be 
interesting.
To the best of our knowledge, Ca$_{2}R_{2}$Cu$_5$O$_{10}$ ($R$=Dy and Gd)
have been synthesized 
but their physical properties were not studied so far.

\section {Theory}
In addressing the  main experimental findings, 
this section consists of three parts.
A and B are devoted to two different phenomenological simulations
of the detected mid gap, while part C deals with
microscopic aspects and consequences of the observed large magnon dispersion.
\vspace{0.3cm}
\subsection{The mid gap-like feature as a quantum effect
from diagonal AFM interchain coupling}\label{SecQuantumGap}
To understand the gap-like behavior around 28~meV for flat chains, we have first
calculated the dynamical spin structure which corresponds to the
experimental INS affected also
by the form factor. The former
is defined as
\begin{equation}
S(q,E) = \sum_\nu
|\langle \psi_\nu |S^\pm_q| \psi_0 \rangle|^2 \delta(E-E_\nu+E_0),
\label{spec}
\end{equation}
where $S_q^\pm$ is the Fourier transform of the spin-flip operator $S_i^\pm$ 
at site $i$ while $| \psi_\nu \rangle$ and $E_\nu$ are the
$\nu$-th eigenstate and eigenenergy of the system, respectively 
($\nu=0$ corresponds to the ground state). Two-chain clusters
($32 \times 2$ sites) were studied by using the Dynamical Density Matrix Renormalization Group (DDMRG) method~\cite{eric02}. 
Open boundary conditions were applied along the chain
direction whereas periodic boundary conditions were applied perpendicularly 
to the chain axis. Then, for an effective two-chain model,
ISs are taken to be $2J_{ac1}$ and $2J_{ac2}$ instead of $J_{ac1}$ and $J_{ac2}$.
\begin{figure}[b]
\includegraphics[width=7.0cm]{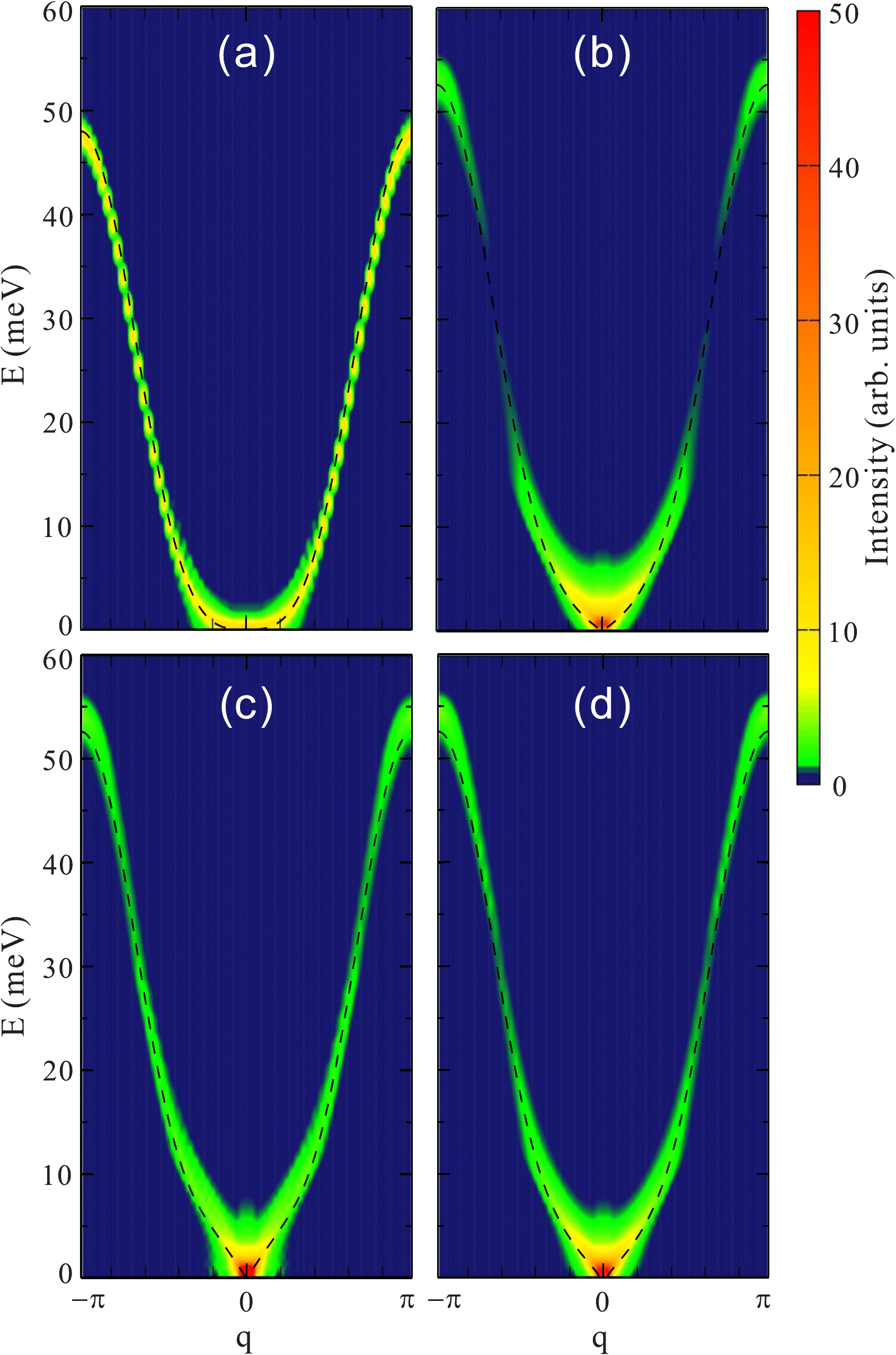}
\caption{DDMRG results of the dynamical structure factor $S(q,E)$ with $J_1=-24$, $J_2$=5.5 meV
for (a) $J_{ac1}=J_{ac2}=0$ meV, (b) $J_{ac1}=2.29$, $J_{ac2}=0$ meV, (c) $J_{ac1}=0$, $J_{ac2}=2.29$ meV,
and (d) $J_{ac1}=1.537$, $J_{ac2}=0.763$ meV. The dotted lines denote 
the magnon dispersions $\omega_q$.
}
\label{DDMRG1}
\end{figure}
The obtained spectra for some sets of 
ICs are shown in Fig.~\ref{DDMRG1}.
The overall dispersion is well
described by LSWT 
with Eqs.\ (2-6)
and $J_{ac1}+J_{ac2} \approx 2.29$ meV~\cite{kuzian}. Especially for $J_{ac1}=2.29$ meV, $J_{ac2}=0$, a gap-like behavior
around $E=30$ meV is clearly seen. This gap position is close to the 
INS-value of 28 meV.
\begin{figure}[]
\includegraphics[width=6.0cm]{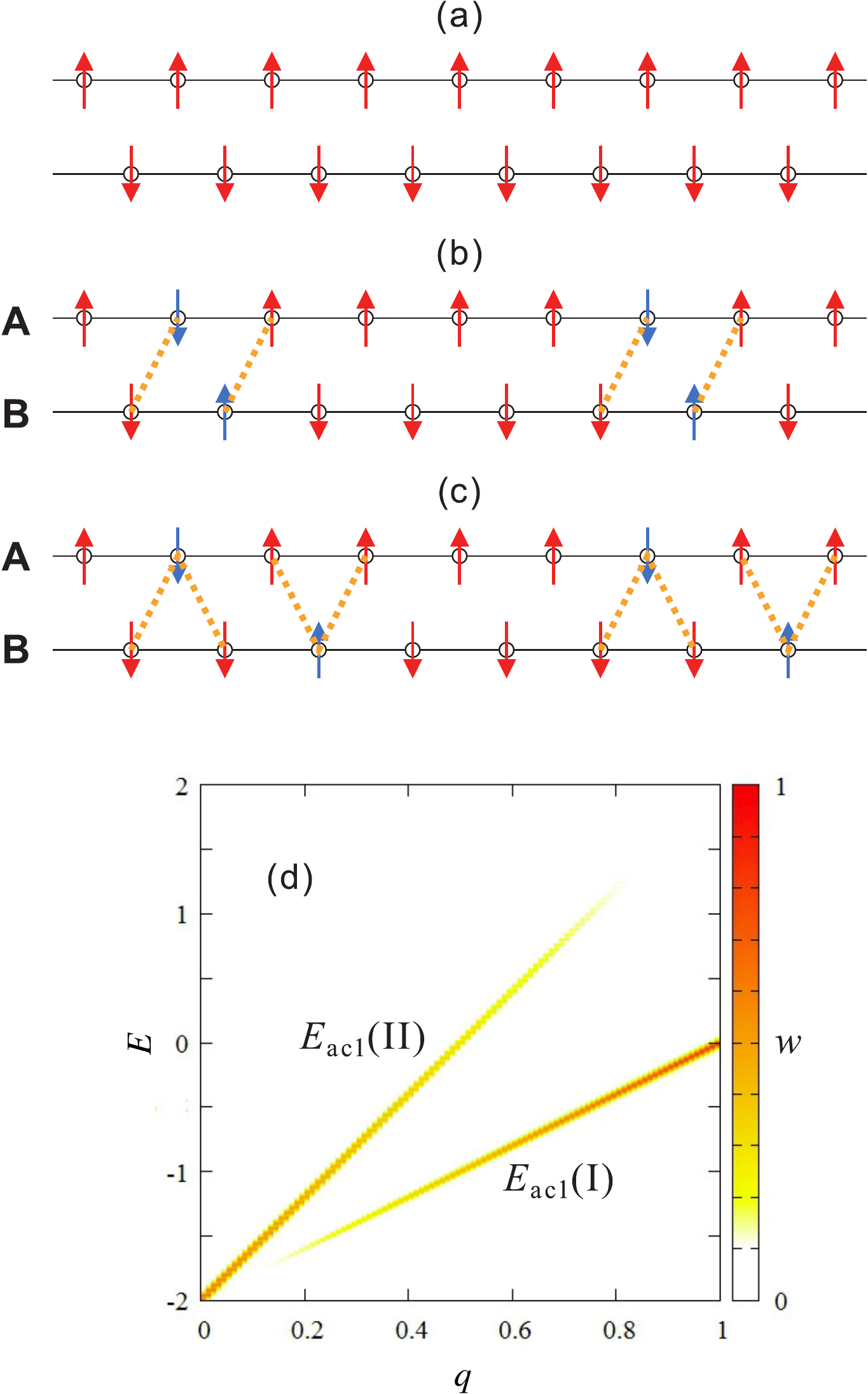}
\caption{Schematic spin configurations of (a) the ground state and (b)(c) excited states with $q=2\pi/5$. Parallel spins are
connected by dotted lines and the numbers of the parallel spin pairs are different between (b) and (c). (d) Energy contributions by IC $J_{ac1}$ and their weights in the excited states. $q$ is in a unit of $\pi$.
}
\label{DDMRG2}
\end{figure}
\begin{figure}[]
\includegraphics[width=5.4cm]{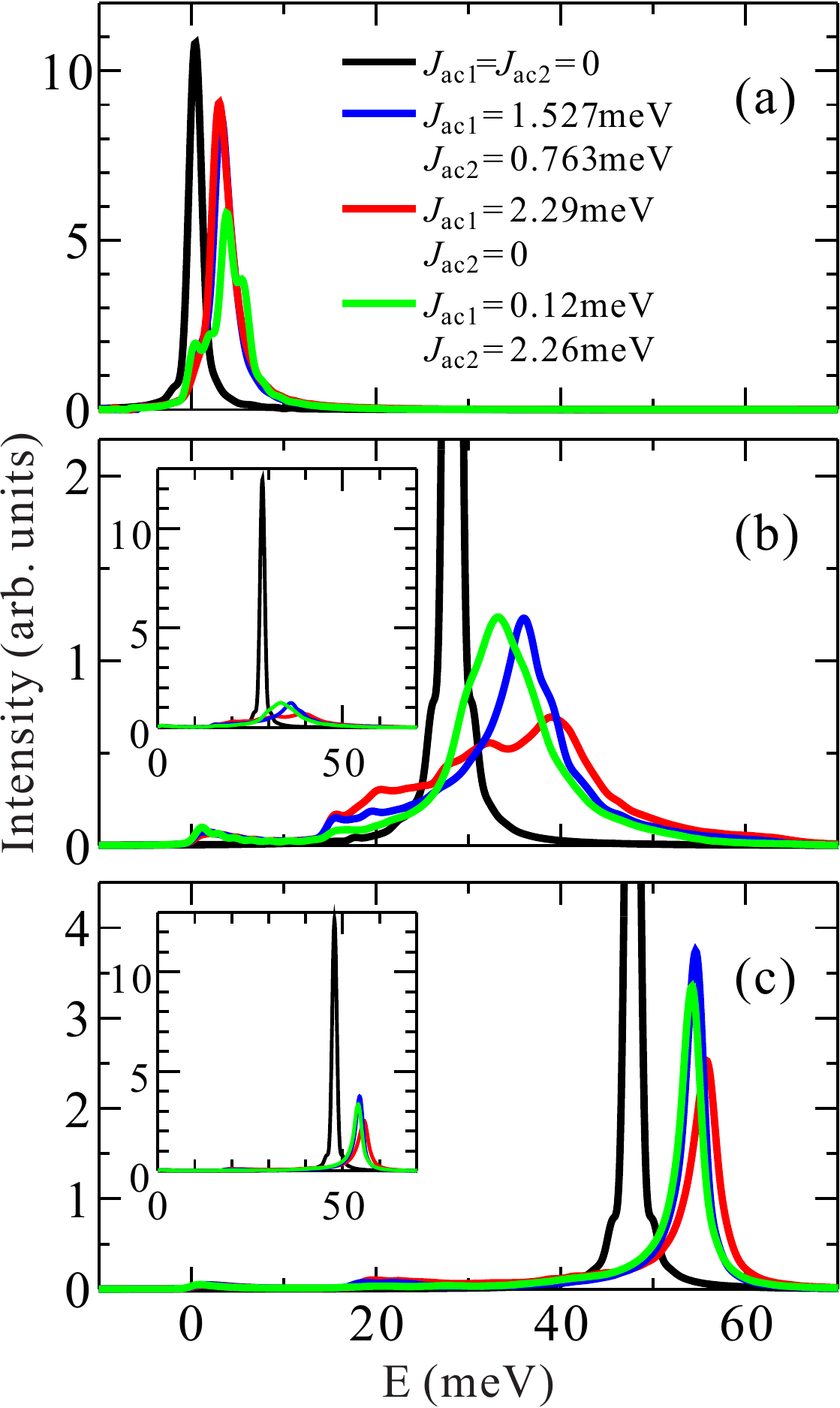}
\caption{DDMRG results of the dynamical correlation functions $S(q,E)$ at (a) $q=0.15\pi$, (b) $0.67\pi$,
and (c) $0.97\pi$. Insets display the entire range of the intensity.
}
\label{DDMRG3}
\end{figure}
The gap-like feature near 28 meV can be understood as a splitting 
of the excitation levels at an intermediate momentum $q
\sim \pi/2$, induced by finite ICs. Let us qualitatively 
illustrate this by considering the spin configurations at
$q=2\pi/5$. For simplicity, we take two chains coupled by $J_{ac1}$ and 
employ an Ising-like picture. A representative snapshot of the
ground state $| \psi_0 \rangle$ is schematically described in 
Fig.~\ref{DDMRG2}(a), where the spins are 
ferromagnetically 
aligned
along the chain and 
antiferromagnetically 
between adjacent chains within the $ac$-plane. Roughly speaking, 
the operators $S^\pm_{q=2\pi/5}$ flip spins on every
5th site on each chain. Thus, spin configurations like {\bf A} 
and {\bf B} in Figs.~\ref{DDMRG2}(b) and (c) are created in the
excited states $| \psi_\nu \rangle$ when $S^\pm_q$ is applied to the 
ground state $| \psi_0 \rangle$. Note that the energies 
differ between Figs.~\ref{DDMRG2}(b) and (c). The energy difference 
comes from interchain contributions which depend on the
relative positions of the flipped spins. On the other hand, the intrachain contributions 
are the same. For Figs.~\ref{DDMRG2}(b) and (c) the
interchain 
contributions per site are 
$E_{ac1}({\rm I})=-\frac{2}{3}J_{ac1}$ and $E_{ac1}({\rm II})=-\frac{1}{3}J_{ac1}$,
respectively. This gives a splitting of the excited energy levels $E_\nu$ 
in Eq.\ (\ref{spec}). The ratio of the probability weights
is 2:1 since it is 
proportional to 
the number of possible combinations of 
ICs. For arbitrary $q$ we obtain
$E_{ac1}({\rm I})=\left(\frac{q}{\pi}-1\right)J_{ac1}$ and $E_{ac1}({\rm II})=\left(\frac{2q}{\pi}-1\right)J_{ac1}$, weighted by
$w({\rm I})=\frac{q}{\pi}$ and $w({\rm II})=1-\frac{q}{\pi}$, respectively. 
In Fig.~\ref{DDMRG2}(d) we plot $E_{ac1}$ and $w$ vs.
$q$. The splitting is zero at $q=0$ and increases with increasing $q$. 
Although the splitting is largest 
near $q=\pi$,
it would be less represented in the spectral functions due to the polarized 
weights, i.e., $w({\rm I}) \gg w({\rm II})$. As a
result, such 
a splitting is most visible
around 
intermediate $q$.
To confirm this, we plot the dynamical correlation functions 
$S(q,E)$ at $q=0.15\pi$, $0.67\pi$, and $0.97\pi$ 
for several sets of the ICs in Fig.~\ref{DDMRG3}. Without 
ICs 
($J_{ac1}=J_{ac2}=0$) {\it no} splitting is seen for any $q$ [Fig.~\ref{DDMRG1}(a)].
But for finite ICs, 
the splitting is clearly confirmed at the intermediate momentum $q=0.67\pi$ [Fig.~\ref{DDMRG3}(b)]. 
This feature 
is most obvious for $J_{ac1}=2.29$ meV and $J_{ac2}=0$ [Fig.~\ref{DDMRG1}(b)].
The splitting 
causes a continuum by quantum fluctuations and it appears as a broadening of the 
intensity in the spectrum. At lower 
($q \approx 0$) and higher ($q \approx \pi$) momenta, the peak height is 
reduced by the ICs, 
{but they are still sharp, only with a 
slight broadening} [Figs.~\ref{DDMRG3}(a) and (c)]. The significant broadening at 
the intermediate momenta ($q \approx \pi/2$) provides a lack of $q$ 
integrated intensities at intermediate 
energies $E$, which corresponds to the experimental
dip of 
the $Q$ integrated intensity. The broadening 
is less pronounced for other ICs,
pointing to a
gap-like behavior 
most 
pronounced for a larger $J_{ac1}/J_{ac2}$ ratio. In fact, the gap-like 
feature is less obvious 
for $J_{ac2}>J_{ac1}$
[Figs.~\ref{DDMRG1}(c) and (d)].

\subsection{Gaps from inhomogeneous CuO$_2$ chains}
Here we briefly illustrate where gaps in the 
magnon curve
can appear
within the adopted period-10 scenario 
for the cationic Ca/Y chain system, ignoring thereby ICs
for 
simplicity. Let us assume that the lattice modulation leads to small 
deviations from the
ion positions in the flat homogeneous chain, i.e.\ $R+s\approx na$ ``in average''.
We consider a single chain where the structural modulations
cause an alternation of the exchange couplings. The spin-Hamiltonian
of a chain with a basis reads 
\begin{align}
\hspace{-0.3cm} \hat{H}_{ch} & =\frac{1}{2}\sum_{R,s,r_{s}}
\left[J_{r_{s}}\hat{\mathbf{S}}_{R+s}\hat{\mathbf{S}}_{R+s+r_{s}}\right. 
\left. +D_{r_{s}}\hat{S}_{R+s}^{z}
\hat{S}_{R+s+r_{s}}^{z}
\right],\label{eq:Hch}\\
& \approx
\sum_{R,s}\left[\varepsilon_{R+s}a_{R+s}^{\dagger}a_{R+s}
+
\frac{1}{2}\sum_{r_{s}}J_{r_{s}}a_{R+s}^{\dagger}a_{R+s+r_{s}}\right],\label{eq:Htb}\\
& \varepsilon_{s} \equiv\frac{1}{2}\sum_{r_{s}}\left(J_{r_{s}}
+D_{r_{s}}\right).\label{eq:epss}
\end{align}
where $R$ 
counts
\begin{figure}[b]
\includegraphics[width=1.00\columnwidth]{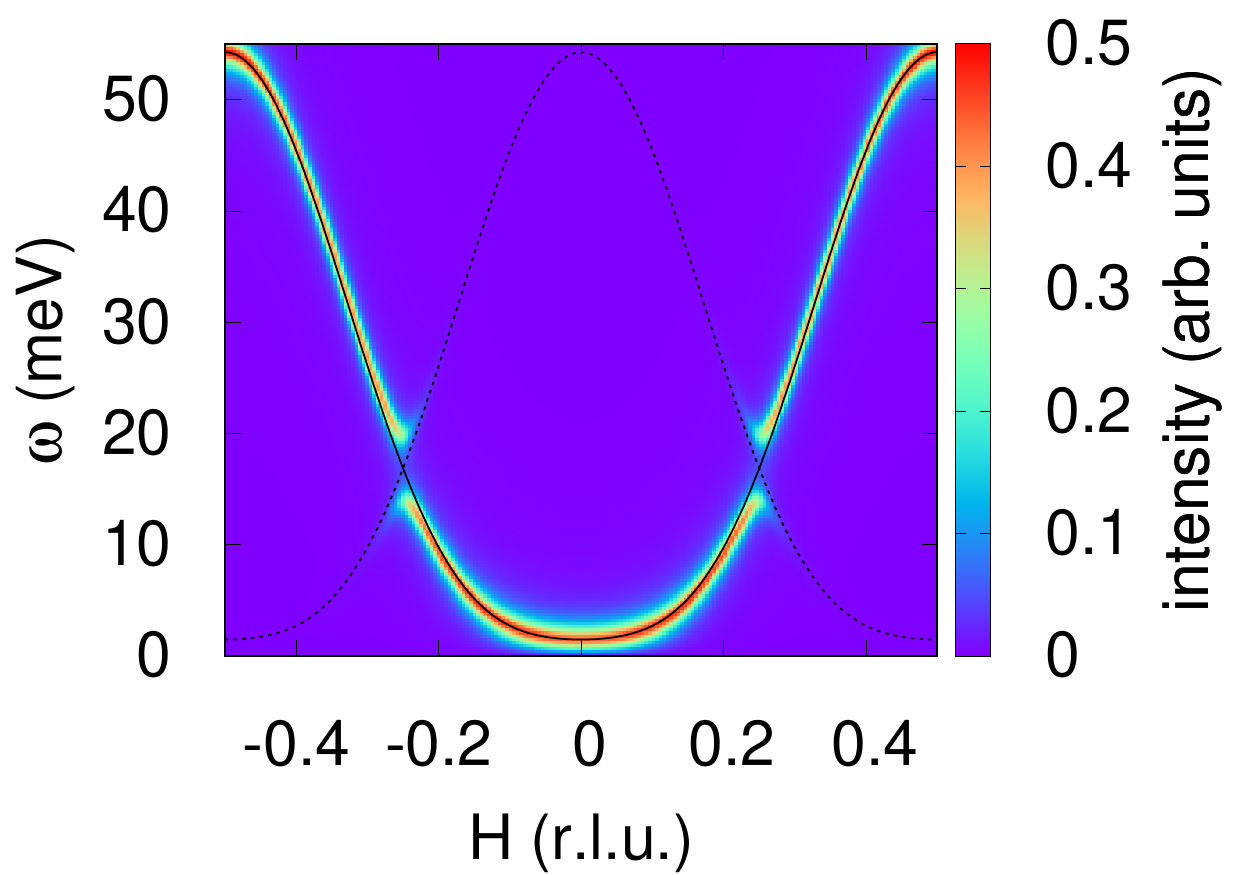} 
\caption{(Color online) The dynamic structure factor $S(q,\omega)$ for
the $J_1$-$J_1^{\prime}$-$J_2$ 
 model. $J=-26.38$, $J_2=5.5$, $\delta =3$, and $D=-1.5$~meV.
 Thin line:  the two-fold supercell
 with $\delta=0$ (see analogous features in Fig.\ S1 of SM \cite{Suppl} for the
 case of a five-fold supercell).
}\label{per2} 
\end{figure}
the cells and $s$ the sites within
the cell, $r_{s}$ defines
the neighbors coupled with the site
$s$;
in Eq.
(\ref{eq:Htb}), we expressed 
the spin-Hamiltonian 
in terms of spin deviation operators using the Holstein-Primakoff
bosonization assuming  a FM ground state of the chain.

We have calculated the structure factor in the 
large Brillouin zone $-\pi/a<q\equiv2\pi h<\pi/a$
\begin{align}
S\left(q,\omega\right) & \propto-\mathrm{Im}\left(\left\langle \left\langle \hat{S}_{q}^{x}|\hat{S}_{-q}^{x}\right\rangle \right\rangle _{\omega}+\left\langle \left\langle \hat{S}_{q}^{y}|\hat{S}_{-q}^{y}\right\rangle \right\rangle _{\omega}\right)\label{eq:Sqw}\\
 & \approx-\frac{1}{2}\mathrm{Im}\left(\left\langle \left\langle a_{q}|a_{q}^{\dagger}\right\rangle \right\rangle _{\omega}+\left\langle \left\langle a_{q}|a_{q}^{\dagger}\right\rangle \right\rangle _{-\omega}\right),\nonumber 
\end{align}
\vspace{-6mm}
where 
\vspace{-0mm}
\begin{align*} 
a_{q} & =\frac{1}{\sqrt{Nn}}\sum_{R,s}\mathrm{e}^{-iq\left(R+s\right)}a_{R+s}
\equiv\frac{1}{\sqrt{n}}\sum_{s}\mathrm{e}^{-iqs}a_{q,s},
\end{align*}
where $N$ is the number of cells, $n$ is that
of sites per
cell.

In Fig.\ \ref{per2} we show
how a 
two-fold
SS of the $J_1$-values (compatible with a period-10 SS)
affects the intensity of the calculated LSWT
dynamical spin structure factor $S(q,\omega)$.
Notice the absence of shadow bands and the slightly changed
dispersion visible in the height of the maxima
(lowered here by 
$\sim$5~meV)
probably due to the omitted 
ICs.
In order to "fit" the observed dispersion and the main 
gap 
near 28 meV, 
the remaining
couplings have to be changed too. Thus, the SS does not only
open gaps as expected, but it also changes the dispersion also 
far from the gap \cite{remarkKlyushina}.

\subsection{
Theoretical aspects of 
large 
FM $J_1$-values}
Despite some exceptions, 
including CuGeO$_3$ 
(all due to  large $\Phi$ and the presence of
 strong crystal fields),
 $J_1$ is usually FM (see Table I in SM \cite{Suppl}).
$J_1 = -24$~meV 
in CYCO 
is remarkable. It
 exceeds $J_1$ of LICO
($-19.6 \pm 0.4$~meV)
and also that of Li$_2$ZrCuO$_4$ slightly ($-$23.5~meV)
 \cite{drechsler}. 
Since a highly dispersive magnon gives dynamical evidence for a strong 
FM NN $J_1$, it deserves a phenomenological and microscopical analysis and verification
by other data. 
We start with an analysis of the magnetic susceptibility
$\chi(T)$ and then continue  with microscopic aspects
of the closely related LICO with a simpler but similar 
averaged structure as CYCO.
The validity of the large $J_1$-regime is  
also confirmed 
by several DFT+$U$ calculations for LICO \cite{lorenz}.
\subsubsection{Consequences for the magnetic susceptibility} \label{SecHTE}
Ignoring the tiny FM couplings along the $b$ axis, 
 CYCO is a 2D N\'eel system with a relatively large 
 FM Curie-Weiss (CW) temperature: 
 \begin{equation}
 \Theta_{\mbox{\rm \tiny CW}}\approx-\frac{1}{2}\left[J_{a1}
 (1-\alpha)+z(J_{ac,1}+J_{ac,2)} \right]\approx 80\ \mbox{K},
 \end{equation}
 where 
 $2z$ measures the number of NN and NNN sites on the adjacent chains. 
 Without the ICs one would arrive at
 $\Theta_{\mbox{\tiny CW}}\approx 107$~K. 
 \begin{figure}[b]
 \vspace{-0.5cm}
 \includegraphics[width=0.9\columnwidth]{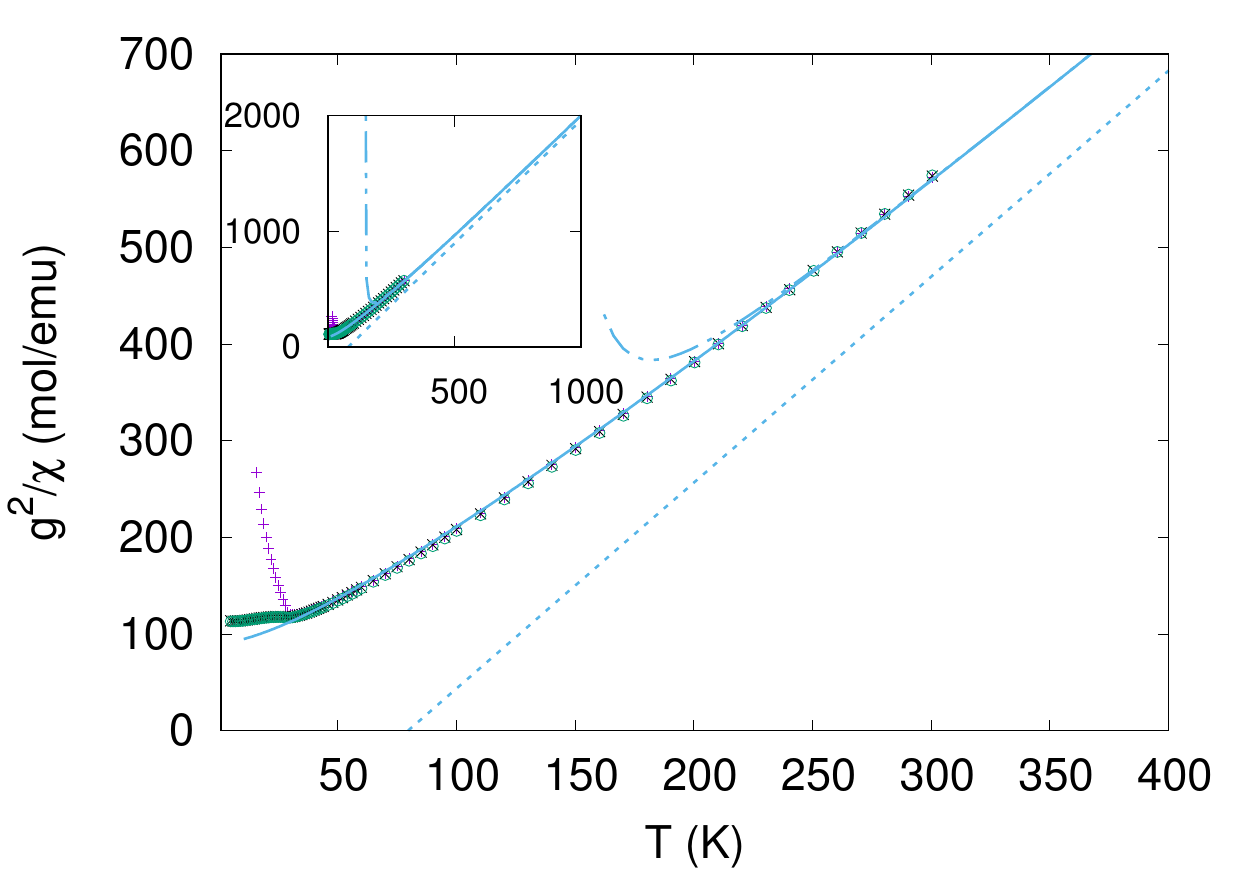}
\caption{The inverse spin susceptibility
for a magnetic field along the 
$a$ ($\times$), $b$ ($+$), and $c$ ($\circ$) 
axes of CYCO
(from Ref.\ \cite{kudo}). Dashed double-dotted line: fit
 by the 10th-order HTE
Eq.\ (\ref{eq:fitchi}); solid line: [5,5] Pad\'{e} approximation.
Short-dashed line:
the exact 
CW asymptotic curve.
Inset: Extended $T$-range 
up to 1000~K.}
\label{invcyco}
 \end{figure}
To show that the exchange values determined from our refined 
INS measurements are fully
compatible with the $\chi (T)$-data, we reproduce in Fig.\ \ref{invcyco} 
the data from Ref.\ \onlinecite{kudo}.
Similarly to our recent 
analysis of $\chi (T)$ for
LICO
\cite{Kuzian2018},
we have fitted the data in the range $240<T<300$~K with the expression 
\begin{equation}
\chi (T) = \frac{5N_Ag^{2}\mu_{B}^{2}}{k_{B}}\chi _{10}(T), \quad
\chi _{10}(T) = \sum_{n=1}^{10} \frac{c_n}{T^n},
\label{eq:fitchi}
\end{equation}
where $\chi _{10}(T)$ 
is the 
10th-order HTE \cite{Lohmann14,hte10}.
$N_A$ is the Avogadro number (one mole of CYCO contains $5N_A$ spins),
$\mu_{B}$ is the Bohr magneton, and $g$
is the gyromagnetic ratio.
A small anisotropy of the couplings as well as the tiny inter-plane couplings ($J_b/k_B,J_{ab}/k_B < 1$~K) is 
unimportant for the $\chi(T)$-analysis and was ignored here.
Evidently, the HTE series, Eq.\ (\ref{eq:fitchi}), fits well 
the data for 
$g_a\approx 2.04$, $g_b\approx 2.28$, $g_c\approx 2.02$
above
$T\sim 240$~K.
We recall that an
ESR study on powder samples of CYCO reports 
 $g_b\approx 2.31$ and $g_{\perp}\approx 2.03$ \cite{Okubo2006}.
The [5,5] Pad\'{e}
approximation fits the curve down to $T\gtrsim T_{\rm \tiny N}$.
Fig.\ \ref{invcyco} also shows the 
CW asymptotic curve,
reached only at $T\gtrsim 1000$~K, as shown in the inset.
  From this comparison the 
 strong IC in 
 CYCO
 is evident,
 explaining the {\it absence}
 of criticality, which manifests itself in a strong upshift of $\alpha_c$ 
 (see\ SM \cite{Suppl})
 and in the large 
 moments seen in neutron diffraction in the quasi-2D N\'eel state below 
 the high $T_N\approx 30$~K.
 In contrast, LICO, Ca$_2$Nd$_2$Cu$_5$O$_{10}$, and
 CuAs$_2$O$_4$ might be closer to the 
 1D
 $\alpha_c$ due to
 $\alpha >1/4$ and a 
 much weaker AFM IC,
 which is even there necessary to stabilize a 
 FM alignment.
 For LICO 
 the analogous AFM IC $J_{bc,1}+J_{bc,2}$ 
 is only 9~K (0.8~meV) per bond,
 where the $b$ axis
 is the chain direction. 
 
To avoid such
strange results found often in the literature
from improper
 $\chi(T)$ fits (e.g.\ $\Theta_{\rm \tiny CW}$=2~K \cite{Kargl2006}
for CYCO),
low-$T$ INS (or RIXS) 
probing the 
magnon
dispersion, 
combined with magnetization data up to the saturation field, allows
to extract more reliable 
couplings. 
The former should be used to cross-check 
any exchange 
set derived from $\chi(T)$, if multiple $J'$s 
are involved. 
Noteworthy, for 
La$_6$Ca$_8$Cu$_{24}$O$_{41}$ \cite{Carter1996} containing
both {\it undoped} TLL and CuO$_2$ 
chains just like in CYCO but with a {\it smaller} $\Phi$,
a tiny 
$\Theta_{\rm \tiny CW}$=21$\pm$1~K
has been reported \cite{Carter1996} from linear fits of  $1/\chi(T)$
data
up to 300~K only. 
But from that $\Phi \approx 91.6^{\circ}$~K,
according to Eqs.\ (14,16,18),
even a {\it larger} FM $J_1$ is
expected.
Then, with a smaller AFM  IC
due to
the 
low $T_{\rm \tiny N}$$\approx$12.5~K and
 a similar misfit 
 from La$^{3+}$ replacing 
 Y$^{3+}$, as in CYCO,
 a much {\it larger} $\Theta_{\rm \tiny CW}$$\sim$100$\gg$21~K
is requested.
 Another related issue 
 is the 
 incorrectly predicted
 critical $h$-doping when
  $\Theta_{\rm \tiny CW}$ changes its sign, i.e., for weak $h$-doping
  a 
  sizable
  magnon dispersion and hence $\Theta_{\rm \tiny CW}>0$ are
  still expected, at odds with the opposite sign
  provided so far in the literature
 from improperly fitted asymptotics.
  Our
  dispersion-law 
  (with a
  generalization for frustration
  perpendicular to the
  AFM 2nd adjacent chain \cite{Matsuda2003} as 
  proposed for La$_5$Ca$_9$Cu$_{24}$O$_{41}$)
  allows 
  to separate
  chains  from  ladders  dominant 
  above their
  spin gap \cite{Rice1993}.
  
Since frustrated FM systems 
are of general
interest in the field of quantum magnetism and statistical 
physics \cite{mueller,haertel,Iqbal2016}, hopefully, our work
will initiate
further work. 
In particular, systematic studies of critical systems like
CuAs$_2$O$_4$ 
with $\alpha \approx 0.27$ \cite{Caslin2014,Caslin2016,Caslin2015}
will give
insight into the role of quantum fluctuations
(see Table I in SM \cite{Suppl}).
In this context the study of other thermodynamic properties
as the specific heat and thermal conductivity might be useful to check
the coupling constants derived here.
Other examples of 
weakly $h$-doped
compounds \cite{yamaguchi,Hayashi1998}
will be addressed elsewhere.
\begin{figure}[b]
\includegraphics[width=5.3cm]{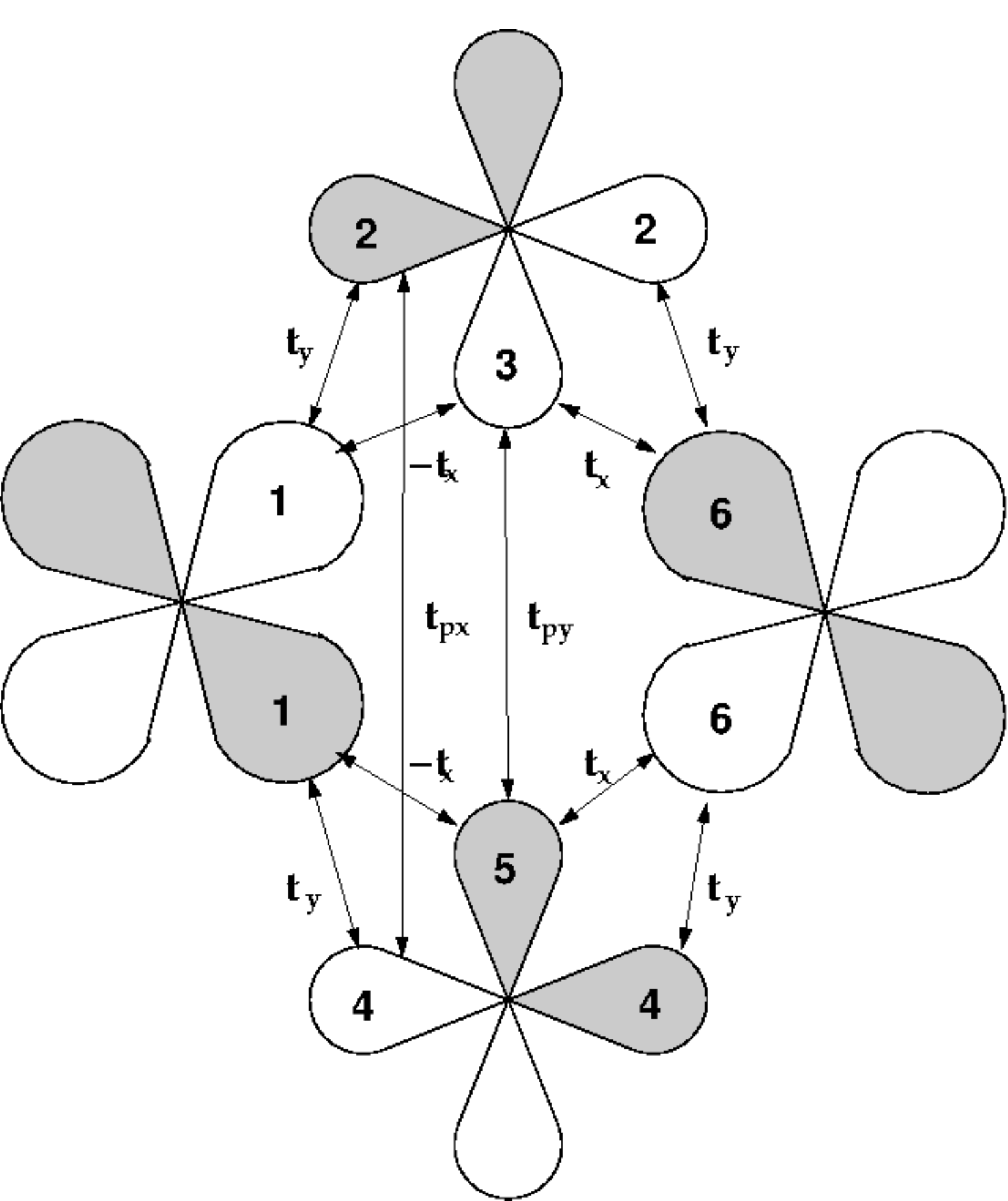}
\caption{$p$ and $d$ orbitals and transfer integrals of a CuO$_2$-cluster treated exactly within a 
planar Cu 3$d$ O 2$p$ five-band Hubbard model: Cu 3$d_{xy}$ 1 (left) and 6 (right); 
intermediate O 2$p_{x,y}$ 
2,3 (upper) 4,5 (lower). The chain is along the horizontal direction ($x$ axis).
For the sake of clarity the hopping $t_{dd}$ between two Cu-sites along the $x$ axis is not
shown.
}
\label{orbitals}
\end{figure}
\subsubsection{QC and DFT analysis
for LICO -- comparison with Mizuno {\it et al.} \cite{mizuno}}
The most notable theoretical finding 
of the present work
with respect to an empirically 
large FM $J_1$ value is several microscopic FM 
intersite couplings behind the spin-chain model,
obtained by QC and DFT-based analysis.
Given the complex
real structure of CYCO, we will present theoretical studies
for 
LICO
since its structure is very 
close to the {\it averaged idealized} structure of CYCO.
In fact, their C-O-Cu bond angles $\Phi$ differ by $\approx 0.1^{\circ}$, 
whereas the Cu-Cu distances $d_{\rm \tiny Cu-Cu}$ by 0.04 \AA\ and 
the Cu-O bond length $d_{\rm \tiny Cu-O}$ by $\approx$ 0.025~\AA\ 
only \cite{Sapina,Gotoh}. 
We therefore believe that the NN
 results
 for LICO
can be transferred to CYCO with an uncertainty of only a few percents. A semi-quantitative
general analysis including  several cuprates will be given elsewhere. Providing
refined theoretical results for
LICO
very much simplifies the modeling
and allows a critical check
of the parameters {\it adopted} in Ref.\ \onlinecite{mizuno}
for LICO.
It is convenient to
decompose the total
$J_1$ into a FM and an AFM contribution: 
\begin{eqnarray}
\hspace{-0.9cm} J_1 &=& J^{\rm \tiny FM}_1 + J^{\rm \tiny AFM}_1 , \quad \quad \quad \mbox{ with}\\
\hspace{-0.9cm} J^{\rm \tiny FM}_1 &\approx & J_1(K_{pd},K_{pp})+J_1(J_H) -K_{dd} \quad \quad \mbox{ and}
\label{fm}
\\
\hspace{-0.9cm}J^{\rm \tiny AFM}_1 &\approx & J_{\Phi}+\frac{4t^2_{dd}}{U_d-V_{dd}}, 
\label{jafm}
\\
\nonumber
\end{eqnarray}
\begin{figure}[t]
\includegraphics[width=4.0cm]{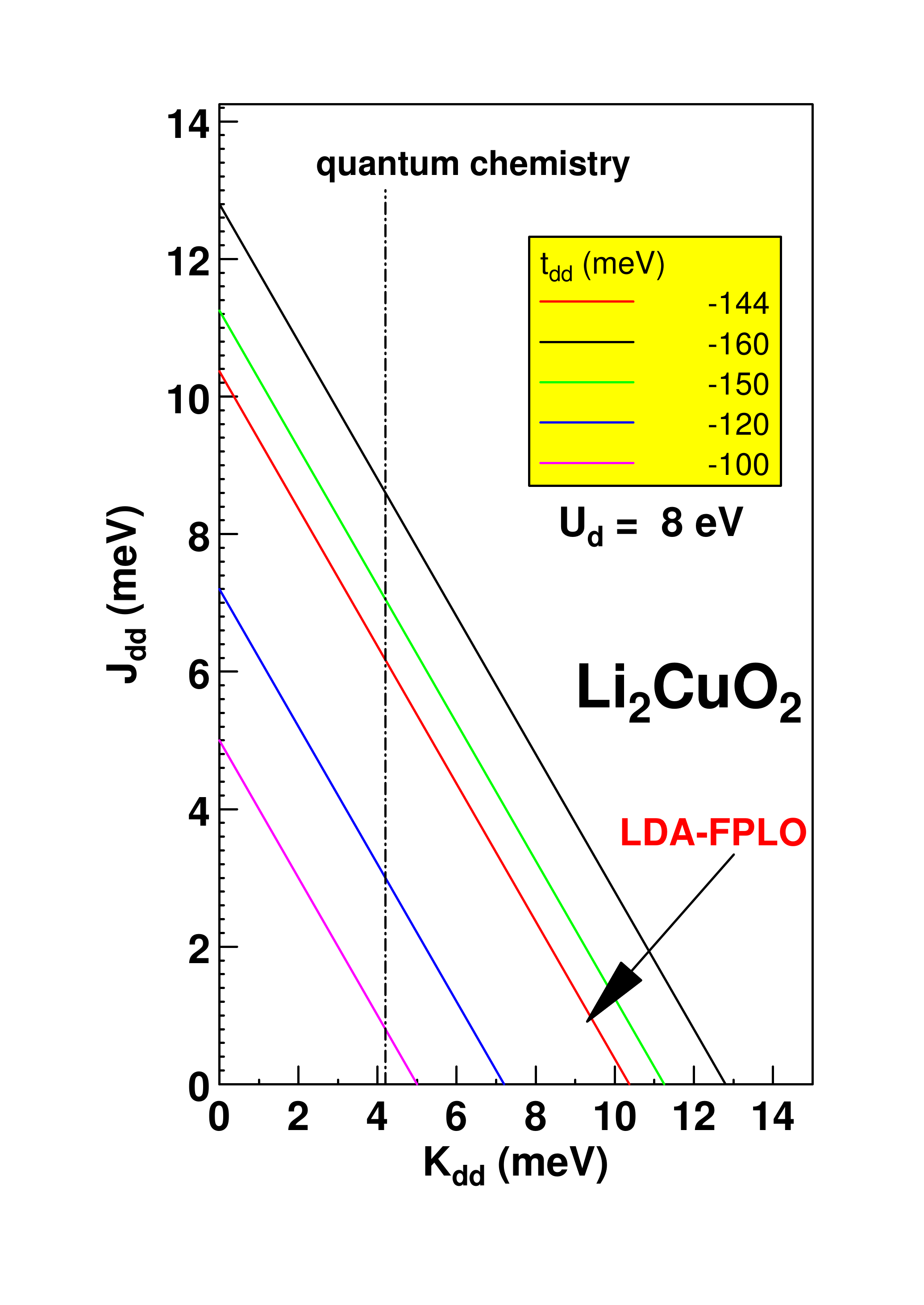}
\includegraphics[width=4.0cm]{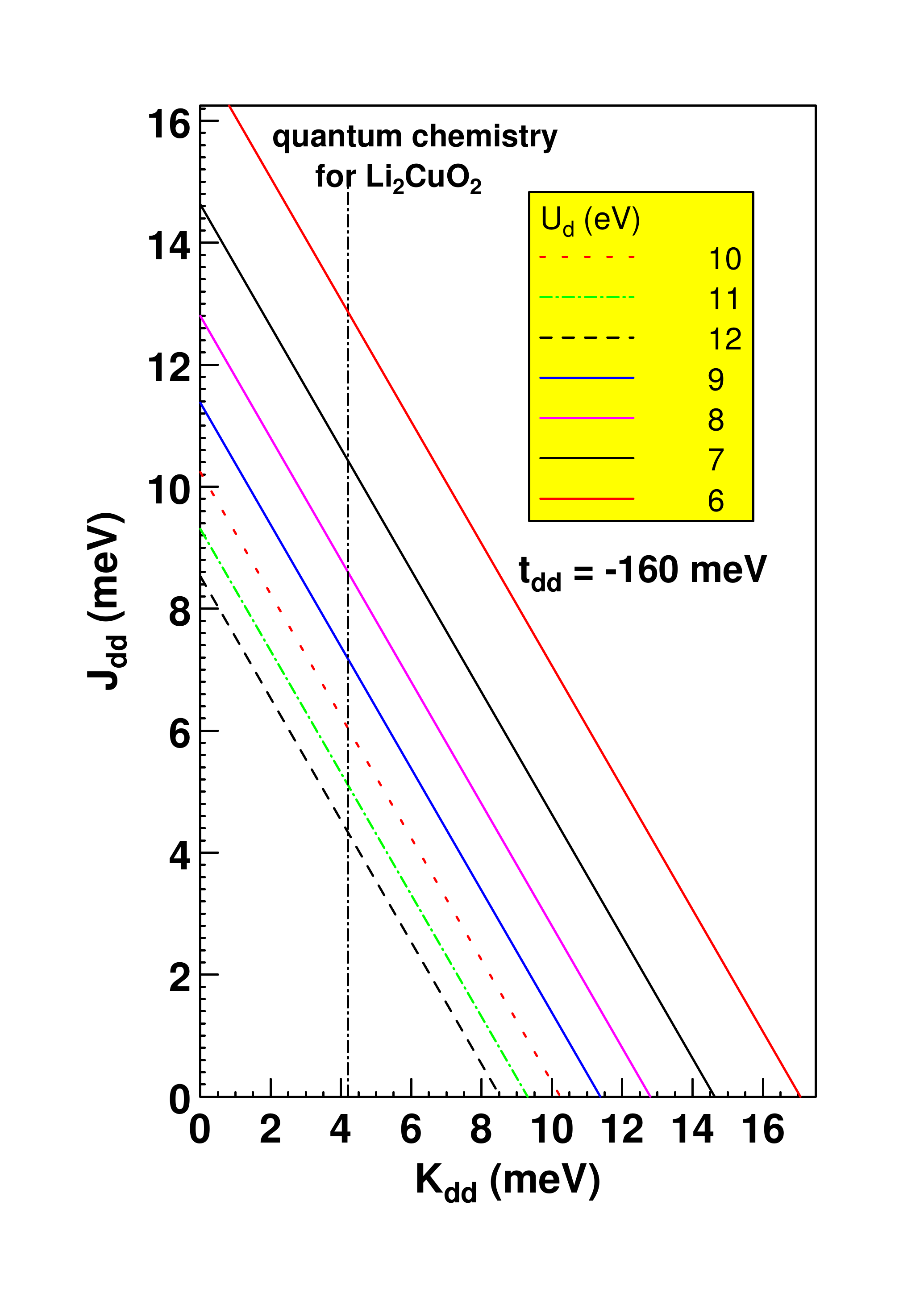}
\vspace{-0.5cm}
\caption{Superexchange from the $dd$ channel $J_{dd}$ vs.
the direct intersite FM
exchange $K_{dd}$ for various direct Cu-Cu hoppings 
$\mid t_{dd}\mid$, $e.g.$ $t_{dd}$ = $-$144 ($-$160) meV by LDA (QC) mapping
(both ignored in Ref.\ 15; for details, see Sec.\ C in SM \cite{Suppl}) ({\bf left})
and adopted
onsite Hubbard $U_d$ values ({\bf right}). Vertical d}ashed-dotted
lines are
   the QC result
for LICO. 
\label{jddabs}
\end{figure}
\noindent
where $K_{pd},K_{pp}$, and $K_{dd}$ denote direct
FM
{\it intersite}
Coulombic (Heisenberg) exchange integrals and $J_H$ is the indirect {\it onsite}
Hund's-rule coupling from each of the bridging O's. For the corresponding
$pd$-Hamiltonian, see e.g.\ Refs.\ \onlinecite{mizuno,Malek,Monney2013,Johnston2016}
and Fig.\ 13.
The generalization including also the Cu-Cu intersite terms $K_{dd}$, $V_{dd}$, 
and the hopping $t_{dd}$ (all ignored there)
is straightforward. For enlarged $K_{pd}$ and $K_{pp}$ in the 2D geometry of  La$_2$CuO$_4$,
see Refs.\ \cite{Hybertsen1990,Hybertsen1992}.

Turning to the various FM sources
in $J^{\rm \tiny FM}_1$ entering Eq.(\ref{fm}), we note that in general
all fundamental FM
 exchange couplings are known 
by order of magnitude only:
1~meV $< K_{dd} \ll K_{pp} \ll K_{pd} \ll J_H < $1.5~eV.
 $K_{pd}$, $K_{dd}$, and $K_{pp}$ are reported here for the first time 
 for chain cuprates. 
Refining the estimates given above, our
QC result is $K_{dd}\approx 4.2$~meV, indeed much smaller than
various $K_{pp} < 20$~meV, and
$K_{p_xd}\approx K_{p_yd}\approx$ 105~meV
(for details see\ Sec.\  C in SM \cite{Suppl}).
 $J_H$ 
 somewhat
exceeds the value of 600~meV adopted in Ref.\ \onlinecite{mizuno}. According to direct
calculations and various empirical 
estimates, in particular, for related superoxides \cite{Solovyev2014}, 
a bit enhanced slightly screened
$J_H$-value $\sim $0.7~eV is even more realistic 
\cite{drechsler2019}.
$K_{pd}$ turns out to be the most important
FM microscopic interaction for $J_1$ since
a weak enhancement of a moderate $J_H$ cannot explain
the more than twice as large $J_1$ at fixed $K_{pd}=50$~meV universally adopted
by Mizuno {\it et al.} \cite{mizuno} (including also CYCO) 
who found $J_1=100$~K for LICO at odds
with $\mid J_1 \mid \geq $
230~K derived
without fully probed by INS \cite{lorenz}
magnon-dispersion leaving therefore some room for further refinements in future
if the same sophisticated intensity analysis as in Sec.\ IV for CYCO could be applied
there too.

Our DFT and
QC analysis succeeded also in the determination of a remarkable direct
$dd$ transfer integral $\mid t_{dd}\mid, 144$ and 160~meV, respectively,
due to the short NN Cu-Cu distance $d_{\rm \tiny Cu-Cu}\approx 2.86$ \AA . 
A finite $t_{dd}$ (ignored in Ref.\ \onlinecite{mizuno})
provides an {\it additional} AFM contribution if
\begin{equation}
\mid t^{c}_{dd}\mid > 0.5\sqrt{\left(U_d-V_{dd}\right)K_{dd}} 
 \ .
\end{equation}
Typically $t_{dd}$ is of the order of +5~meV (Fig.\ 14), yielding
further arguments for a
  larger $K_{pd}$. With this AFM $dd$ channel alone,
$J_1\approx 165$~K would be achieved when fixing $K_{pd}$.
The precise value of 
$U_d$ (actually unknown) affects strongly
the efficiency of the $dd$ AFM superexchange,
as shown in Fig.\ 14,
especically for FFESC cuprates with a short Cu-Cu
distance. 
It markedly exceeds that 
for typical charge transfer insulators
($U_d\gg \Delta_{pd}$), where it is negligible
due to the tiny $t_{dd}$.
Also a 
realistic intersite Cu-Cu Coulomb interaction 
$V_{dd}$ of 0.3 to 0.5~eV 
\cite{remarkIC,Higashiya2008,Drechsler2010,Neudert1999}
might be  
relevant (compared to\ Eqs.\ (\ref{Dimer}), (\ref{single1}),
and (S24) and Sec.\ J in SM \cite{Suppl}).
The 
$dd$ channel 
is of interest for further
theoretical 
studies since it
gives
a new Cu-Cu-O-Cu exchange path in addition to the known
Cu-O-O-Cu one, with possible modifications of $J_2$.

\vspace{0.2cm}
\subsubsection{Comparison with other cuprates and general trends}
 
Looking for an empirical support, we
will 
 compare our $J_1$-values
with those of other cuprates and 
provide thereby a critical analysis for the exclusive  
attempt to present a general
description of edge-sharing chain cuprates and ladders performed 20 years ago
\cite{mizuno}, when almost no detailed microscopic studies were known. 
We will show that it is timely
to reconsider not only
LICO and CYCO 
since our criticism concerns 
the
unjustified
use of a systematically  {\it underestimated, universal},
and {\it isotropic} direct FM Cu-O exchange $K_{pd}=50$~meV~\cite{mizuno} with 
serious consequences 
for the FM $J_1$.
It is in fact of general interest for all
cuprates with edge-sharing elements, 
including
infinite CuO$_2$
chains, ladders, coupled Cu$_n$O$_{n+2}$ 
($n= 2, 3, 4$ ...), and 
other finite edge-sharing CuO$_4$ units
(see\ Table\ I of SM \cite{Suppl}).
This provides
 the basic picture of the interactions in cuprates
 and the minimal stage with the
 5-band Hubbard $pd$-model in terms of which their  
 fundamental physical properties must be discussed.

The magnitude of the NN exchange is important for any
quantum magnet, in particular,
for cuprates with edge-sharing elements present in single and double
(zigzag) chains, since it determines or strongly affects the role of frustration 
 measured here 
 by $\alpha$ in Eq.\ (1).
 As mentioned above,
 $\alpha \gg 1$ can be treated as an
 effective AFM system with slightly renormalized
 $J_2$-values
 ignoring the finite $\alpha^{-1}$-value at all. 
 Unbiased 
QC and DFT studies allow insight into the 
 magnitude of $J_1$, despite 
 some uncertainty due to 
certain correlation 
 and spin-orbital effects
 ignored here.
 With this 
 in mind, we select and comment on available data for various 
 cuprates in Table\ I of SM \cite{Suppl}.
Naturally,
$\Phi$ near 90$^{\circ}$, the Cu-Cu
$d_{\rm \tiny Cu-Cu}$ and the Cu-O distances within the generic
CuO$_4$ plaquettes as well as the strength of the crystal field, 
strongly affected 
by the charge and  position of the surrounding cations near 
the bridging O,
are important physical ingredients.

 Similar
 or even larger $J_1$-values have been observed or
 predicted by theoretical studies \cite{Schmitt2014,Bordas,Calzado2000} so far
 only in (i) ladder compounds ($J_1$ corresponding to the interladder
 couplings),
(ii) double-corner-sharing (zigzag) chain compounds 
 with predicted $J_1$ values of $-28$ to 55~meV,
 and (iii) alternating FM-AFM chain systems Li$_3$Cu$_2$SbO$_6$ \cite{Koo2016}
 and Na$_3$Cu$_2$Sb(Te)O$_6$ \cite{Schmitt2014}
 with $J_1=-23.56$~meV.
 Li$_3$Cu$_2$SbO$_6$ \cite{Koo2016} and 
 Na$_3$Cu$_2$SbO$_6$\cite{Schmitt2014,Miura2006,Miura2008},
 with similar AFM couplings but  very different relatively large 
 FM $J_1$ values, are particularly
 striking. In view of its large $\Phi = 95.27^{\circ}$, an AFM or very weak FM
 coupling would be expected
 according to Ref.\ \onlinecite{mizuno}. Hence, an enlarged
 FM interaction and/or a strongly suppressed AFM
 exchange via the bridging O must be responsible for the resulting  
 FM 
 $J_1=-17.8~$meV, twice of $-$8.6~meV estimated for $\Phi = 93.97 ^{\circ}$
 (the case of LICO)
 \cite{mizuno}. 
 The smaller 
 value of
  $-$12.7~meV derived from INS data \cite{Miura2008}
 is probably
 due to
 the 
too small limiting energy of 14.7~meV probed there,
 similarly
 as for CYCO previously.
 There the 
 CuO$_2$ chains are distorted by 
 a similar
 cationic misfit, as in CYCO, in the combined 
 TLL and chain system La$_6$Ca$_8$Cu$_{24}$O$_{41}$.
 Due to the
 $\Phi$ closer to 90$^{\circ}$ also somewhat larger
 $J_1$ values are expected. Unfortunately, its 
 chain magnon component is not yet fully
 understood, hampered by the dominant two-spinon contribution and the 
 $h$-doping
 in some cases \cite{remarkdoublechain}.
Our present $J_1$-value 
strongly 
exceeds 
an earlier
estimate of $-$2.15~meV
for 
CYCO
and that of $-$8.6~meV 
for LICO
\cite{mizuno}.
Furthermore, it 
is at odds with
$\alpha=2.2$ \cite{mizuno}
and puts it 
close to 
$\alpha_c$.
The results from
QC analysis
for LICO
show a markedly enlarged 
$J_1$ by 42\% as compared with 
Ref.\ \onlinecite{mizuno} but  
not enough
when compared with the empirical and DFT-derived values with a still
 larger 130\% enhancement. For the TLL
 SrCu$_2$O$_3$ a less dramatic 
 but also 
 enhanced value by about 15\% was predicted which however
 is caused by the smaller $\Phi$ 
(see also Ref.\ \onlinecite{remarkdoublechain}).
 For 
 $\Phi=90^{\circ}$, generalizing an expression for $J_2$ by the account of
 a moderately enhanced
 direct FM $pd$ intersite $h$-exchange $K_{pd}$, $\alpha \approx $ 3 to 4
 is estimated, as shown below. Then based 
 on the empirical
 $J_2\approx 166$~meV, $J_1\approx -55$ to $-$41.5~ meV
 can be estimated for SrCu$_2$O$_3$ as an upper bound and we would arrive
 at $J_1\approx -38 \pm 1$~meV as a 
 realistic estimate. Note that  again it
 exceeds the value from Ref.\ \cite{mizuno}, now by $\approx$ 25\% and
 $J^{\rm \tiny FM}_1= -38$~meV like in
  LICO.
 Thus, one has enough reasons to doubt the 
 results
 given there for various chains and ladders.
 Then 
 one may ask:
 "What is the 
 reason for the {\it material dependent} 
 systematic underestimates given in Ref.~\cite{mizuno}?".
The numerous examples discussed above
require a strong material dependence which is
unlikely for $J_H$.
But its efficiency is lowered for split O onsite energies
due to strong crystal field effects.
In addition the usual indirect superexchange $J_{\Phi}$ 
via the bridging O increases and one can easily arrive at
small $J_1$ values $\sim$
80-100~K, as realized e.g.\ in linarite with a moderate $\Phi\approx 93^{\circ}$.
Then $K_{pd}$ remains the main microscopic source for FM $J_1$ values 
in cases as LICO since only a weak material dependence of $J_H$ 
(governed mainly by 
$U_p$) 
is expected.
 The  {\it two} interacting bridging O ions cause a slower convergency of
 standard perturbation theory, used previously to
 the effect of $K_{pd}$ and $J_H$. Hence, exact diagonalizations on
 small clusters are used to study this point,
 as shown in Figs.\ 15 and 16.
\subsubsection{Description in the five and single band Hubbard models}
Now we will show semi-quantitatively 
that
a realistic
microscopic scenario for such large FM $J_1$-values
well exceeding $-$200~K can be proposed.
Our arguments will be expressed in terms of 
the most natural and vivid multiband 
Cu 3$d$ O2$p$ Hubbard model with
five magnetically active orbitals in the $xy$-plane containing
ideally flat
CuO$_2$ chains, 
namely,  the single Cu 3$d_{xy}$
orbital and the two O $2p_x$ and $2p_y$ orbitals for each of the two bridging
O (Fig.\ 13), where $x$ is the chain-axis ($a$ axis in Fig.\ 1)
and the $y$-axis corresponds to the crystallographic $c$ axis (see also SM \cite{Suppl}).
Since this model contains already a large set of partly not precisely known
interactions,
we consider below also an effective single-band model 
with a reduced number of parameters.
The first $pd$ term of the AFM contribution
according to standard 4th-order perturbation 
theory \cite{Tornow,Geertsma96} reads
\begin{widetext}
\vspace{-0.3cm}
\begin{eqnarray}J_{\Phi}
&\approx&
\frac{4t^2_{p_xd}}{\left(\Delta_{p_xd}+V_{p_xd}\right)^2}
\left[\frac{2t^2_{p_xd}}{\Delta_{p_xd}+V_{p_xd}}+ \frac{1}{U_d}\right]
-\frac{4t^2_{py_d}}{\left(\Delta_{p_yd}+V_{p_yd}\right)^2}
\left[\frac{2t^2_{p_yd}}{\Delta_{p_yd}+V_{p_yd}}+ \frac{1}{U_d}\right] \ ,
\\
&\approx&
\frac{4t^2_{pd}\sin^2 \Phi/2}{\left(\Delta_{pd}+V_{pd}\right)^2}
\left[\frac{2t^2_{pd}\sin^2\Phi /2}{\Delta_{p_xd}+U_p}+\frac{1}{U_d}\right]
-\frac{4t^2_{pd} \cos ^2\Phi /2}{\left(\Delta_{p_yd}+V_{pd}\right)^2}
\left[ 
\frac{2t^2_{pd}\cos^2\Phi /2}{\Delta_{p_xd}+U_p}+\frac{1}{U_d}
\right] \ ,
\nonumber \\
&\approx& -
\frac{4t^2_{pd}}{\left( \Delta_{pd}+V_{pd}\right)^2
}\left[\frac{2t^2_{pd}}{\Delta_{pd}+U_p}+\frac{1}{U_d}\right]
\cos \Phi \ , \nonumber\\
\nonumber
\label{JPhi}
\end{eqnarray}
\vspace{-0.7cm}
\end{widetext}
ignoring for shortness the  O-O NN hopping terms  
 (compare to\ Eqs.\ (S46) and (S47) in SM \cite{Suppl}  
for the case $\Phi = 180^{\circ}$,
where the last Eq.\ is obeyed in the isotropic limit). 
It vanishes for $\Phi=90^{\circ}$, if one
ignores the slightly
different O 2$p_x$ and 2$p_y$ onsite energies due to the 
weak crystal field \cite{Neudert1999}.
In contrast, in cases of strong crystal fields or the presence of ligands,
even at $\Phi=90^{\circ}$ there is a significant AFM contribution that reduces
the total value of $J_1$.
For the experimental value of $\Phi $ and by adopting the
parameters
of
Ref.\ \onlinecite{mizuno}, i.e.\ ignoring first of all
the intersite Coulomb interaction $V_{pd}$, one has
$J_{\Phi} \approx 200$~K. Since
$J^{\rm \tiny AFM}_1$ 
depends
markedly
on $\Phi$,
the total $J_1$ may change its
sign at $\Phi$ values far enough from $90^{\circ}$, which happens in fact in several 
cases different from LICO
and CYCO
(see\ Table I in SM \cite{Suppl}).

$K_{pd}$ and $J_H$
occur in reverse orders of the 
($t_{pd}/\Delta_{pd}$)-perturbation 
theory 
affecting their weight
and $n_p$ on the two O sites which interact by hoppings and
FM $K_{pp}$. 
In the spirit of
this approach for the five-band Hubbard
model sketched in Refs.\ \onlinecite{Geertsma96,remarkGeertsma} 
for the case of edge-sharing plaquettes with two common O, 
Eq.\ (\ref{fm})
can be 
approximated by
\begin{widetext}
\begin{equation}
J^{\rm \tiny FM}_1\approx -K_{dd}-8Z\left( \frac{t_{pd}}{\Delta_{pd}}\right)^2K_{pd} 
-\frac{4Z^2J_H}{(1+U_{pp}/\Delta_{pd})}
\left( \frac{t_{pd}}{\Delta_{pd}}\right)^4
\approx-K_{dd}-4\left( \frac{t_{pd}}{\Delta_{pd}}\right)^2K_{pd} 
-\frac{J_H}{(1+U_{pp}/\Delta_{pd})}
\left( \frac{t_{pd}}{\Delta_{pd}}\right)^4.
\label{perturbation}
\end{equation}
\end{widetext}
where the renormalization factor $Z(t^2_{pp}/U_p,K_{pp},\Delta_{pd}) < 1$ has been introduced. 
It 
contains higher-order corrections due to various
O-O hoppings and direct FM 
couplings $K_{pp}$ taken from
our DFT and
QC analysis.

\begin{figure}
\includegraphics[width=6.3cm]{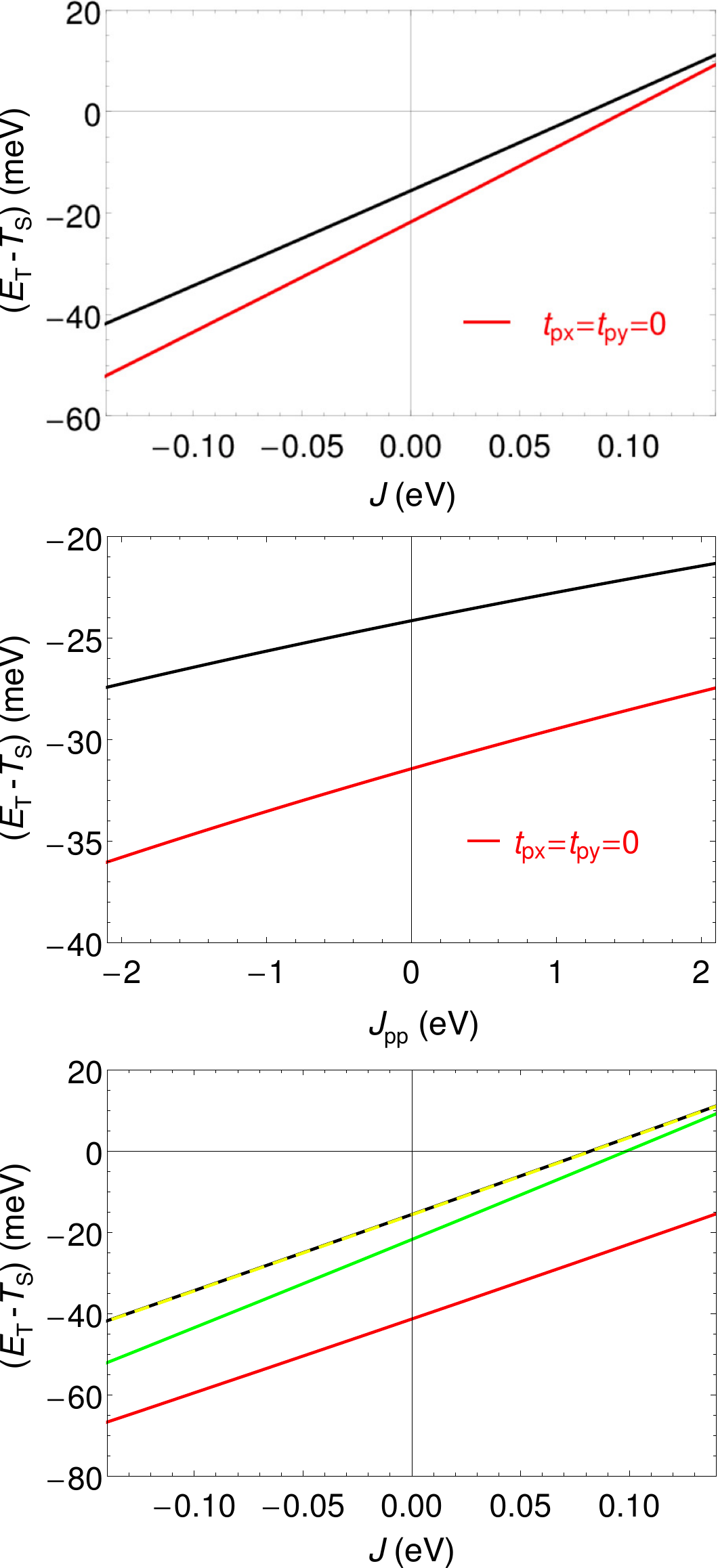}
\caption{The NN exchange $J_1$=$E_T-E_S$ 
from exact diagonalizations
similarly to
Eq.\ (\ref{perturbation}). 
The suppression of the intersite interaction between the
 "upper"
and lower O for $U_p=\infty$ (green curve) and $t_{p_yp_y}
=t_{p_xp_x}=0$ (red curve) as compared to the set described above (black curve).
Dotted curve: including also the weak FM direct exchange 
$K_{pp}$ ($K_{p_yp_y}=18.4$, $K_{p_xp_y}=13.4$, and $K_{p_xp_x}=8.7$~meV).}
\label{cluster}
\end{figure}

For $\Delta_{pd}=3.5$~eV, 
$Z\approx 0.48 $. 
A  quasi-linear 
law for $J_1$, $K_{pd}$, and $J_H$, like in Eq.\ 
(\ref{perturbation}) (with slightly changed 2nd and 3rd 
coefficients due to additional interactions)
holds, also beyond $\left( t_{pd}/\Delta_{pd} \right)$-perturbation theory as confirmed
  by the exact treatment of 
Cu-O$_2$-Cu dimers (see\ Fig.\ \ref{cluster} and SM \cite{Suppl}) 
as well as for larger clusters
with small finite size effects 
within the effective single-band
Hubbard model.
Then the estimated ratio $\rho_{\rm \tiny HK}$ of the 
FM 
on- and intersite
contributions to $J_1$ reads
\cite{Geertsma96}
\begin{equation}
\rho_{\rm \tiny HK}=\frac{J_1^{\rm \tiny H}}{J_1^{K_{pd}}}
\approx\frac{J_1^H}{2K_{pd}\left(1+(U_p-2J_H)/
\Delta_{pd})\right)}\left(\frac{t_{pd}}{\Delta_{pd}}\right)^2,
\label{hundratio}
\end{equation}
where an often used aproximation 
defines $J_H$:
\begin{equation}
J_H=0.5\left( U_p-U_{pp}\right).
\label{HundO}
\end{equation}
\begin{figure}
\includegraphics[width=6.5cm]{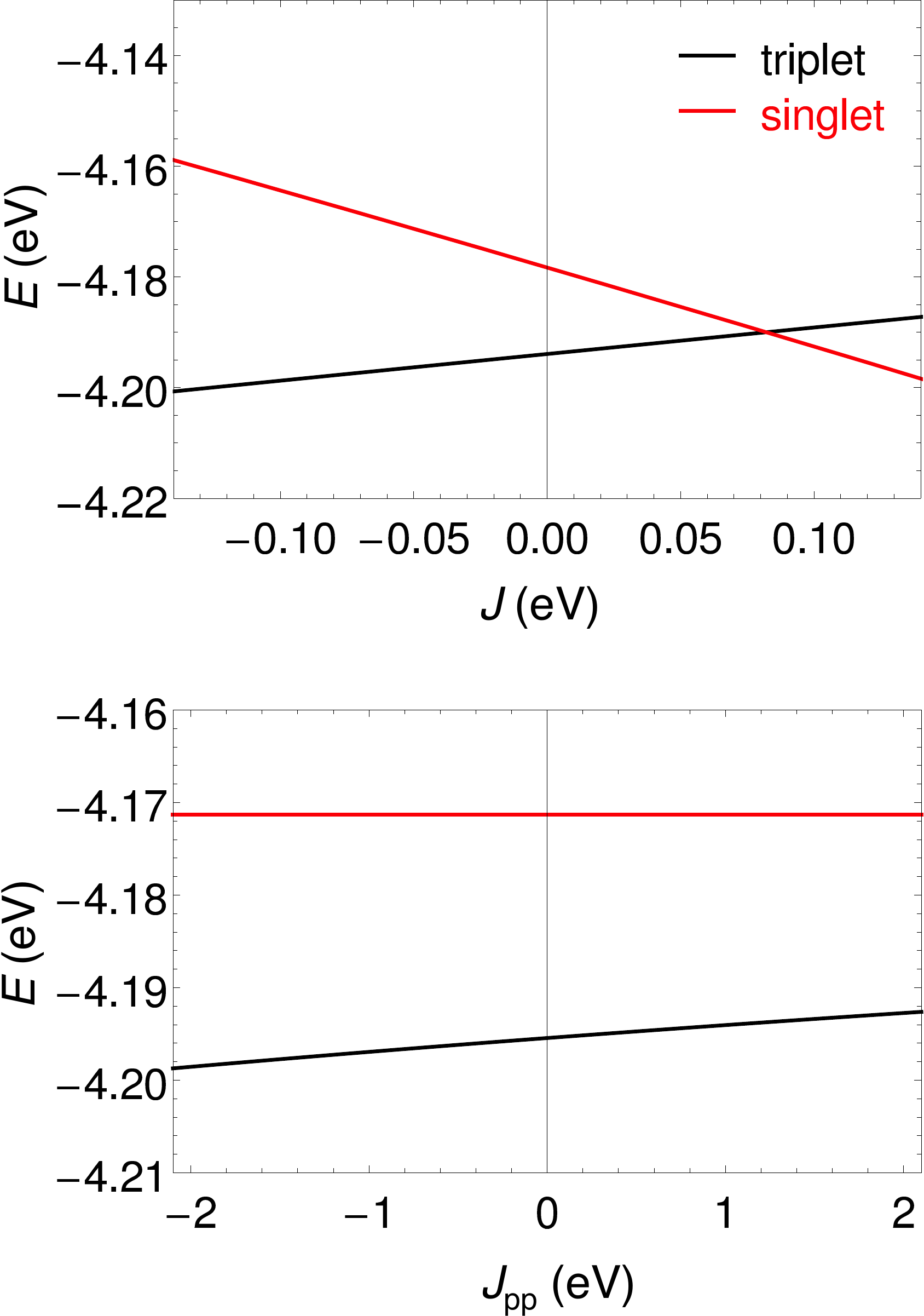}
\caption{Lowest singlet and triplet energies
of 
a $\Phi =90^o$ 
CuO$_2$Cu cluster as in Fig.\ 13
 vs.\ the main FM exchanges
for  
a set
close to
Ref.\ \onlinecite{mizuno},
where 
 a O$_2$-CuO$_2$Cu-O$_2$ 
 cluster 
was used.
($t_{p_xd}=t_{p_yd}=0.7155$, $t_{dd}=0$, $t_{p_x}$=0.17, $t_{p_y}=0.69$,
 $V_{pd}$=$V_{dd}$=$J_{dd}$=0, $\Delta_{pd}$=3.2, $E_{p_x}$=1.75, $E_{p_y}$=1.45, $U_d$=8.5,
 $U_p$=4.1, $K_{pd}$=0.05, and $J_H$=0.6; all in hole notation and in units of 
 $t \equiv t_{pd}\approx 1$~eV)
$J$=$-K_{pd}$ ({\bf upper plot} for 
$J_H$=0.6)
and $J_{pp}$=$-J_H$ at $K_{pd}=0.05$ ({\bf lower plot}).}
\end{figure}
In case of ideal two-leg ladders, 
i.e.\  
$\Phi=90^{\circ}$ and $U_p=4.4$~eV, Rice {\it et al.} \cite{Rice1993}
 ignoring $K_{pd}$,
derived 
a 
useful expression for the FM interladder coupling:
\begin{equation}
J^{H}_1=\frac{2t^4_{pd}}{\Delta^2_{pd}}
\left[\frac{1}{E_T+2\Delta_{pd}}-\frac{1}{E_S+2\Delta_{pd}}\right] \ ,
\label{jladder} 
\end{equation}
where $E_{T}=7.3$~eV ($E_S=1.8$~eV) denote the energy of the Zhang-Rice triplet
(singlet)
state, respectively,  for $t_{pd}=1.3$ and $\Delta_{pd}
=3.3 $~eV.
Inserting these 
numbers into Eq.\ (\ref{jladder})
one arrives at $J^H_1=$24.7~meV and a 
frustation ratio of 6.7
using the experimental value $J_2=J_{\rm \tiny leg}= 166$~meV.
The QC result can be confirmed
if the direct FM  coupling and the residual AFM
couplings 
from $\Phi \neq 90^{\circ}$ and that from the 
$dd$ channel
are taken into account
assuming equal leg and rung AFM exchange integrals. 
Experimentally, however, they differ slightly: $J_{\rm \tiny leg}/J_{\rm \tiny rung}
 \approx 1.1$ for SrCu$_2$O$_3$,
which is caused by different O 2$p$ onsite energies.
We ignore this small difference $\sim$ 15~meV 
 and use for the double-chain
problem the experimental value of $J_{leg}=J_2$.
From Eq.\ (\ref{jladder}) one estimates
$-J^H_1\approx J_2/7 \approx $0.629~eV. Using Eq.\ (\ref{hundratio}) one obtains
1/3 for the set in Ref.~\cite{mizuno}
and $\approx$
1/4 for $K_{pd}\sim 100$~meV. Adopting nearly the same value for the 
ladder compound value as for LICO
obtained here, one would arrive
at $\alpha <$ 4 to 5 
in 
accord
with
a QC prediction for SrCu$_2$O$_3$ \cite{Bordas2006} (see also
SM \cite{Suppl}).
High-energy spectroscopy
and more theoretical studies 
are desirable to put 
material specific upper limits on important 
$U_d$ and $J_H$.  
Without the $dd$ chanel,
the INS data \cite{lorenz} were 
described at $J_H$=0.6 ~eV  
by already enlarged values  $K_{pd}$=81~and 96~meV, 
in accord with 
our 
optical conductivity, EELS, and RIXS spectra for LICO
\cite{Drechsler2010,Drechsler2010a,Monney2013,remarkKPD}.
Figs.\ 15 and \ref{cluster} clearly show
the larger sensitivity of the singlet-triplet separation to $K_{pd}$ 
than
to $J_H$, where only the triplet state is slightly
affected. This 
confirms
$K_{pd}$ as the 
key FM source.

To summarize,  a 
precise general microscopic assignment of the origin
of the observed large $J_1$-values is still difficult since
$K_{pd}$, $J_H$, and the AFM $J_{dd}$ are 
involved.
But without doubt $K_{pd}$ is the leading FM term.
The  short $d_{\rm \tiny Cu-Cu}$ leads to a
 sizable direct intersite AFM superexchange, negligible in 
 corner-sharing cuprates with $\approx$$\sqrt{2}$
 larger $d_{\rm \tiny Cu-Cu}$.
 The FM $K_{pp}$ somewhat reduces the generic AFM $J_2$.
 Cuprates are usually classified as 
 charge transfer insulators, which is not strictly valid here since the $J_1$
 is certainly  affected by the additional superexchange governed 
 by the Cu $U_d$  and the hopping $t_{dd}$ [Eqs.\ (13-15)]
 as in standard Mott insulators.

Within a much simpler effective
single-band extended Hubbard model, larger clusters can be treated exactly. 
Here we include, in addition to the NN transfer integral $t\equiv t_1$,
the Hubbard onsite repulsion $U$,
and a NNN counter part $t_2$ to the former,
the NN and NNN intersite Coulomb interactions
$V_1$ and $V_2$, respectively, two external exchange couplings $\tilde{J}_{1}$
(to allow for a FM NN exchange) and $\tilde{J}_{2}$ to account for the corresponding FM
contribution to $J_2$ arising from 
$K_{pp}$ between O sites
(see \ SM \cite{Suppl}).
For a dimer, an exact analytical expression
is available \cite{Schumann2017} beyond
the  Hubbard-model:
\begin{equation}
J_1=\frac{\sqrt{16t^2+\tilde{U}^2} -\tilde{U}}{2} +\tilde{J}_{1} \quad ,
\label{Dimer}
\end{equation}
where $\tilde{U}= U-V+\frac{3}{4}\tilde{J}_{1}$.
For 
$U \gg |t|, |\tilde{J}_1|$
in Eq.\ (\ref{Dimer}) simplifies to
[used in Eqs.\ (14,15) for the $dd$ channel]
\begin{equation}
J_1\approx 
\frac{4t^2}{\tilde{U}} +\tilde{J}_{1} \quad .
\label{single1}
\end{equation}
The 2nd, "external", term on the right hand side
is 
FM 
and it 
overcompensates the first one, which represents the
non-negligible AFM superexchange.
The dimer model provides also a direct tool for materials
like Li$_3$Cu$_2$SbO$_6$,  Na$_3$Cu$_2$Sb(Te)O$_6$, and other alternating FM-AFM chain compounds
with a dominant FM NN exchange.

\vspace{-0.25cm}
\section {Summary}
 A large magnon
 dispersion 
 up to 53~meV was
observed 
in 
CYCO. 
It is caused mainly by a large 
extracted FM NN
coupling
$J_{a1}=-24$~meV 
exceeding that 
for 
LICO \cite{lorenz} and represents the 
highest value detected
so far for any FFESC cuprate. 
From our experience with 
CYCO, 
a successful search of the 
{\it full} dispersion, up to 
$\sim$45~meV 
would 
refine the
$J_1$ and $J_2$ 
in LICO
and Li(Na)Cu$_2$O$_2$.
The NNN AFM 
$J_{a2}$=5.5 meV
puts CYCO near
to 
criticality
($\alpha$$\sim$0.23)
for a 1D chain but not for
strongly 
enough AFM coupled
NN chains shifted by half a Cu-Cu distance. This
chain structure
causes an O mediated stable FM
alignment of magnetic moments along 
the chain 
in a 
stacked structure of quasi-2D-N\'eel commensurate collinear magnetic 
ordering.
From our analysis and microscopic arguments for
systems with similar 
chains, we expect
highly dispersive magnons for La$_6$Ca$_8$Cu$_{24}$O$_{41}$,
Ca$_2$Nd(Gd)$_2$Cu$_{5}$O$_{10}$, and SrCa$_{13}$Cu$_{24}$O$_{41}$ 
including 
the slightly $h$-doped systems.
Then the available magnon dispersion-law 
should be helpful.

Gap-like features are observed at $\sim 11.5$~meV and $\sim 28$~meV. 
The smaller gap at 11.5 meV is ascribed to 
phonon-magnon coupling 
while the gap at 28 meV is ascribed
to quantum effects due to the 
AFM IC as well as to the non-negligible
inhomogeneous cuprate chain structure caused by the 
misfit with the NN 
Ca/Y chains
generic for the composite symmetry of its two subsystems in CYCO
and in the cases mentioned above.
Both effects are most cooperative for a 
lock-in SS
consisting of 
Ca$_2$Y$_2$Cu$_5$O$_{10}$ and 
Ca$_4$Y$_4$Cu$_{10}$O$_{20}$ domains.

The large $J_1$ values in 
CYCO and LICO
provide deep insight
into the microscopic exchange
pointing
 to a dominant 
direct FM 
interaction
$K_{pd}\sim 105$~meV,
more important
than the indirect O 2$p_xp_y$ exchange mediated by the
Hund's coupling $J_H$.
Noteworthy, we found that the usually ignored direct
Cu-Cu superexchange may somewhat reduce the effect of
FM 
couplings. $J_H$ and $U_d$ 
should be studied systematically \cite{drechsler2019}, especially in view of 
smaller empirical values of $J_H \leq$ 
0.6~eV 
reported for superoxides \cite{Solovyev2014}.
Although the issue of a large FM
$J_1$-value is now almost unraveled, 
we are still left with a new question:
given such natural
$K_{pd}\sim 100$~meV,
what is the reason for the markedly { \it lower} values 
for linarite and some
FFESC materials? Presumably,  ligand effects
lowering the efficiency of $J_H$ and raising the O mediated superexchange.
Thus, seemingly well understood
"classical" systems,
studied already for many
years, are still sources of surprises,
deserved to be studied in more detail 
to elucidate
the 
interactions 
behind the 
exchanges described by various spin-Hamiltonians
and their interplay with structural details.
\begin{acknowledgments}
We
used resources at the High Flux Isotope Reactor 
and Spallation Neutron Source, DOE Office of Science User Facilities 
operated by the Oak Ridge National Laboratory.
Support by the SFB 1143 of the DFG
 is acknowledged (SN).
 We thank U.\ Nitzsche for technical assistance and S.\ Johnston,  
D.\ Miloslavlevic, A.\ Tsirlin, O.\ Janson,
J.\ M\'alek,  
R.\ Klingeler, W.E.A.\ Lorenz, T.\ Schmitt,
C.\ Monney, J.\ van den Brink,  
U.\ R\"o{\ss}ler,
A.S.\ Moskvin, D.\ Khomskii, A.M.\ Oles, 
G.\ Sawatzky, K.\ Wohlfeld, A.\ Yaresko,  
G.\ Roth, and J.\ Thar for discussions.
Special thanks to
GR and JT for providing
figures of the
CuO$_2$ chains in CYCO according to their structure model
as well as to DT, AT, OJ, SL, and UN for providing unpublished 
theoretical results included partly in Table I of
SM \cite{Suppl}. We also thank
J.\ Richter and A.\ Hauser for
help with their 
HTE
package code used here. 
\end{acknowledgments}


\begin{thebibliography}{90}
\bibitem [{rem({\natexlab{d}})}]{remarkdoublechain}%
  \BibitemOpen
  \href@noop {} {}\bibinfo {howpublished} {{{}Another relevant case is that of
  ferromagnetically coupled corner-sharing double (zigzag)
  chains as in SrCuO$_2$ and various ferromagnetically coupled ladder systems
  with a huge NNN AFM coupling $J_2 \gg -J_1$ which dominates
  their high-$T$ magnetic and dynamical response,
  hindering the extraction of $J_1$. An analysis of the FM
  $J_1 $ value like that given here below has not yet been performed. The
  reported relatively large $J_1 \approx -30.8$~meV for SrCu$_2$O$_3$ was
  mainly caused in the simulations \cite{mizuno} by a more than three times
  smaller competing $J_{\Phi}$ via the bridging O as compared to LICU and CYCO
  [see\ Eq.\ (\ref{jafm})]. }} ({\natexlab{d}})\BibitemShut {NoStop}%
\end{thebibliography}
\end{document}


{\bf \Large     SUPPLEMENTARY PART TO}
\vspace{0.5cm}
\newline
{\large \bf "{\large \bf \it \large Highly dispersive magnons with  spin-gap like features  
in the $s=1/2$
frustrated ferromagnetic  chain compound} Ca$_2$Y$_2$Cu$_5$O$_{10}$ 
{\large \bf \it  detected by inelastic neutron scattering}"}
\newline  
  \hspace{10cm} {\it by}
\newline
{\bf \small M.\ Matsuda, J.\ Ma, V.O.\ Garlea, T.\ Ito, H.\ Yamaguchi, K.\ Oka, 
S.-L.\ Drechsler, R.\ Yadav, L.\ Hozoi,\\
 H.\ Rosner, R.\ Schumann, R.O.\ Kuzian, 
S.\ Nishimoto}
\renewcommand{\theequation}{S\arabic{equation}}
\setcounter{equation}{0}
\renewcommand{\thefigure}{S\arabic{figure}}
\setcounter{figure}{0}
\vspace{0.0cm}
\subsection{Dynamical structure factor and gaps in the magnon dispersion for inhomogeneous cuprate chains}
Here we provide 
details of the calculation of the dynamical structure 
factor
$S(\omega,q)$ 
for the 1D spin-Hamiltonian given by Eq.\ (5) in the MT
in terms of Green's functions $G$ (GF) and provide the results for 
the simplest case where within the  
period 5 approximation
for the Ca-Y cationic chain system opens gaps in the magnon curve
in addition to  FIG.\ 11 in the MT for the case of a 
doubling of the CuO$_2$ chain compatible with a period 10
superstucture for CYCO. 
The commutator two-time retarded GF
for the 
operators $\hat{X}$
and $\hat{Y}$ is defined as 
\[
\langle\langle\hat{X}|\hat{Y}\rangle\rangle\equiv-
\imath\int_{t^{\prime}}^{\infty}\!\!dte^{i\omega(t-t^{\prime})}
\left\langle \left[\hat{X}(t),\hat{Y}(t^{\prime})\right]\right\rangle ,
\]
where the expectation value $\langle\ldots\rangle$ denotes the ground
state average, the time dependence of an operator $\hat{X}(t)$ is
given by $\hat{X}(t)=\mathrm{e}^{it\hat{H}}\hat{X}\mathrm{e}^{-it\hat{H}}$.
It is convenient to introduce the notations
\begin{equation}
G_{q} = \equiv\left\langle \left\langle a_{q}|a_{q}^{\dagger}\right\rangle \right\rangle _{\omega}.
\quad  \mbox{and} \quad
G_{q,s} = \equiv\left\langle \left\langle a_{q,s}|a_{q}^{\dagger}
\right\rangle \right\rangle _{\omega}.
\label{eq:Gs}
\end{equation}
Here we consider the 
$J_{1}-J_{1}^{\prime}-J_{2}$ model
We have $n=2$, $s=a,2a$ and introduce
$\varepsilon_{a}=-\frac{1}{2}\left(J_{1}^{z}+J_{1}^{z\prime}\right)-J_{2}^{z}=
\varepsilon_{2a}\equiv\varepsilon $
The equations of motion for the GF
 are
\begin{align}
\left(\omega-\varepsilon-J_{2}\cos2qa\right)G_{q,a} & =
\frac{\mathrm{e}^{iqa}}{\sqrt{2}}+\frac{1}{2}\left(J_{1}\mathrm{e}^{-2iqa}
+J_{1}^{\prime}\right)G_{q,2a},\label{eq:G1}\\
\left(\omega-\varepsilon-J_{2}\cos2qa\right)G_{q,2a} & =
\frac{\mathrm{e}^{2iqa}}{\sqrt{2}}+\frac{1}{2}\left(J_{1}\mathrm{e}^{2iqa}
+J_{1}^{\prime}\right)G_{q,a}.\label{eq:G2}
\end{align}
\begin{equation}
\mbox{Introducing the notations} \quad J_{1} = \equiv J+\delta, \quad , 
J_{1}^{\prime} = \equiv J-\delta, \quad \mbox{and} \quad 
\varepsilon = \equiv-J-J_{2}-D,
\end{equation}
\begin{equation}
\mbox{\rm we obtain} \quad G_{q}=\frac{\mathrm{e}^{-iqa}}{\sqrt{2}}G_{q,a}+\frac{\mathrm{e}^{-2iqa}}{\sqrt{2}}G_{q,2a}
=\cfrac{1}{\omega-\varepsilon-J_{2}\cos2qa
 -J\cos qa-\cfrac{\delta^{2}\sin^{2}qa}{\omega-\varepsilon-J_{2}\cos2qa+J\cos qa}}.
\end{equation}
FIG.\ 11 in the MT  shows the structure factor for this case.
\vspace{-0.cm}
\begin{figure}[b!]
\includegraphics[width=0.32\columnwidth]{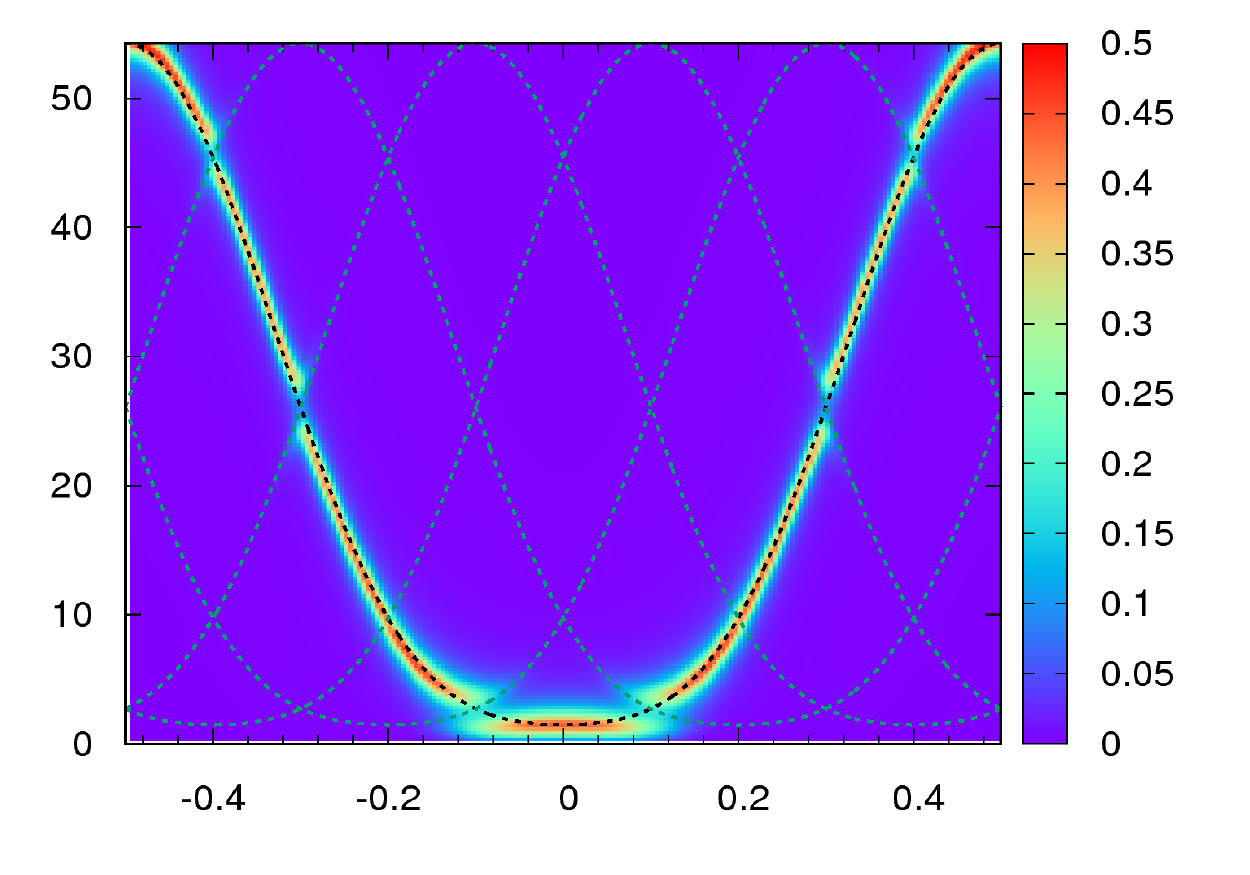}
\includegraphics[width=0.32\columnwidth]{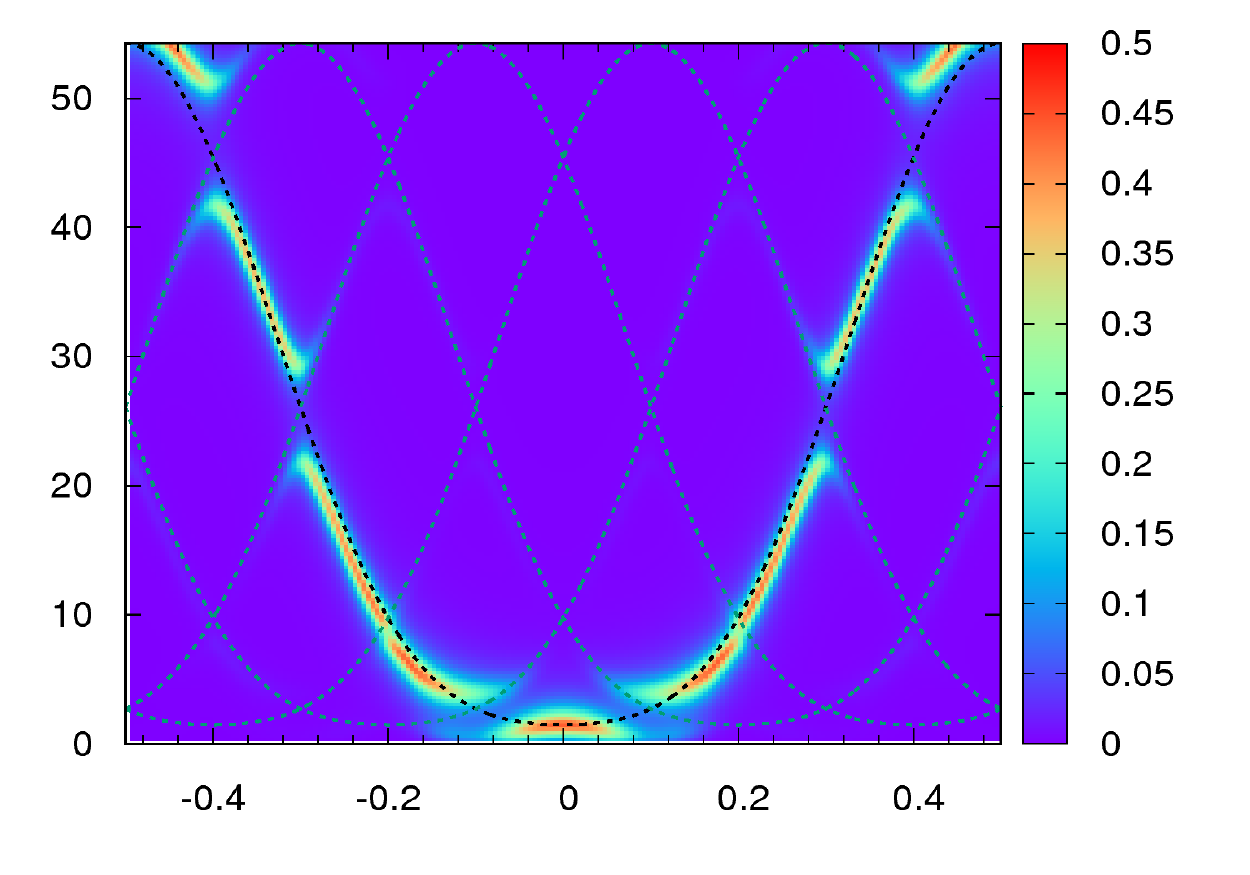}
\includegraphics[width=0.32\columnwidth]{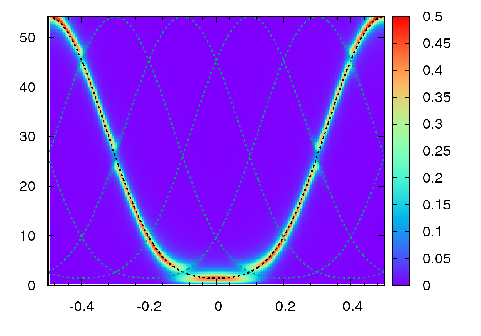}
\caption{The structure factor for a
lattice with a  period 5 and the 
coupling constants of the spin Hamiltonian
shown in FIG.\ 1 of the MT with the relations
between the parameters from
Eqs.\ (S5') and (S5"),
$J=-26.38$, $J_2=5.5$, $D=-1.5$~meV.
\textbf{Left:}
$\delta =\gamma =2$; 
\textbf{Middle:}
$\delta =5$, $\gamma =2$
\textbf{Right:}
$\delta =0$, $\gamma =2$. 
Thin lines: the spectrum for the 5-fold supercell
with $\delta =\gamma =0$.
}\label{per5} 
\end{figure}
For the 1D model depicted in FIG.\ 1b in the MT 
with
$n=5$ 
we have a system of 5 linear Eqs.\
for $G_{q,s}$ 
solved numerically. The result is shown in
FIG.\ S1
for the parameters:
\begin{equation}
J_1^{\prime}=
 J,\quad J_1=J+2\delta ,\quad J_1^{\prime \prime}=J-\delta ,  \ (S5') \hspace{2cm}
J_2^{\prime}=
I,\quad J_2=I+\gamma ,\quad J_2^{\prime \prime}=I-2\gamma \ (s5")  \ .
\nonumber
\end{equation}
\vspace{-0.0cm}
\subsection{FM and AFM contributions to the NN and NNN exchange within
single and 5-band Hubbard models}
Here we sketch the derivation of the isotropic part of the 1D spin-Hamiltonian
given by Eqs.\ (1-5) in the MT within  
the simplest approach of an effective  
generalized extended single band-Hubbard model (ESBHM) 
at half filling and
as well as we
present a general rigorous
analytical solution of the $\Phi=90^o$
problem generic for FFESC 
within the multi-band O$2p$Cu$3d$ model
\cite{Schumann2017}. The former 
includes 4
terms:  kinetic contributions given by the NN and NNN hoppings with the
transfer integrals $t\equiv t_1 \equiv t$ 
and $t_2$, resp., the onsite and  
NN, NNN intersite Coulomb repulsions $U$, $V_1$ and $V_2$, resp.,
and the corresponding  external
intersite exchange 
integrals $\tilde{J}_1$ and $\tilde{J}_2$
introduced to allow for FM NN-spin spin couplings and sizable AFM
NNN exchange couplings, resp., missing in the simple Hubbard model.
\small
\begin{equation}
H=\sum_{i,\sigma }
t\left(c^{\dagger}_{i+1}c_i+h.c. \right)+ t_2\left(c^{\dagger}_{i+2}c_i+h.c. \right)+
Un_{i,\uparrow}n_{i,\downarrow} +V_1n_{i+1}n_{i}+V_2n_{i+2}n_{i}
+\tilde{J}_1c^{\dagger}_{i+1,\uparrow}c_{i,\uparrow}c^{\dagger}_{i+1,\downarrow}c_{i,\downarrow} +
+\tilde{J}_2c^{\dagger}_{i+2,\uparrow}c_{i,\uparrow}c^{\dagger}_{i+2,\downarrow}c_{i,\downarrow} \quad .
\label{eq: extHub}
\end{equation}
\normalsize 
where $n_i$ is the density
operator.
With respect to edge-sharing CuO$_2$ chain compounds and LICU in particular, the ESBHM
parameters obey the following hierarchy of inequalities generic for strong coupling
(strong correlations):
\begin{equation}
U > 3V_1,  V_1 \geq 5V_2, V_2\geq 2t_1\sim 2t_2, t \gg \mid \tilde{J_1}\mid \gg \tilde{J_2} \ ,
\end{equation}
which allows one to expand tedious expressions.
Noteworthy,  relative small interactions as the NN intersite
Coulomb interaction $V_2$ might cause visible effects. [s.\ e.g.\ FIG.\ 13 of
 the MT and Eq.\ (S13).]
For the trimers and dimers  
the ESBHM 
[Eq.\ (\ref{eq: extHub})]
can  be exactly diagonalized even in an analytic form for any parameters. 
 To start with the dimer, we deal with 16 eigenstates in the ESBHM. Among them there are 
 a singlet (S) and 1  triplet (T), 
 state, i.e.\ 
  in total 4 spin states
  relevant here.
 The energy of the triplet and singlet states are respectively given by
 \begin{equation}
 E_{T}= \tilde{J}_1/4 + V_1, \quad \quad E_S=-0.5\left(\sqrt{16t^2 +\tilde{U}^2}-\tilde{U}\right) 
 +V_1-3\tilde{J}_1/4 \quad  , \quad \mbox{with} \quad \quad \tilde{U}_d=U-V_1+3\tilde{J}_1/4  \ .
 \end{equation} 
 For a FM $\tilde{J}_1<0$ the ground state (GS) is formed by the triplet
 and for an AFM $\tilde{J}_1$ by the singlet state.
 The difference of the eigenenergies 
 yields  the NN exchange integral $J_1$
 of the corresponding spin-Hamiltonian
 Thus, we have {\it exactly}
 \begin{equation}
 J_1-\tilde{J}_1 =0.5\left(\sqrt{16t^2 +\tilde{U}^2}-\tilde{U}\right) 
 \approx 4t^2/\tilde{U} \ .
 \label{J1dimer}
 \end{equation}
 The latter Eq.\ is valid for 
 $U\gg t$ relevant here and for moderate $V_1\ll U$ values, in particular
 for $V_1=0$ adopted in Ref.\ \onlinecite{mizuno}.
 Eq.\ (\ref{J1dimer}) can be expanded for small $t/\tilde{U}_d$
 For $\tilde{J}_1=0, V_1=0$ one has $J_1=0.5(\sqrt{16t^2+U^2}-U)$.  
 At strong coupling
 it yields the 
 AFM superexchange
 4$t^2/U$ of the Hubbard model.
 Turning 
 to the asymmetric dimer with 2 holes on 3 orbitals:
 1 Cu and 2 planar O orbitals, we  note that there is an exact very tedious
 analytical solution for {\it any} ESBHM parameter values available 
 \cite{Schumann2017}.
 That spin 
 model has $2^3=8$
 states, among them: a quadruplet  with the energy $E_Q$ and total spin $S=3/2$ as well as
  2 doublets with 
 energies $E_{D_1}$, $E_{D_2}$, and spin $S=1/2$. These energies provide 
 the mapping Eqs.\ for the exchange
 parameters we are looking for:
 \begin{equation}
 E_Q=J_1/{2}+J_2/{4}, \quad E_{D_1}=-3J_2/{4}, \quad E_{D_2}=-J_1+J_2/{4} \ .
 \quad \quad
 \mbox{\rm As a result we have }
 \label{eq: doublet1}
 \end{equation} 
 \vspace{-0.0cm} 
 \begin{equation}
  \quad J_1=2\left(E_Q-E_{D_2} \right)/3 \ \mbox{and} \quad  \  
 J_2=-J_1/2+E_Q-E_{D_1} \ . 
 \mbox{The quadruplet energy in terms of the ESBHM
 reads }
 \end{equation}
 \vspace{-0.0cm}
 \begin{equation}
 E_Q= 2V_1+V_2 +\tilde{J}_1/{2}+ \tilde{J}_2/{4}\ \quad \quad 
 \mbox{Then Eqs.\ (\ref{eq: doublet1} -\ref{eq: doublet3}) can be rewritten as:} 
 \label{eq: doublet3}
 \end{equation}
 \vspace{-0.0cm}
\begin{equation}
J_1=\tilde{J}_1-2\varepsilon/3
J_2=\tilde{J}_2+\left(\varepsilon -\epsilon\right)/3, \ \mbox{where} \ 
 E_{D_1}=\epsilon +2V+V_2 -3\tilde{J}_/4 .
\end{equation}
The exact analytical expressions for  
the ESBHM
are very cumbersome to be given here. 
The shifted eigenvalues $\varepsilon$ and $\epsilon$ have been introduced
to find approximations for the 2 corresponding ones of the doublet
states ar strong coupling.
The energies of the doublets $D_1$ and $D_2$ are given by the lowest 
 eigenvalues of their two 4$\times$4 matrices, respectively: 
\begin{eqnarray}
 \hat{D}_1&=& \left(\begin{array}{cccc}
-\frac{3}{4}\tilde{J}_2+2V+V_2&t&-2t_2&t\\
t&U+V_2-t_2&t&0\\
-2t_2&t&U+2V&-t\\
t & 0 & -t & U + 2V +t_2 \\
\label{eq: D1}
\end{array}\right)  \ , 
   \\
 \hat{D}_2&=&
\left(\begin{array}{cccc}
-\tilde{J}_1+\frac{1}{4}\tilde{J}_2+2V+V_2 & \sqrt{3}t &0&-\sqrt{3}t\\
\sqrt{3}t&U+V_2+t_2&t&0\\
0&t&U+2V&-t\\
-\sqrt{3}t & 0 & -t & U + 2V -t_2 \\
\label{eq: D2}
\end{array}\right)   \ .
\end{eqnarray}
Both Eqs.\ can be rewritten to get analytic expressions
in {\it any} order of the perturbation theory by an iterative procedure:
\begin{equation}
\mbox{i.e.\ the eigenvalue problem of Eq.\ (\ref{eq: D1}) can be rewritten as} \quad
 J_1-\tilde{J}_1=-\frac{2}{3}\varepsilon=
\frac{2t^2\left[\frac{1}{z_1-\varepsilon}+\frac{1}{z_3-\varepsilon}\right]}
{1+\frac{t^2}{z_2-\varepsilon}
\left[\frac{1}{z_1-\varepsilon}-\frac{1}{z_3-\varepsilon}\right]} \ .
\end{equation}  
\begin{equation}
\varepsilon=
-\frac{3t^2\left[\frac{1}{z_1-\varepsilon}+\frac{1}{z_3-\varepsilon}\right]}
{1+\frac{t^2}{z_2-\varepsilon}
\left[\frac{1}{z_1-\varepsilon}-\frac{1}{z_3-\varepsilon}\right]}
 \ , \ \mbox{or more explicitly} \ \varepsilon^{(n+1)}=
-\frac{3t^2\left[\frac{1}{z_1-\varepsilon^{(n)}}+\frac{1}{z_3-\varepsilon^{(n)}}\right]}
{1+\frac{t^2}{z_2-\varepsilon^{(n)}}
\left[\frac{1}{z_1-\varepsilon^{(n)}}-\frac{1}{z_3-\varepsilon^{(n)}}\right]} ,
\end{equation}
$$
\mbox{\rm where} \quad Z_1=\tilde{U}_2-\left(2V-V_2\right)+t_2+\tilde{J}_1 -\tilde{J}_2,
 \quad Z_2=\tilde{U}_2+\tilde{J}_1 -\tilde{J}_2, \ 
 Z_3=\tilde{U}_2-t_2+\tilde{J}_1 -\tilde{J}_2 \ .
$$
The exact solution is reproduced for $n \rightarrow \infty$.
Anyhow, at strong coupling, 1 or 2 iterations
are sufficient for an insight. 
Our recursion formula provides 
analytical expressions in any even order
of the perturbation theory  in terms of 
 $\left(t/U)\right)^{2n}, \  n =0,1,2 ...$.  
 Notice the asymmetry between the $t$ and $t_2$ expansions, the latter contains also
 {\it odd} terms of $t_2/U$.
 Iterating  Eq.\ (17), we have  in 0$^{th}$ and 1$^{st}$ order  
$\varepsilon^{(0)}=0$ or $J_1=J_1^{(0)}=\tilde{J}_1$, resp., according to Eq.\ (S16) and
 \vspace{-0.0cm}
 \begin{equation}
 \hspace{-0.1cm}\mbox{introducing} \quad U_J=U_2+\tilde{J}_1-\tilde{J}_2=U-V_2+\tilde{J}_1-\tilde{J}_2/4,
 \quad \
 U_2=U-V_2+3\tilde{J}_2/4,\quad \mbox{in the next step we arrive at} 
 \end{equation} 
\begin{equation}
\hspace{-0.2cm}
J^{(1)}_1-\tilde{J}_1=
2t^2\frac{\frac{1}{Z_1}+\frac{1}{Z_3}}
{1+\frac{t^2}{Z_2}
\left[\frac{1}{Z_1}-\frac{1}{Z_3}\right]}
=\frac{4t^2}{U_1}
\left[1+\frac{1}{2}\left(\frac{v_1}{1-v_1} +\frac{v_2}{1-v_2} \right) \right]
\left[1+\frac{t^2}{\left(U_1-V_2\right)U_1}\left(\frac{v_1}{1-v_1}-\frac{v_2}{1-v_2}\right) \right]^{-1},
\end{equation}
\begin{equation}
\mbox{\rm where } \quad U_1=U+\tilde{J}_1-
\tilde{J}_2/4,  \quad 
 v_1=
 \left(2V-t_2\right)/
 U_1
 , \ \mbox{and} \ v_2=
 \left(V_2+t_2
 \right)/U_1
 \ .
\end{equation}
After the 2nd 
iteration step 
one arrives
at a similar expression,  where 
the large $U_1 \gg J_1-\tilde{J}_1$ is slightly renormalized:
\begin{equation}
\tilde{U}_1=U_1+
3\left(J^{(1)}_1-\tilde{J}_1\right)/2 \quad \mbox{or} 
\quad \tilde{U}^{(n+1)}_1= U_1
 +
 3\left(J^{(n)}_1/2-\tilde{J}_1\right) .
\end{equation}
Thus, the recursion results in a fast 
converging renormalization of 
$U_1$.
 The recursion for $E_{D_1}$, i.e.\ $\epsilon$ reads
 \begin{equation}
 \epsilon=-
 \frac{
 4t_2^2\frac{1}{
 \tilde{U}_{22}-\epsilon }
 +
 t^2
 \left[ 
 \frac{1}{\tilde{U}_{11}-\epsilon}+\frac{1}{\tilde{U}_{33}-\epsilon}+4t_2\frac{1}{U_{22}-\epsilon}
 \left(\frac{1}{\tilde{U}_{11}-\epsilon}-\frac{1}{U_{33}-\epsilon }\right)
 \right]
 -4t^4\frac{1}{(U_{11}-\epsilon )(U_{22}-\epsilon )(U_{33}-\epsilon )}
 }
 {1-t^2
 \frac{1}{U_{22}-\epsilon }
 \left[
 \frac{1}{U_{11}-\epsilon}
 +\frac{1}{U_{33}-\epsilon}
 \right]
 } 
 \ ,
 \label{eq: epsilon}
 \end{equation}
 \begin{equation}
 \mbox{\rm where} \quad U_{11}=U-2V-t_2+
 3\tilde{J}_2/4 , \quad U_{22}\equiv U_2=U-V_2+
 3\tilde{J}_2/4 ,
 \ \mbox{and} \quad U_{33}=U-V_2+t_2+
 3\tilde{J}_2/4 .
 \end{equation}
 Note the  dominant AFM term triggered by $t_2$ and
 the weak FM 4th order term corresponding
 to N\'{e}el fluctuation with $\parallel$ NNN spins. The 2nd order term
 vanishes for $V=t_2=0$ due to the compensation with a
 2nd order term from $\varepsilon /3$.
 Since $\tilde{U}_2 \gg -\epsilon $, an  iteration procedure 
 applied to Eq.\ (\ref{eq: epsilon}) converges fastly.
  Then
 for $J_2$ we arrive at
 \small
 \begin{eqnarray}
 J^{(1)}_2-\tilde{J_2}&=&\frac{4t^2_2}{\tilde{U}_{22}-\epsilon^{(0)}}+
 t^2\left[ \frac{1}{U_{11}-\epsilon^{(0)}}+\frac{1}{U_{33}-\epsilon^{(0)}}+t_2\frac{4}{U_{22}-
 \epsilon^{(0)}}
 \left(\frac{1}{U_{11}-\epsilon^{(0)}}-\frac{1}{U_{33}-\epsilon^{(0)}}\right)
 \right]
  +\nonumber \\
 &&-t^2
\left[\frac{1}{\tilde{U}_2+t_2-2\left(V-V_2\right)+\tilde{J}_1-\tilde{J}_2}
+\frac{1}{\tilde{U}_2-t_2+\tilde{J}_1-\tilde{J}_2}\right]
\nonumber\\
&\approx& \frac{4t_2^2}{U_{22}}+t_2\frac{4t^2}{U_{11}U_{22}U_{33}}
\left(2V-V_2\right)+
t^2\left(2t_2+\tilde{J}_1-\tilde{J}_2\right)\left[1/\left(U_{11}Z_1\right)
+1/\left(U_{33}Z_3\right)\right], 
\\
 \epsilon^{(0)}&=&-0.5\left(\sqrt{U^2_{22}+16t_2^2}-U_{22} 
  \right)\approx -4t^2_2/U_{22} .\\
  \nonumber
 \end{eqnarray}
 \normalsize
 For $t\equiv t_1=0$, i.e.\ if for $\Phi=90^\circ$ one ignores the 
 different onsite energies and $t_{dd}$ or 
 due to an accidental compensation of the $dd$ and the $pd$
 channels.
Then the central spin
 is magnetically coupled only by the external couplings $\tilde{J}_1$
  and we arrive at a dimer-like formula for $J_2$ replacing
  $V_1$ by $V_2$ [s.\ Eq.\ (\ref{J1dimer}).], i.e.\
  $V_1$ dropes out:
 \small
 \begin{equation}
 J^{(0)}_1=\tilde{J}_1 \quad  \mbox{and} \quad 
 J^{(0)}_2-\tilde{J}_2=8t_2^2/\left(\sqrt{16t_2^2+U^2_2}+U_2\right)\approx
  4t_2^2/\left(U_2+4t_2^2/U_2\right)\approx
 4t^2_2/U_2 .
\end{equation}
\normalsize
Turning to $J_2^{(1)}$ and $J_2^{(2)}$, we note the appearance of 
3 FM contributions at most.
The 4th order term $-4t^4/\left(U_{11}U_2U_{33}\right)$
is independent of $\tilde{J}_1$. It
is indeed small, although it can be significantly enhanced as compared with the simple
Hubbard model [s.\ FIG.\ S3] $\sim$ by the factor $U_2/\left(1-2V+V_2-t_2\right)$
for strong intersite Coulomb repulsions.
 The FM terms induced by
the auxiliary FM couplings as well as for $t_2 <0$ and for $2t_2 < \tilde{J}_2-\tilde{J_1}$
occur already in the 2nd order of $t/U$.
Neglecting the 2nd term $\frac{4t^2_2}{U_2} \approx 6.4$~meV in the last denominator
of Eq.\ (S26)
introduces a small error for $J^{(0)}_2-\tilde{J}_2$ 
$\sim$
1.6$\cdot 10^{-3}$, only.
Anyhow, ignoring the weak $\tilde{J}_2$ using $U_2\approx 4$ eV and $t=80$~meV
we would arrive at $J^{0)}_2=6.39$~meV in accord with the INS data. Taking into account
a further slight enhancement by the 2nd order effect from the finite $t$ there is some room
for a small FM $J_2$-value of the order of 0.1$\tilde{J}_1$. Thereby a non-negligible
intersite Coulomb repulsion might be helfpul according to Eq.\ (S26).
Without more sophisticated microscopical studies we will 
ignore a sizable FM contribution to $J_2$ in accord with our QQC
$K_{pp}$ and adopt hereafter 
$\tilde{J}_2\approx 0$. 

 Now we may follow the weak evolution of $J_1$ with 
 increasing cluster size as well
 as provide a first estimate for $J_2$, too. 
 First, in the simple Hubbard model extended by $\tilde{J}_1$, only,
 we arrive  at:
 \begin{equation}
 J^{(0)}_1=\tilde{J}_1, \quad
 J^{(1)}_1-\tilde{J}_1=\frac{4t^2}{U+\tilde{J}_1}, \quad 
 J^{(2)}_1-\tilde{J}_1=\frac{4t^2}{U+\tilde{J}_1+6t^2\frac{1}{U+\tilde{J}_1}}, \quad
 J^{(2)}_1-\tilde{J}_1=\frac{4t^2}{U+\tilde{J}_1+6t^2\frac{1}{U+\tilde{J}_1+6t^2/\left(U+\tilde{J}_1\right)}} .
 \end{equation}
 Thus, the AFM contribution to $J_1$ is only very slightly enhanced:
 ongoing from the dimer to the trimer
 \begin{equation}
 \frac{J_{\rm 1, trimer} -\tilde{J}_1}
 {J_{\rm 1, dimer}-\tilde{J}_1}
 =\frac{1+\frac{3}{4}\frac{\tilde{J}_1}{U}}
 {1+\frac{\tilde{J}_1}{U}}
 \approx 1-\frac{\tilde{J}_1}{4U} \approx 1.002.  \end{equation}
  adopting $\tilde{J}_1\sim -25$~meV from the INS data and $U\approx 4$~eV in 2nd
  order and 1.0028 in 4th order of $t/U$, respectively. Similar very small changes
  occur for larger clusters. Hence, a simple dimer-trimer approach 
  is justified.
 
 Next, switching on $t_2$ and $\tilde{J}_2$ we arrive also at a very small 
 change of $J_1 \sim 4\times10^{-4}$ 
 or $t_2=80$~meV and $U\approx 4$~eV:
 \begin{equation}
 J^{(1)}_1-\tilde{J}_1=\frac{4t^2}{U_1}\left[1+\frac{t_2^2}{U^2_1-t^2_2}\right].
 \end{equation}
 Finally, we will consider the more interesting and important 
 case of finite NN and NNN intersite Coulomb repulsions $V$
 and $V_2$. We restrict ourselves to the
lowest order expansion in the strong coupling limit $|t|/U \sim 2.5\times 10^{-2} \ll 1$ 
relevant here. The AFM contribution to both exchanges are monotonously increasing functions
of $V$ and $V_2$. For small $V \ll U$ it can be linearized. For $t_2>0$ as considered
here there is an absolute enhancement for $V> 0.5t_2$ (ignoring the weak $V_2$ which further reduces
this critical value).
 \begin{figure}[b]
\includegraphics[width=6.0cm]{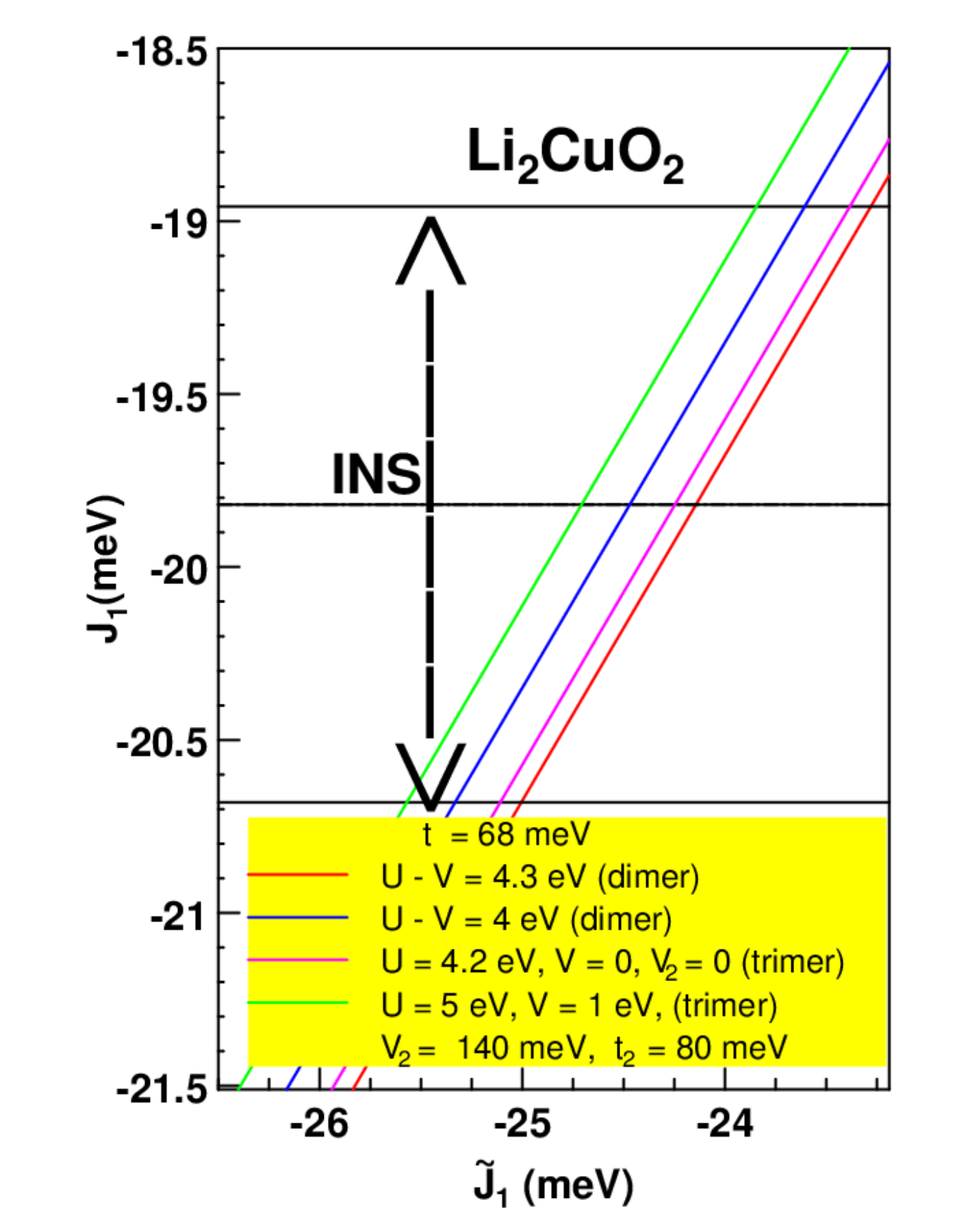}
\includegraphics[width=5.8cm]{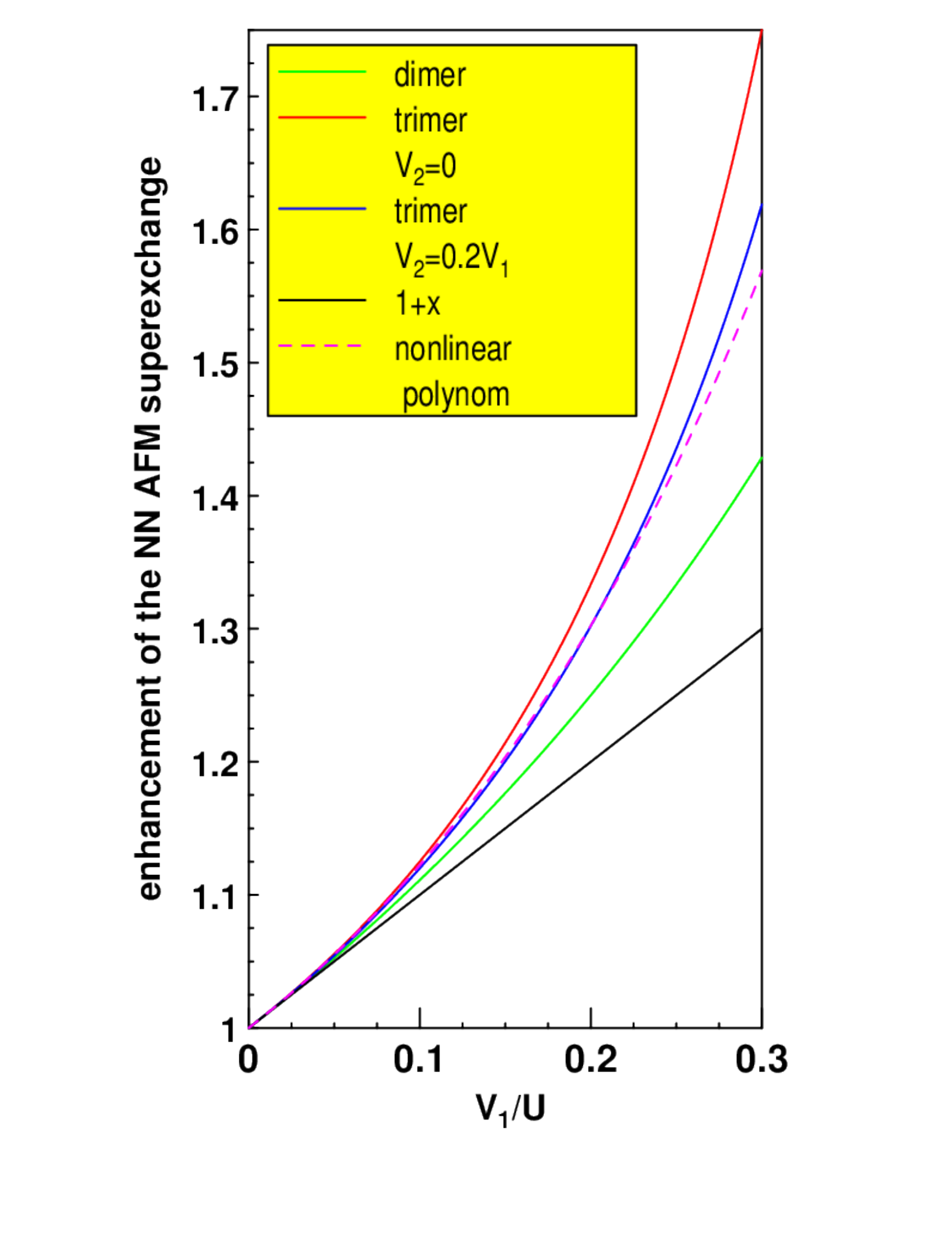}
\caption{{\bf Left:} The empirical NN exchange $J_1$ from the INS \cite{lorenz} 
LSWT analysis including the experimental error bars
vs.\ various possible phenomenological FM contributions $\tilde{J}_1$
of the extended one-band Hubbard model (E1BH).
{\bf Right:} relative enhancement of the AFM contribution to $J_1$ as a function of the
NN intersite Coulomb repulsion $V=V_1$ for a trimer and a dimer cluster as well as
the NNN $V_2$ for a trimer within the E1BH-model based on Eq.\ (S24).
}
\label{V1}
\end{figure}
Then one arrives in 2nd order of $t/U$ at
 \small
 \begin{equation}
 J_1^{(1)}-\tilde{J}_1=
 \frac{4t^2}{U_1}
\left[1+\frac{1}{2}\left(\frac{v_1}{1-v_1} +\frac{v_2}{1-v_2} \right) \right] \ .
 \end{equation}
 \normalsize 
 At variance with the cases considered above, a relative strong intersite 
 Coulomb interaction $V\sim U/4 \sim 1$~eV, i.e.\ $v_1 \sim 0.5$, can in principle
 considerably enlarge the AFM contribution
 to $J_1$  (See\ FIGs.\ S2 and S3.) and this way also add a non-negligible FM
 contribution  via $-J_1$ to $J_2$ according to Eq.\ (S16) which is only
 partially compensated by its own induced AFM contribution.
 This increase leads to an enhancement of the phenomenological FM
 $\tilde{J}_1$-value which roughly scales with 
 $K_{pd}$. 
 The influence for realistic $V_2 \approx t_2\sim
 160$~meV by 
 the 2nd factor in Eq.\ (S34) is very weak. The increase
 in the trimer 
 stems from the 
 different
 energy denominators $\approx U-2V$ vs.\ $U-V$ in the dimer case with one
 bond, only. Notice that this 
 $V$-dependence cannot be described
 in terms of a small $V/U$ expansion, 
 instead
 the analytic expression with the correct account of the denominator $\sim
 U-2V$ in Eq.\ (S30) is essential.

Notice that in both small superexchange terms $\propto t^2$ of Eqs.\ S17 and S18,
the small terms $V_2$ and $\tilde{J}_2$
might be ignored. Indeed, for the former one estimates
$V_2/V_1  \sim 0.5/\varepsilon_{\infty} \sim 1/6 \div 1/7 $, 
where $\varepsilon_{\infty} \approx 3 \div 3.5$ is the dielectric constant.
For the latter value we estimate $-\tilde{J}_2/\tilde{J}_1 <K_{pp}/K_{pd}
\sim 1/10 \div 1/5$.
\narrowtext
\begin{figure}[b]
\includegraphics[width=6.6cm]{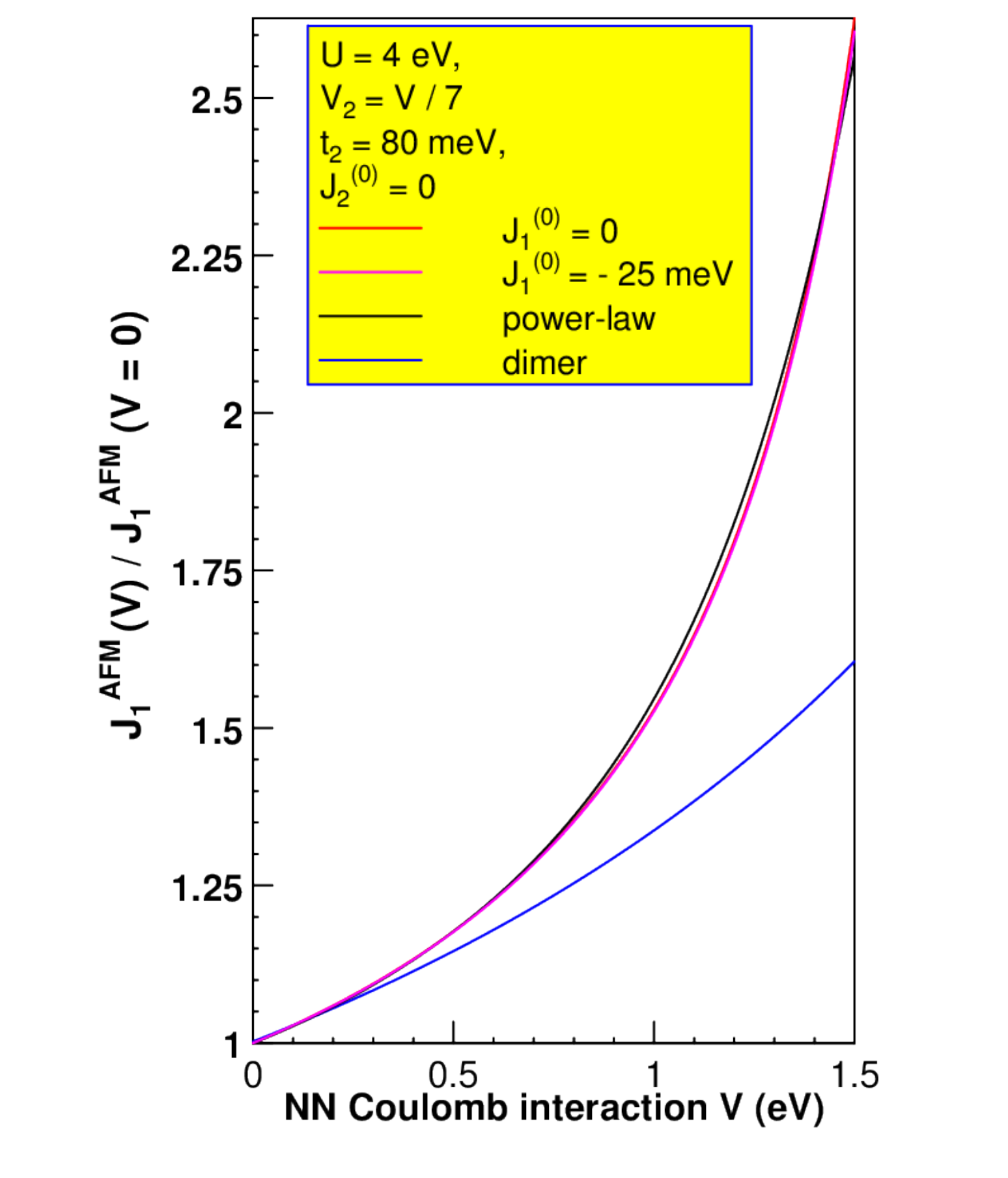}
\includegraphics[width=6.0cm]{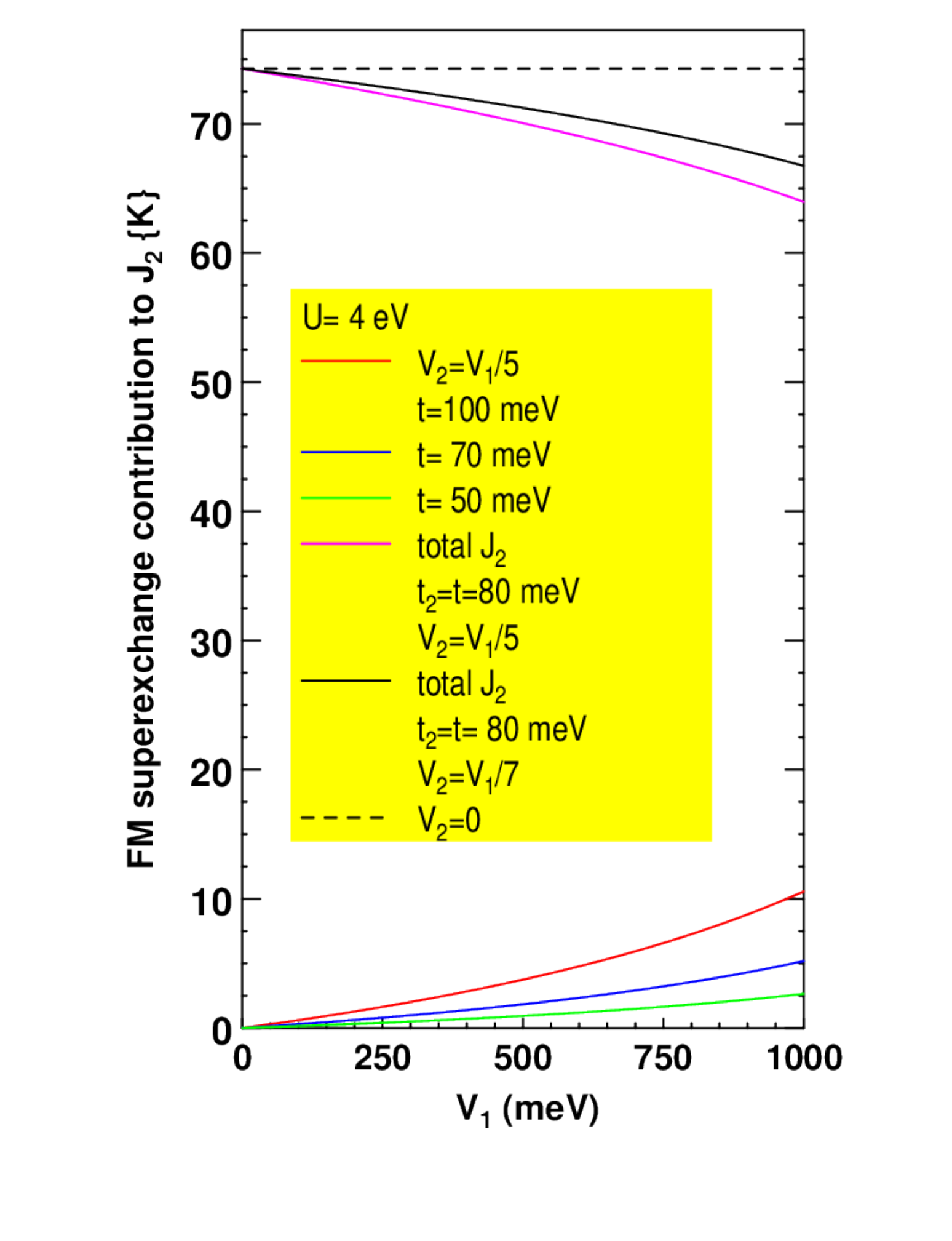}
\caption{Left: relative enhancement of the AFM contribution to $J_1$ vs.\ the
intersite Coulomb interactions $V=V_1$ ((NN) and $V_2$ (NNN).
Right: enhancement of the FM contribution to $J_2$ (lower curves) and total $J_2$ (upper curves)
up to the constant shift due to a finite FM $\tilde{J}_2$ (from $K_{pp}$)
 vs.\ $V_1$ and two typical values of $V_2$ adopting various dielectric screenings. }
\end{figure}
Notice that in both expressions the 2nd order higher terms in $t$ may compete (slightly reduce)
with the dominant first FM term and with the NNN hopping $t_2$ induced AFM 
superexchange (second term). 
In particular, the occurrence of the dominant AFM contribution
from $t_2$ (i.e.\ the O-O derived transfer integral is decisive for the frustration
in the edge-sharing CuO$_2$ chains, i.e $J_2$ is always AFM in nature.
In FIG.\ S3 the influence of the NN and NNN intersite Coulomb interactions $V_1$ and
$V_2$ on the NN AFM superexchange $\propto t^2$ is depicted.
Notice the pronounced nonlinear dependence caused by the formal divergence
of the expression in brackets at $V_1=(U+V_2+\tilde{J}_1-\tilde{J}_2/4)/2$.
[For the FFES CuO$_2$ chain systems under consideration
$V_1/U {\vspace{0.8cm} \small \stackrel{<}{\sim}} 1/4$ 
is expected which is still not too close to that point
of artificial divergency
justifying the lowest order in the expansion used in Eq.\ (S24).] 
Notice the dominant influence of a small but finite $V_2$ for the 
FM superexchange contribution to $J_2$
in the present lower order approximation. 
%
Finally, we turn to the case
of 2 holes and $\Phi=90^\circ$ 
It is usually considered
in low orders of the perturbation theory in terms of $t_{pd}/\Delta$ 
\cite{Geertsma96,Tornow} which however
exhibits in the presence of 
$K_{pd}$, intersite 
Coulomb interaction $W$ and especially $t_{pp}$ a  
slow convergency behavior. Hence, we present here an exact analytic
expression taking fully into account the first 3 interactions and postpone the third 
interaction relevant for the intra-plaquette O-O hopping for a future study.
Thus, we deal with a 4-site (orbital) problem with two Cu 3$d_{xy}$ sites
and one O-site with two orbitals in between.
Within the planar case we have the following parameters:
the two onsite repulsions $U_d > U_p$, the onsite interorbital
repulsion $U_{pp}$ affecting
the Hund's rule coupling $J_H$ according to Eq.\ (21) in the MT, 
the intersite repulsion $W=V_{pd}$, the direct FM exchange $K_{pd}$ and the hopping
$t_{pd}$. The   O and Cu onsite energies $\Delta=\varepsilon_p-\varepsilon_d$
were measured as $\pm \Delta /2$, respectively. Then the GS is given by the 
lowest eigenvalue from the triplet matrix:
\vspace{0.0cm}
\begin{eqnarray}
 \hat{T}_{GS}&=& \left(\begin{array}{ccc}
W-\frac{1}{4}K_{pd}&\sqrt{2}t&\sqrt{2}t\\
\sqrt{2}t&\Delta -\frac{1}{4}J_H&0\\
\sqrt{2}t&0&-\Delta\\
\end{array}\right)  \ , 
   \\
 \hat{S}_2&=&
\left(\begin{array}{ccc}
W+\frac{3}{4}K_{pd}& \sqrt{2}t &\sqrt{2}t\\
\sqrt{2}t&\Delta +\frac{3}{4}J_H&0\\
\sqrt{2}t & 0 & -\Delta  \\
\end{array}\right)   \ ,
\end{eqnarray}
\vspace{0.0cm}
For simplicity we show here the eigenvalues 
ignoring the intersite interaction 
$W$.   
Then the triplet energy reads
\small
\begin{eqnarray}
E_T&=&\frac{1}{12}\left[K_{pd}+J_H-2\sqrt{(J_H-K_{pd})^2-K_{pd}J_H-12J_H\Delta+48\Delta^2+192t^2_{pd}} 
\cos \left[\frac{1}{3}\arccos (\phi_T) \right]\right], \nonumber \\
\phi_T&=&\frac{(J_H-K_{pd})(72\Delta^2 -K_{pd}^2+K_{pd}J_H+2J^2_H-36J_H\Delta-288t^2_{pd})}
{2\left((J_H-K_{pd})^2+K_{pd}J_H+12J_H\Delta+48\Delta^2+192t^2_{pd}\right)^{3/2}}
\end{eqnarray}
\normalsize
\vspace{-0.0cm}
\begin{figure}[b]
\includegraphics[width=7.cm]{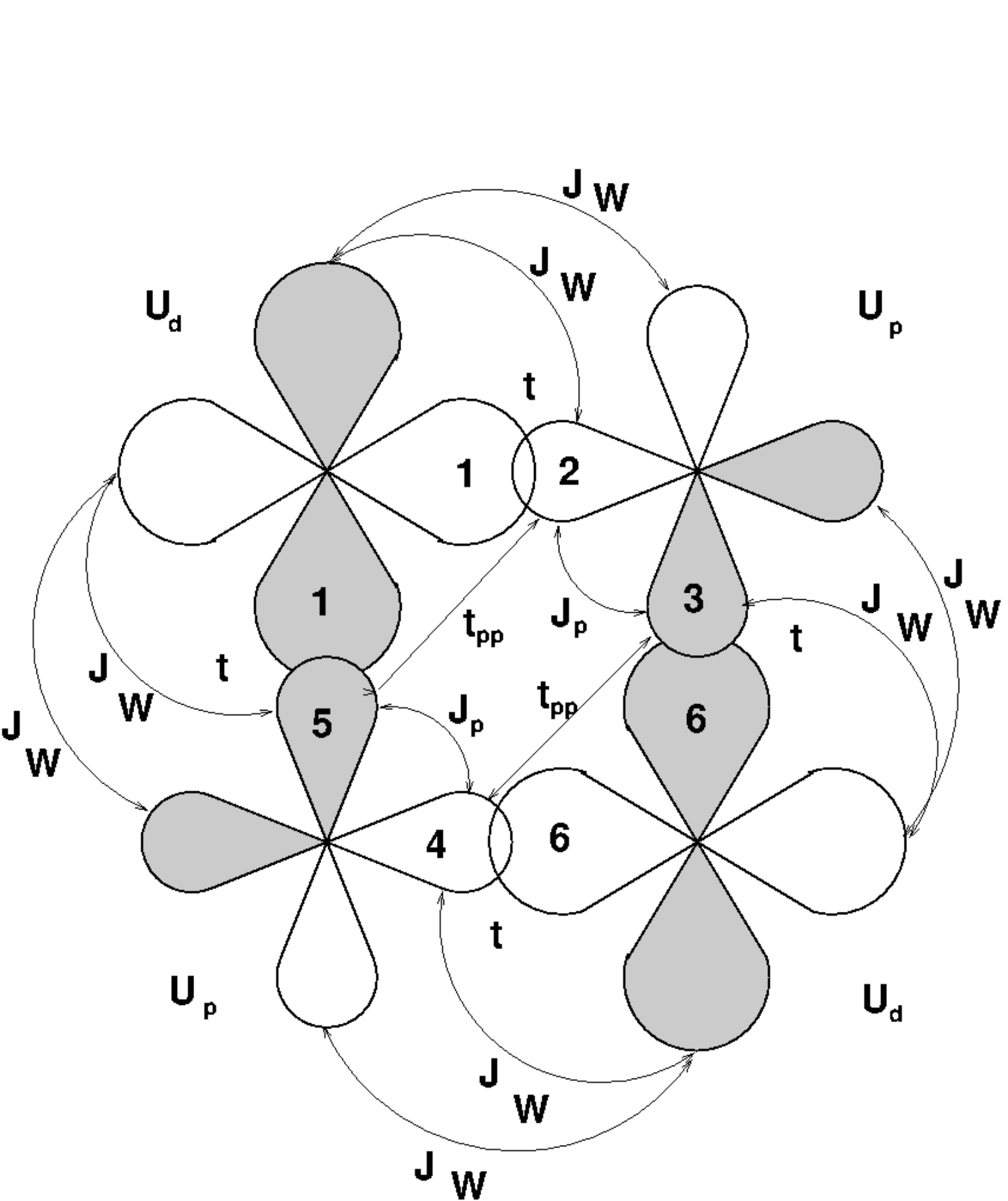}
\includegraphics[width=6.0cm]{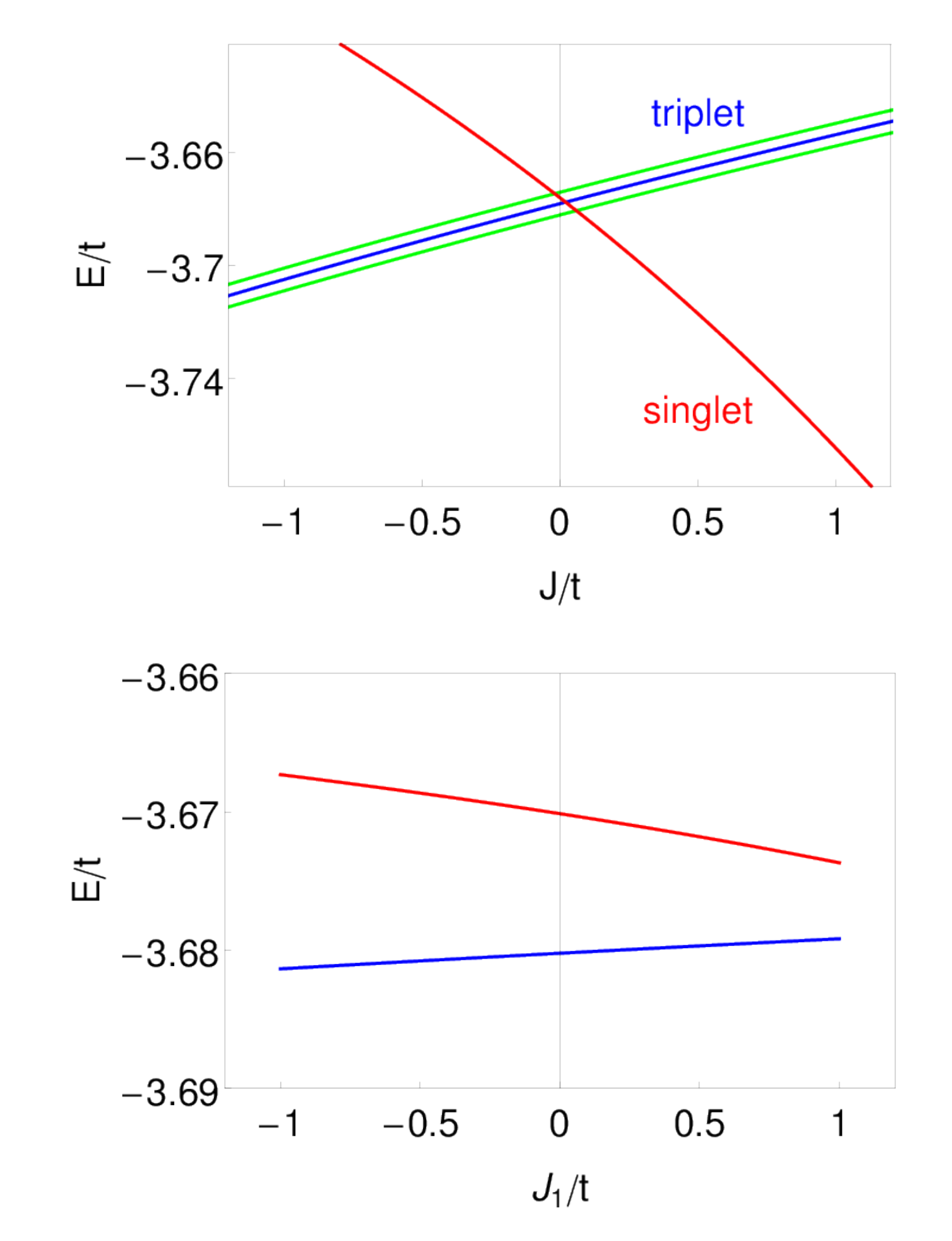}
\caption{{\bf Left:}Orbitals and transfer integrals 
of a CuO$_2$Cu-cluster treated  exactly in the 
planar Cu 3$d$ O 2$p$ 5-band Hubbard model: Cu 3$d_{xy}$ 1 (left) and 6 (right); 
intermediate O 2$p_{x,y}$ 
2,3 (upper)  3,4 (lower). Here at variance to FIG.\ 13 in the MT 
a representation of the O 2$p$ states $\parallel$ and $\perp$ to the Cu-O bonds
has been chosen and
the chain axis is taken along the descending diagonal.
For the 3 exchange couplings the 
notations
$J=-K_{pd},\ J_p=-J_H,\ J_{dd}=-K_{dd}$ have been used. $W$ is the intersite
(Cu-O) Coulomb interaction. The interorbital Coulomb repulsion at O 
$U_{pp}$
is included via $J_H$ and Eq.\ (21) in
 the MT. 
{\bf Right:}The lowest triplet and singlet states as a function of  $K_{pd}$
($J\leq 0$. {\bf Upper:} without intersite Cu-O Coulomb interaction $W=0$
and finite $V_{pd}=0.8t_{pd}$ (lower) for $\Delta=3.2$ and weak 
$J_H=0.5t_{pd}$ and $J=-K_{pd}$= 0.1. 
In both cases $U_d$= 8, $U_p$=4, and $\Delta_{pd}$=3.3 in units of $t$. 
To show the lifting of the 3-fold degenerated triplet state
an external magnetic field (green lines) with $h= 0.002$ was applied. (For
$g$=2 this corresponds to 
17.3 T.)
The $dd$- channel is switched off (i.e.\ $t_{dd}$=$J_{dd}$=$K_{dd}$=$V_{dd}$=0).
}
\label{bond90}
\end{figure}
The corresponding expression for the lowest singlet state is
given by
\small
\begin{eqnarray}
E_S&=&\frac{1}{4}\left[K_{pd}+J_H-2\sqrt{(J_H-K_{pd})^2-K_{pd}J_H-12J_H\Delta+48\Delta^2+192t^2_{pd}} 
\cos \left[\frac{1}{3}\arccos (\phi_S) \right]\right], \nonumber \\
\phi_S&=&
\frac{3\sqrt{3}(J_H-K_{pd})(-16\Delta^2 -K_{pd}^2+K_{pd}J_H+2J^2_H-36J_H\Delta-32t^2_{pd})}
{2\left((J_H-K_{pd})^2+K_{pd}J_H+12J_H\Delta+48\Delta^2+64t^2_{pd}\right)^{3/2}}
\end{eqnarray}
\normalsize
Ignoring the intra-plaquette O-O transfer $t_{pp}$ and the weak direct FM
exchange $K_{pp}$, the FM NN exchange  is given by twice the
singlet-triplet energy difference. [s.\ the remark after Eq.\ (S9) for a dimer
here simply generalized by the factor 2 
due to the account of 2 exchange paths via the 2 noninteracting O.]
$J_1=-2\left(E_S-E_T\right)$.
Anyhow, their exchange
governed
mainly by $t_{pp}$ and $U_p$, has been taken 
fully into account in our solutions shown in FIG.\ 16 of the MT.
These hoppings were 
taken from {\it modified} Slater-Koster integrals. 
based on atomic wave functions 
in Refs.\ \onlinecite{mizuno,Braden1996}.
However, in a real ionic solid their absolute values and the 
ratio are affected by the crystal field.
In this context we mention that $t_{p_y} =$ 0.59~eV in Refs.\ 1 and 3,
whereas it is $\approx$ 1~eV in our DFT-calculations. The Cu-O transfer integrals are a
bit closer differing by about 20{\%}, only. In view of these uncertainties it is often
useful to determine the main hoppings from fitting experimental
optical, EELS and RIXS spectra. As a result we arrived in the past at intermediate values 
close to those of the DFT.
The expressions given below  illustrate the complex interplay
of various Hamiltonian parameters hidden 
by the
simple
quasi-linear behaviors. Thus, 
expanding Eqs.\ (S33) and (S34) in powers of $K_{pd}/t_{pd}$, we get 
in first order
\small
\begin{eqnarray}
J_1&=&-K_{pd}\frac{t^2_{pd}}{\Delta^2+t^2_{pd}}+
\frac{1}{32}K^2_{pd}\frac{t^2_{pd}}{\left(\Delta^2+
t^2_{pd}\right)^{3/2}}\left[1+\frac{3\Delta^2}{\Delta^2+t^2_{pd}}\right]+ \nonumber \\
&&-\frac{1}{8}J_H\left(1-\frac{8\Delta}{\sqrt{\Delta^2+t_{pd}^2}}+\frac{7\Delta^2}{\Delta^2+t^2_{pd}} 
+K_{pd}\frac{t^2_{pd}+4\Delta\left(\Delta-\sqrt{\Delta^2+
t^2_{pd}}\right)}{2\left(\Delta^2+t^2_{pd}\right)^{3/2}}\right)
+\frac{5}{148}J_H^2\frac{t^2_{pd}}{\left(t^2_{pd}+\Delta^2\right)^{5/2}} \ .
\end{eqnarray}
\normalsize
For $\left(t_{pd}/\Delta\right)^2$ the 3rd term yields a small 2nd order dependence
$\sim -\frac{1}{16}J_H\frac{t^2_{pd}}{\Delta^2}$ but $\approx -9.3$~meV, only, adopting 
$\mid t_{pd}\mid =0.5\Delta$. Thus, 
for $J_H=0.6$~eV  and $K_{pd}=0.1$~eV 
the Hund's rule term yields $\approx$
1/3 of 
$J^{\rm \tiny FM}_1$, i.e.\ it is still significant but not dominant.
Note that its contribution is further reduced inspecting the 2nd order term which yields
a tiny AFM contribution
$\sim J_H^2t^2_{pd}/\Delta^5$. Also the 2nd order quadratic term in $K_{pd}$ 
is negligible. 
Thus, the quasi-linear dependence of $J_1$ on the two main FM
interactions employed in the MT has been verified.
The corresponding plots (s.\ Fig.\ S4) are shown for the 
set used in Ref.\  
\cite{mizuno} and for a doubled $K_{pd}$,
i.e.\ $\approx$
to $J/t_{pd}=-0.05, -0.1$, respectively.
From Figs.\ S4 and 16 in the MT it is clear 
that $J_H$ contributes less than $\sim$ 25\% to 30\%  
to the singlet-triplet
separation, i.e.\ to  $J^{\rm \tiny FM}_1$ entering Eq.\ (14) of the MT.
\vspace{-0.0cm}
\subsection{Quantum chemistry and DFT calculations: computational details}
Among the various pieces of information we used for the analysis of the different contributions
to the effective magnetic couplings is the strength of direct exchange between spins belonging to
NN Cu$^{2+}$ ions.
The strength of this direct exchange on two edge-sharing plaquettes was determined on the basis
of {\it ab initio} Hartree-Fock calculations \cite{Helgaker}.
Given the complicated incommensurate lattice structure of 
CYCO
\cite{Gotoh},
we employed however the simpler lattice configuration of the closely related 
LICU \cite{Sapina}.
The material model consists in this case of two NN magnetic centers (Cu$^{2+}$ sites), the six
O ligands coordinating these two Cu ions, and the nearby sixteen Li atoms.
The plaquette plane coincides with the $xy$ plane, with the $x$ axis pointing along the chain.
The remaining part of the crystalline lattice was modeled as a finite array of point charges
fitted to reproduce the ionic Madelung field in the cluster region.
All-electron basis sets of quadruple-zeta quality were used for Cu \cite{Balabanov} while basis
sets of quintuple-zeta quality were applied for the two bridging O's \cite{Dunning}.
For the non-bridging O ligands and for the Li ions, basis functions of triple-zeta \cite{Dunning}
and single-zeta quality \cite{Dovesi} were used, respectively. 
The direct FM Cu-Cu exchange integral $K_{dd}=-4.1$~meV was determined from the splitting between
the singlet and triplet states associated with the $d^9$--$d^9$ electron configuration as obtained
from a restricted open-shell Hartree-Fock (ROHF) 
calculation \cite{Helgaker,Calzado2000}.
Further, to gain some insight as concerns the magnitude of direct hopping between two NN Cu sites
($t_{dd}$), we considered the same two-plaquette cluster described above but removed one electron
(giving a $d^9\!-\!d^8$ configuration for the NN Cu ions).
Orbitals as obtained in the ground-state $N$-electron calculation, with no further optimization,
were used to this end.
The hopping between NN Cu sites was then computed as half of the energy 
separation between ROHF
states having the $d_{xy}$ electron either in the two-site molecular-like orbital (MO) of gerade
symmetry or in the MO of ungerade symmetry \cite{Calzado2000,Sanz93}.
The resulting hopping is --174 meV.
Similarly, --160 meV is obtained from the splitting of the bonding and antibonding Cu
$d_{xy}$ MO's in the GS $N$-electron configuration.
The computations were performed using the quantum chemistry (QC) package {\sc molpro}~\cite{Werner}. 
A more detailed QC analysis of exchange mechanisms in Li$_2$CuO$_2$, beyond the
present results for the 
 direct $dd$ exchange ($K_{dd}$) and for the $t_{dd}$ hopping integral, will
be provided elsewhere.

We additionally performed calculations within the local density approximation (LDA) and the general
gradient approximation (GGA), using the full-potential local orbital FPLO code \cite{fplo1,fplo2,fplo3},
version fplo14.00-49. 
For the exchange-correlation potential, the LDA and GGA parametrizations of Perdew-Wang \cite{PW} and
Perdew-Burke-Ernzerhof \cite{PBE} were chosen, respectively.
Both exchange-correlation potentials yield essentially the same band structure.
To obtain precise band-structure information, the final calculations were carried out on a well
converged mesh of 1920 $k$-points (16$\times$12$\times$10 mesh).
The LICU crystal structure of Ref.~\cite{Sapina} was used.
The tb-hoppings and the respective on-site energies were evaluated on the basis of Wannier
functions (WF's).
The single-band model implies a single Cu-centered 3$d_{xy}$ WF.
For the $pd$-model, a minimal 5-band construction (Cu 3$d_{xy}$, O 2$p_x$, 2$p_y$) and an 11-band
model (all Cu and O orbitals) were derived  (with $xy$ for the plaquette plane and $x$ for the chain
direction).
A 12 $k$-point mesh was used for the WF fit (while the WF grid subdivision was 30$\times$30$\times$30
in the =.wandef file).
We obtained for the effective single-band approximation small NN and NNN transfer integrals
$t\!\equiv\!t_1\!=\!0.068$~eV and $t_2\!=\!0.080$~eV.
Within the  multiband $pd$-model we computed Cu-site energies $\varepsilon_d= -2.42$~eV 
(in electron ($el$)-notation) and a level splitting of 194~meV for the relevant O 2$p$ orbitals, with
$\varepsilon_{p_x}=-3.986$ and 
$\varepsilon_{p_y}=-3.792$~eV.
Larger hopping integrals $t_{pd_{Ox}}=0.933$,  $t_{pd_{Oy}}=0.8129$, $t_{p_xp_x}=0.7676$ eV were
found for the multiband model, with $t_{p_yp_y}=-0.084$~eV
and 
$t_{p_yp_y}=1.035$~eV (compare with 0.69~eV in the Slater-Koster type parametrization \cite{Harrison} 
used in Ref.\ \onlinecite{mizuno} and
denoted as $t_{p_y}$ in Fig.\ 13 of the MT of the intra plaquette
transfer integral between the lower and the upper O 
as discussed elsewhere). A transfer integral $t_{dd}=-143.8$~meV was computed,
similarly
 to the ROHF estimates given above. With respect to the $pd$ and $dd$ channels,
we note that a  
straightforward
extraction from DFT band structure analyzed in terms of a single-band extended
TB-model by
 fitting its antibonding Cu-O band
is somewhat dangerous
 due to the {\it 
 linear}
 interference with
the indirect $pd$ kinematic contributions. In contrast
 the
superexchange (SE)
one looks for
 depends mainly on {\it even} powers of each component
 resulting in a {\it larger} SE.
Anyhow, a separation of the indirect {\it pd} channel from the {\it dd}
one is also necessary to
estimate  correctly the 
effective 
$U_{\rm \tiny eff}\approx \Delta_{pd} +V_{pd}$ which enters the $pd$ channel, only.
Similarly, the different contributions from the two $p_xd$ and $p_{yd}$
channels must be also separated since it 
strongly affects the FM $J_H$ 
on $J_1$ from the two bridging O considered below
(s.\ Fig.\ 16 in the MT). These problems 
are almost 
gone
using a multi-band
WF-function analysis or an orbital specific basis
as we did here.
\vspace{-0.0cm}
\subsection{No criticality and sizable quantum 
fluctuations for chains with strong 
diagonal
 AFM interchain couplings} 
Here we demonstrate the significant up-shift of the 
 quasi-clasical critical point (CCP) between a quasi-2D Ne\'e'l sate
with collinear FM chain ordering and a quasi-2D-spiral state with a non-collinear 
ordering 
\cite{remarkBalents,Balents2016}
for the quasi-2D model
 of  AFM 
  interacting chains adopted in the MT illustrated by
 two simple examples. Thereby we have changed only one of the main inchain interaction 
 parameters and leave the
 other 4 parameters fixed (s.\ FIG.\ S1.) In case of reducing the AFM IC
 despite some 
 flattening of the dispersion 
there are no qualitative changes showing the absence of criticality.
In general, allowing the change of all parameters, one might arrive at a critical
point yielding smaller 
$\alpha$-values being closer to that of 
single chains with $\alpha_c=1/4$. 
But the 
systematical study of 
such a
"critical" 
surface in the 5D parameter space 
is very time consuming
 and even difficult to present the large manifold of  critical points in 
 a vivid manner. Model studies of
 reduced Hamiltonians e.g.\ for equal IC in the isotropic 
 case $D=0$ in a 3D parameter space is under progress. 
 Here, we focus on the interplay of $\alpha$
 vs.\ AFM IC.
 Fixing $J_1$ and $J_2$ but adding a 
 weak FM as derived above 
 NNN-IC $J_b$, in the 
 formal
 limit 
 $J_{ac,1}$,$J_{ac,2} \rightarrow 0$ one is left with two
 interpenetrating FM coupled sublattices and the resulting ordering would 
 be determined by a
 {\it quantum} "order by disorder" 
 problem. A quantum critical point (QCP) 
 would occur, only, if the  resulting GS would be a FM
 state at variance with the AFM Ne\'el GS of FM aligned chains. 
 This situation differs
 from that for LICU where the uniform FM inchain ordering 
 is
 induced solely by the weak AFM IC with  shifted adjacent chains 
 (s.\ Fig,\ 1 of the MT)
 \begin{figure}[b]
\includegraphics[width=7.0cm]{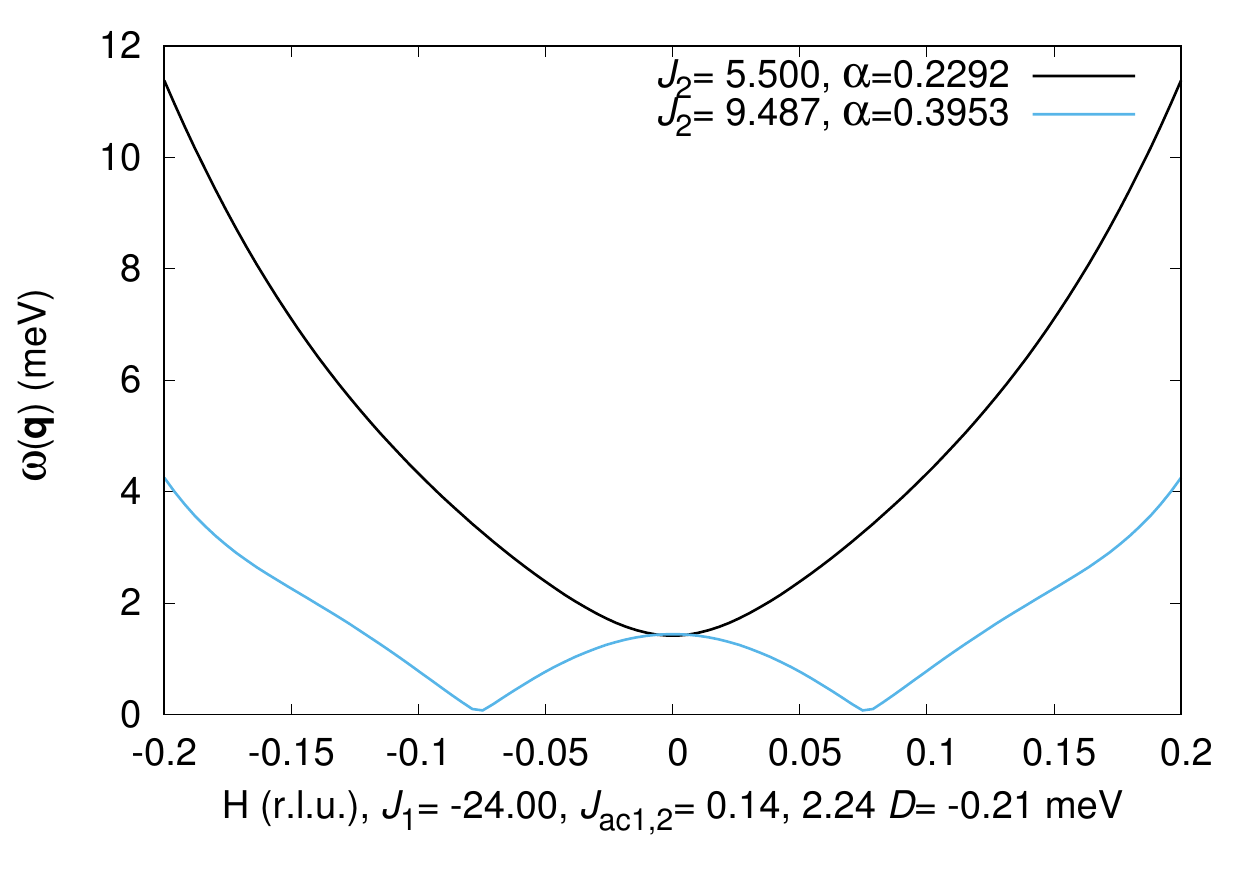}
\includegraphics[width=7.0cm]{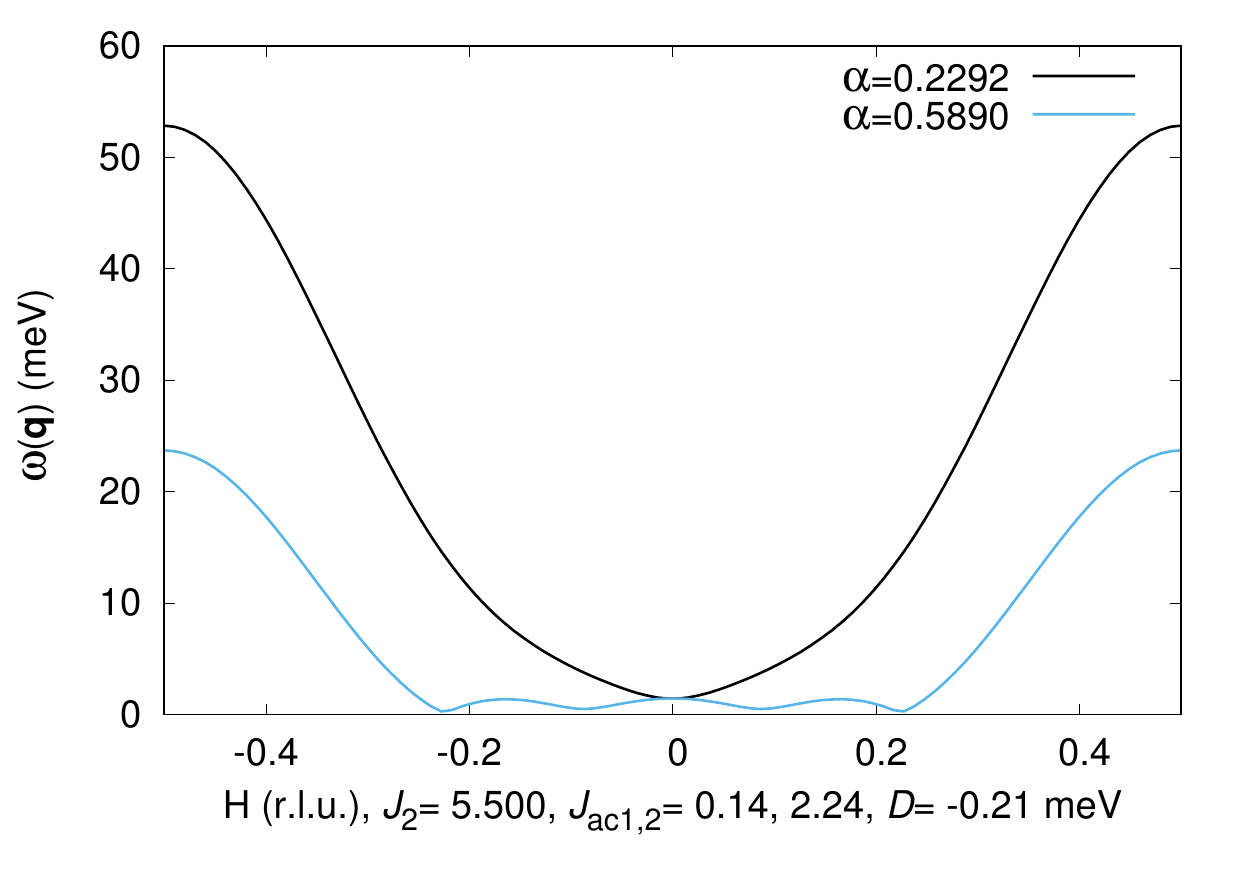}
\caption{Dispersion for hypothetical systems with changed inchain couplings in order to reach
the spiral critical point. (Left): for changing $J_2$, only, leaving all other three 
exchange interactions and the anisotropy parameter $D$ fixed; (Right): the same for $J_1$.
Notice the occurrence of four minima pointing to two competing distorted phases.
}
\label{critj2}
\end{figure}
 Then a 3D-criticality
 would be reached
 for a slightly
 weakened IC by about 4~K  (in total 32~K summing up all NNN)
 a critical point to the ordered spiral states is achieved. Analogously, for fixed IC 
 at 9.2~K a critical point is achieved at an upshifted value of $\alpha=0.39$, only, 
 to be compared with the measured 
 0.332 \cite{lorenz}. 
 
 The recently reported pressure dependence
 of the lattice structure and the stability of the orthorhombic phase below
 $\approx$ 6~GPa, only, for LICU \cite{You2009} is of great interest, because it is
 accompanied  at first by an increase of $\Phi$ up to 97$^o$
 which causes a decrease of $-J_1$ and an increase of $J_2$ resulting in an
 increase of $\alpha$ at almost unchanged AFM IC according
 to DFT+$U$ calculations by U.\ Nitzsche \cite{Nitzsche2011} (s.\ also  
 the Tab.\ I).
 Then, according to these findings an {\it incommensurate} spiral state is expected
 in a {\it classical conventional} scenario
 at intermediate pressure before a transition to a monoclinic phase with
 smaller $\Phi$-and expected strongly increased (decreased) $-J_1$ ($J_2$), 
 resp., which explains naturally the reentrant transition to a 
 commensurate state with FM alligned moments along the chains
 called "accidental" by the authors \cite{Li2011}, provided the AFM IC remains relatively
 weak as suggested by the reported 
 lattice structure and especially if, $\alpha < 1/4$
 in that novel monoclinic phase as suggested by the experimental 
 much closer $\Phi $ to $\pi/2$ than in the orthorhombic phase. 
 \textcolor{black}{
 The absence of critical 
 fluctuations in CYCO and Li$_2$CuO$_2$ at {\it ambient} pressure
 are  seen in the thermodynamic properties, in particular,  
 in the large ordered magnetic moments which are ascribed also to a significant easy-axis
 anisotropy for the latter. Hence, a high-pressure study
 of the ordered magnetic moment for CYCO and LICU as well is highly desirable
 to elucidate the stability of the commensurate chain Ne\'el phases. 
 The stability of the induced FM inchain ordering can be revealed by dynamical
 measurements as the INS considered here. With the AFM IC 
 properly taken into account there is a critical point nearby which separates
 a commensurate and collinear Ne\'el phase with unusual FM ordering along the
 chain direction from a non-collinear spiral state. Thus, near
 that critical point at {\it finite} pressure for {\it weak}
  diagonal AFM IC  between shifted chains 
 there is now indeed room 
 for an interesting
 "{\it order from disorder}" specific quantum mechanism 
 proposed for LICU \cite{Xiang2007} originally 
 already for the case of ambient pressure.
 There, however, it seems to be ruled out experimentally  
  by the mentioned thermodynamic data and the
 success of the LSWT. In contrast, in the critical region 
 more sophisticated many-body effects for the magnons
 as renormalizations of the large dispersion and damping effects 
 frequently discussed in the literature
 might be however expected.
  The somewhat smaller IC extracted from the LSWT analysis
 in LICU ( 9~K (0.8~meV) vs.\ 13~ or 11~K) from the LSDA+$U$ calculations
 might be already considered as a related physical effect. 
 Then the question arises to what extent the expected incommensurate spiral
 phase shrinks or it is removed by such quantum order by disorder effects
 forming a modulated FM state \cite{Matsuda1996}?.
 In this context one might speculate that the observed,
 but not yet understood, incommensurate phase in the related slightly $h$-doped}
 La$_5$Ca$_9$Cu$_{24}$O$_{41}$ \cite{Matsuda2003} might be governed by a 
 related mechanism,
 since a  weaker IC is suggested (i) 
 by the lowered $T_{\rm \tiny N} =10.5$~K  as compared with 
 the undoped parent compound La$_6$Ca$_8$Cu$_{24}$O$_{41}$ ( $T_{\rm \tiny N} =12.2$~K)
 both are apparently already much
 less stable than CYCO with the highest 
 $T_{\rm \tiny N} \approx 29.5$ for all FFESC-systems.
 (ii) The inspection of the 
 $\chi(T)$ shown in Ref.\ \onlinecite{Matsuda1996} 
 (which is determined by the frustrated CuO$_2$ chains due to the large two-leg ladder
 gap) of the undoped parent compound La$_6$Ca$_8$Cu$_{24}$O$_{41}$
 with a maximum position near 20~K 
 pointing to
 a {\it larger} 1D frustration ratio than in CYCO: $\alpha >$~0.3 to 0.4.
 Adopting 
 $\alpha=1/3$ or 0.4, one estimates
 for that peak position (0.04 to 0.06)$\mid J_1 \mid$ 
 \cite{Heidrich2006,Lu2006,Sirker2010}, resp., which yields 
 $-J_1\approx 43$~ to 28~meV according to a full diagonalization of 
 a large cluster with $N=24$
 sites or by the transfer-matrix renormalization group teqchnique.
 Thus, 
 an empirical $J_1$ value $\sim$ 36 $\pm 6$~meV 
 might be estimated. But having in mind also the corresponding $J_2$-value,
 and our geometrical experience with various CuO$_4$ plaquettes 
 and the involved $\Phi$ \cite{drechsler2019},
 a smaller interval, say $-J_1 =30 \pm 2$~meV seems to be more reasonable.
 (s.\ also the discussions 
 in Chapt.\ I with respect to
 Li$_2$ZrCuO$_4$ and LICU). Anyhow, the estimated  $\mid J_1 \mid $
 {\it exceeds}
 the values of CYCO and LICU. 
 Our empirical estimate is certainly again well above the prediction
 of 18.5~meV \cite{mizuno} (s.\ Tab.\ I).
  \footnote{In the 1D isotropic
 $J_1$-$J_2$ model the peak positiion of $\chi(T)$ is a monotonous function of $\alpha$.
 $\chi(T)$ shows
 a divergency at $T=0$ for $\alpha \leq \alpha_c=1/4$.
 \cite{drechsler,Drechsler2007,Heidrich2006} whereas the magnetic 
 specific heat $c_v$ exhibits {\it two} maxima for
  $\alpha \stackrel{<}{\sim} 0.4$. Its 
 lower, sharper peak is field dependent reflecting
 the low-lying FM excitations above the spiral spin liquid GS. The Sommerfeld
 coefficient $\gamma$ is also a monotonous function of $\alpha$, diverging
 for $\alpha \rightarrow \alpha_c$.} 
 Then,
 reanalyzing these data, the main inchain exchanges can be estimated
 from  already measured thermodynamic quantities 
 and the vicinity to an at first glance
 expected spiral
 GS could be checked. However, the observed unusual
 for a spin-1/2 Heisenberg model 
 {\it collinear} and {\it modulated} FM ordering along the chains
 is probably caused by a special spin anisotropy which prevents the 
 non-collinear spiral
 ordering. The slightly $h$-doped
 La$_5$Ca$_9$Cu$_{24}$O$_{41}$ \cite{Matsuda2003} seems to be a real 
 candidate for quantum
 criticality whose frustrated CuO$_2$ chain might be described with stronger inchain
 couplings  and modified weak ICs opposite to earlier
 attempts. The latter might be determined from the known Nee\'l temperatures
 or from the two 
 gaps expected in this system, too, which  depend sensitively
 on the IC, provided the scattering intensity from the ladder excitations could be separated
 from the chain ones.
  Our large dispersive magnon for CYCO should be a good 
 starting point with some modifications to allow a very soft local
 behavior near the wave vector corresponding to the spiral critical point in
 chain direction and natural enhanced damping effects expected in that region.
 Anyhow, we strongly believe that the large
 inchain exchange integrals for CYCO are much less affected by such remnant quantum
 fluctuations 
 and the derived large $J_1$-values for both CYCO and LICU
 from the observed large dispersions do remain valid. 
 In contrast, LICU and probably also Ca$_2$Nd$_2$Cu$_5$O$_{10}$ and
 CuAs$_2$O$_4$ might be closer to the inchain spiral $\alpha_c$ due to
 slightly larger $\alpha$-values and a significantly weaker AFM IC
 which is even there necessary to stabilize the observed modulated FM alignment of 
 the magnetic moments along the chains.
\subsection{Aspects of the phonon mode responsible for the 
low-energy gap-like feature near 11~meV}
To get qualitative insight into
the origin and the nature of the phonon mode responsible
for the first gap-like feature observed near 11~meV at the 
wave vector $H$=0.2, 
we suggest that it should be an optical mode
due to the weak observed dispersion. We assume that it is derived from the cationic
chains with alternating Ca$^{2+}$ und Y$^{3+}$ ions. Having in mind also the sister compound
Ca$_2$Nd$_2$Cu$_5$O$_{10}$, we are confronted with a two-atomic chain model 
with rather different masses $M_2$. We will
measure them in units of the Ca -mass $M_1$. This mass ratio amounts for the general
system Ca$_2$R$_2$Cu$_5$O$_{10}$  including the isovalent substitions for Y by
R = Sc, Y, Nd, and Lu 1.113, 2.21, 3,99 and 4.37, resp. Then in a toy
model approach with a single force constant $K_1$ one has the   following 
phonon dispersion-law
\small
\begin{equation}
\omega_{\mbox{\tiny ph}}=\sqrt{\frac{K_1 \left[ M_1+M_2
 +\sqrt{M_1^2+M^2_2+2M_1M_2\cos(ka^*)}\right]}{M_1M_2}},
\label{phonon}
\end{equation}
\normalsize
 where $a^*$ is the Ca-R  distance related to the Cu-Cu one $a$ of the spin chain $a$ as $a^*=1.25a$.
 Then the attributed experimental crossing point with the magnon appears 
 near $H$=0.2 (s.\ FIG.\ 2 in the MT) which corresponds 
 to $\pi/2a^*$. The calculated dispersion of $\omega_{ph}$
 from Eq.\ (\ref{phonon}) is shown in FIG.\ S6 ({\bf left}).
 Since the observed dispersion is smaller, one might expect 
 that at least NNN neighbor and also further more distant neighbors might 
 be responsible for the requested 
 flattening of the dispersion. In view of their ionic character slowly 
 decaying contributions from 
 long-range ionic forces in addition to the hard-core NN ones resulting 
 from  the finite ionic radii
 are indeed expected.
 Then due to the different involved charges larger NNN Y-Y interaction as 
 compared to the NNN Ca-Ca
 counter parts should be realized. Within a pure point-charge Coulomb picture 
 their ratio should be universally 9/4.
 The absolute values can be estimated from the screened ionic couplings 
 $K_{2,\mbox{\tiny YY}}=9e^2/(4\epsilon a^{*^2} )$ for the Y-Y case and similarly
 as $2e^/(3\epsilon a^{^2})$ for the 3rd neighbor Y-Ca elastic coupling, 
 where $\epsilon \approx 3.5$ denotes 
 the dielectric constant like in LICU \cite{mizuno,Johnston2016}.
 In a realistic picture also various skew couplings all scaled by the 
 inverse square of the 
 corresponding distances on the adjacent and further distant chains 
 are expected to contribute as well as the ionic interactions with O and Cu.
 Since the latter elongations modify the $\Delta_{pd}$, a weak coupling
 to the magnon via the modulation of $J_1$ and $J_2$ is expected in accord
 with the weak mixing mentioned in Sec.\ III. A quantitative consideration of the latter 
 is tedious and beyond of the scope of the present paper. Finally, in case of magnetic rare earth
 elements R=Nd in Ca$_2$Nd$_2$Cu$_5$O$_{10}$ \cite{Wizent,Wizent2} an ordering of these $f$-moments 
 at few K well below the observed high $T_N\approx 24$~K of the Cu derived spin-1/2 moments 
 is expected.
 In fact, the $\chi(T)$ data exhibit a broad peak near 15~K in a field of 1~Tesla
 $\parallel$ to the $b$-axis and a broadened feature around 8~K
 $\parallel$ to  ${\bf c}$ \cite{Wizent2} pointing to a 2nd magnetic transition below
 this upper limit, i.e.\  near 10~K which is too high 
 for a simple weak dipole-dipole coupling.
 In the high-$T_c$ cuprates only in the case of PrBa$_2$Cu$_3$O$_6$
 a magnetic transition at $\approx$ 16~K much higher than in all others rare earth substituted 
 cases has been found. We speculate that here already for the case of Nd, the O-holes
 have some covalence with the Nd $4f$-electrons.
 
 Anyhow, the elongated R-ions of that phonon will interact also via the elastic 
 NNN coupling $K_{2,\mbox{\tiny RR}}$ and affect the O-ions  bearing charge and a 
 magnetic moment . The much heavier Nd mass is expected to downshift
  the phonon mode, even stronger than shown in FIG.\ S6 ({\bf Left }) also due to
 a somewhat weaker NN $K_{1\mbox{\tiny Ca,R}}$ force constant from the smaller 
 ionic radius of Nd$^{3+}$ as compared 
 to Y$^{3+}$. The "hybridization" with the acoustic Cu-O mode might
 strengthen the flattening of the optical phonon mode.
 Anyhow, a detailed study of this phonon mode and its interplay with magnetism
 might provide much more insight into  the magnetism of the two magnetic subsystems 
 and their interplay. 
 \vspace{-0.0cm}
\subsection{Details of the $\chi(T)$ fit using the 10th order high-$T$ expansion
for the magnetic susceptibility}
We use the method and the program packages from
Refs.\ \onlinecite{Schmidt11,Lohmann14,hte10} for the 
10th-order high-$T$ 
expansion (HTE) of 
$\chi (T)$
 for a general spin-model with up to 4 
different exchange parameters $J_1, J_2, J_3, J_4$.
The spin model 
for
CYCO
reads
\begin{equation}
\hat{H}=\frac{1}{2}\sum_{\mathbf{R,r}}J_{\mathbf{r}}
\hat{\mathbf{S}}_{\mathbf{R}}\hat{\mathbf{S}}_{\mathbf{R+r}}, 
\label{Heis}
\end{equation}
depicted in FIG.\ 1 of the MT, we take 
$J_1 =J_{1a}/k_B\approx -278.5 \ \mathrm{K}, $
$J_2 =J_{2a}/k_B\approx 63.82 \ \mathrm{K}, $
$J_3 =J_{ac1}/k_B\approx 1.38 \ \mathrm{K}, $
$ J_4 =J_{ac2}/k_B\approx 26.24 \quad \mathrm{K}$. 
The input for the HTE package is the definition file where all bonds  
within a cluster or a $2L\times L\times L$ (L=10) super-cell of a periodic 
Heisenberg lattice are enumerated with the indication of a corresponding 
value of the exchange interaction. We use an
originally developed C++ program, for the generation of the definition 
file.
The HTE program provides 
\emph{exact analytic} expressions for 
10
coefficients of the HTE 
\begin{equation}
\frac{k_{B}}{5N_Ag^{2}\mu_{B}^{2}}\chi (T) = \sum_{n=1}^{\infty} \frac{c_n}{T^n}
\label{eq:chihte}
\end{equation}
and computes 
the Pad\'{e} (PA)
approximants,ratios of two polynomials:
$\chi (T) \approx [m,n]=P_m(T)/P_n(T)$. 
The PA
extend 
the 
region of 
validity of the HTE \cite{Lohmann14}. 
The analytic expressions for the first 5 coefficients $c_n$ 
for the model (\ref{Heis}) are 
\begin{figure}[b]
\includegraphics[width=5.0cm]{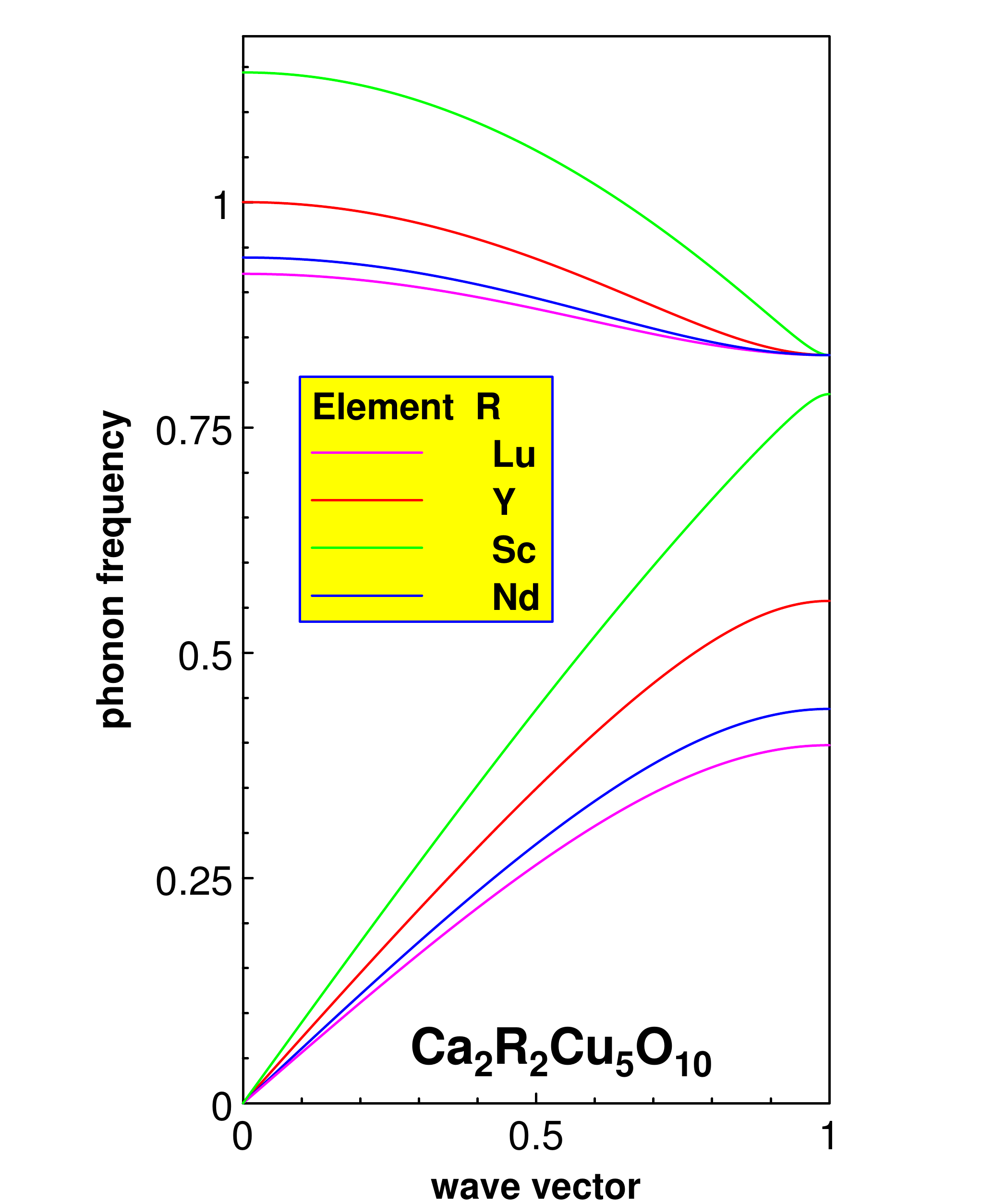}
\includegraphics[width=0.45\columnwidth]{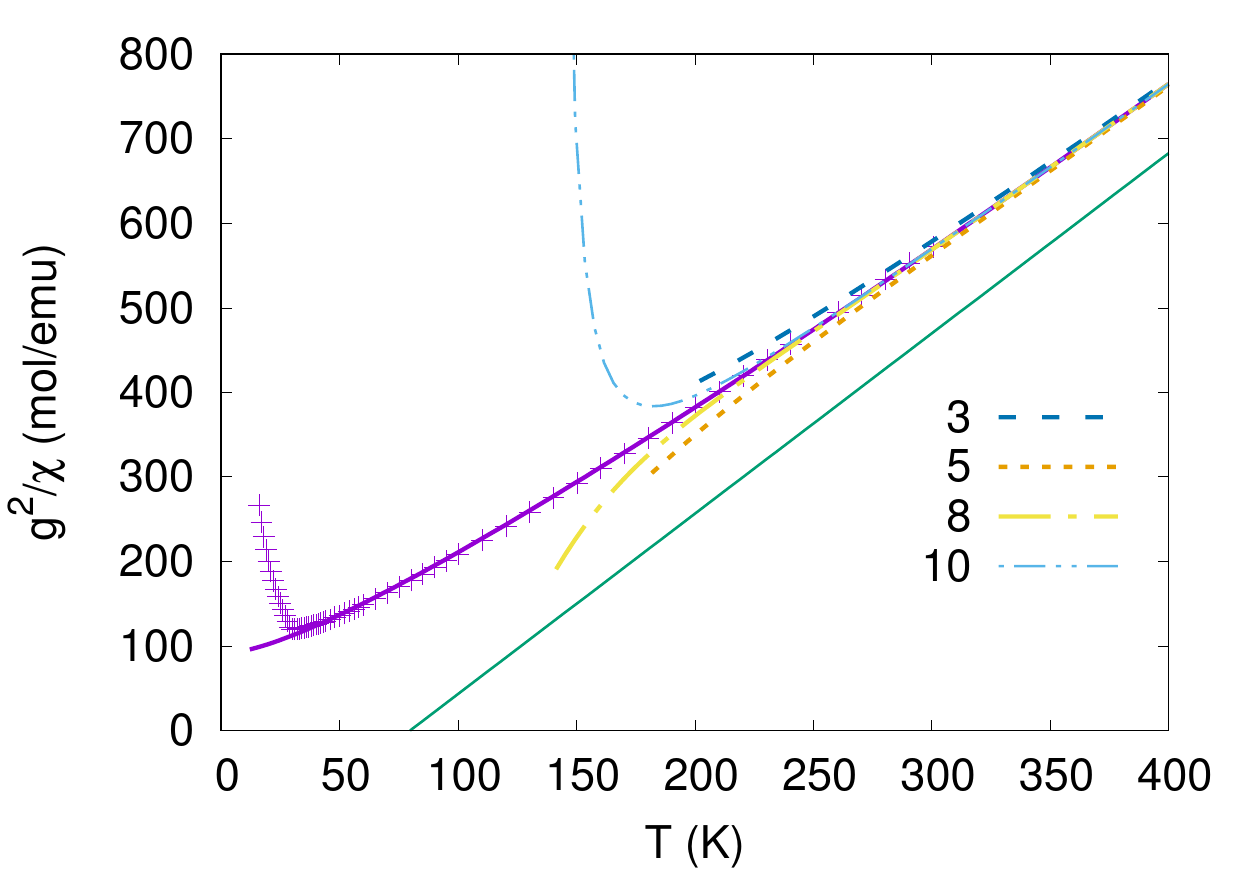}
\caption{{\bf Left:}Phonon dispersion for various elements R in the alternating cationic 
Ca$^{2+}$-R$^{3+}$ chain within  the NN coupling approximation
given by the force constant $K_1$ normalized to the case of Y at the $\Gamma$-point in the BZ
 for the optical phonon mode.
The wave vector 
$k$ is given in units of $\pi/a^*$, where $a^*$ denotes the R-Ca distance and R is a trivalent
element. In these units the 
experimental crossing point with the magnon amounts  0.5.
{\bf Right:}Symbols: the inverse spin susceptibility
for the magnetic field along $b$ ($+$)  
axis of CYCO,
Ref.\ \cite{kudo}. Dashed lines 3,5,8,10 of the 3th, 5th, 8th and 10th HTE orders,
thick solid line: the [5,5] Pad\'{e} approximant for the 10th order,
thin solid line: the exact Curie-Weiss asymptotic.
}
\label{invhte}
\end{figure}
%
\tiny
\begin{align}
\hspace{-1.0cm}c_1 &= +1/4, \label{eq:c1}\\
c_2 &= -(1/8)(J_1+J_2+2J_3+2J_4), \label{eq:c2}\\
\hspace{-1.0cm}c_3 &= (1/8)(J_1J_2+2J_1J_3+2J_1J_4+2J_2J_3+2J_2J_4+J_3^2+4J_3J_4+J_4^2),\\
\hspace{-1.0cm}c_4 &= 4\left[(J_1^3+J_2^3)/24+J_1J_2(J_1-4J_2)/32
-(J_1^2+6J_1J_2+8J_3J_4+J_2^2)(J_3+J_4)/4
 -11J_1J_3^2/16 -19(J_1+J_2)J_3J_4/8-((J_1+J_2)J_4^2+J_2J_3^2)-(J_3^3+J_4^3)/6\right]
 \\
\hspace{-1.0cm}c_5 &= 5J_1^4/1536-23J_1^3J_2/768+J_1^2J_2^2/512
-J_1J_2^3/96
 +(7/64)J_1^2J_2(J_3+J_4)+5J_2^4/1536
 +3/16J_1J_2^2J_3
 +25/256J_1^2J_3^2+3J_1J_2^2J_4/16+(5/32)(J_1^2J_4^2+J_2^2J_3^2 )\nonumber \\
 & 
 +23J_1^2J_3J_4/128+13J_1J_2J_3^2/32
 -(1/48)(J_2^3J_4+J_1^3J_4+J_2^3J_3+J_1^3J_3)+33J_1J_2J_3J_4/32+
 +19J_1J_3^3/384+31J_1J_2J_4^2/64 +33J_2^2J_3J_4/128+89J_1J_3^2J_4/128+\nonumber \\
&  J_2J_3^3/6+5J_2^2J_4^2/32
 +(109/128)J_3J_4(J_1J_4+J_2J_3+J_2J_4)+13(J_3^4+J_4^2)/768
 +49J_3^3J_4/192
  +J_4^3(J_1+J_2)/6+51J_3^2J_4^2/64+J_3J_4^3/3
   \\
 \nonumber
 \end{align}
\normalsize
The coefficients $c_n$, $n>1$ are functions of the exchange parameters $J_i$
given by Eqs.\ (S40-S43). Their explicit form 
depends on the geometry of the bonds encoded in the input definition file. Only the
\emph{single} parameter $g$ was varied during the fit for each direction of the magnetic field.
FIG.\ \ref{invhte}~\textbf{Right} shows the curves of the HTE series
 for various orders and the [5,5] Pad\'{e} approximant for the 10th order 
 that fits the 
data even down to $T_N$ as shown also in Fig.~12 of the MT.
The CW-asymptotics is given by the  \emph{2nd} order
of the HTE series
\begin{equation}
 \frac{k_{B}}{5N_Ag^{2}\mu_{B}^{2}}\chi_{\rm \tiny CW}(T) = 
 \frac{c_1}{T-\Theta_{\rm \tiny CW}}, \quad
 \mbox{and} \quad
 \Theta  _{\rm \tiny CW} = \frac{c_2}{c_1}\approx +80\quad \mathrm{K}.\label{eq:TCW}
\end{equation}
Inserting the expressions (\ref{eq:c1}),(\ref{eq:c2}) into Eq.\ (\ref{eq:TCW}) gives
Eq.\ (21) of the MT. It is an \emph{exact} relation,  which coincides with the mean-field result. 
[s.\ Eq.\ (27) of Sect.\ IV.B in Ref.\ \onlinecite{Schmidt01}.]
$\Theta  _{\rm \tiny CW}$
is determined by the dominant inchain exchange 
($J_1$, $J_2$ in 
case of 
our
chains),
while
the 3D GS and
especially $T_N$ do depend on the
small IC, too.
\vspace{-0.0cm}
\subsection{Curie-Weiss temperature and saturation field -- a comparison with 
CuSb$_2$O$_4$ and NaCuMoO$_4$(OH)}
The difficulty to apply a HTE can be circumvented 
 applying a strong magnetic field (up to saturation at $H_s$ where all AFM 
 couplings are suppressed)
 at very low $T$. In the isotropic limit  [s.\ Eq.\ (S21) in the
 Suppl.\ Mat.\ of Refs.\ \onlinecite{Nishimoto2011},\onlinecite{remarkCW}]
 there is a convenient and accurate relation 
 valid 
 approximately  
 at low-$T$
 \begin{equation}
 g\mu_{\mbox{\tiny B}}H_s+ 2J_b+ 4\Theta_{\mbox{\tiny CW}}=2|J_1|(1-\alpha ) \ .
 \label{satfield}
 \end{equation}
For CYCO
$H_s$ slightly exceeds 
70~Tesla \cite{kudo} whereas 
 our predicted value
 $\Theta_{\mbox{\tiny CW}}$ for CYCO is much larger than the experimental values 
 of 39~K and 26.3~K  for two other frustrated 
 quasi-1D systems: CuAs$_2$O$_4$ \cite{Caslin2014} and CuSb$_2$O$_4$ \cite{Caslin2015}
 with a predominant FM $J_1$ coupling. CuAs$_2$O$_4$ exhibits probably a 
 FM ordering below $T_C=7.6$~K
 near the critical point and the 
 point also to a weak AFM IC whereas CuSb$_2$O$_4$ shows
 a {\it helicoidal}
 ordering below $T_{\rm \tiny N}=1.8$~K 
 pointing to a larger  $\alpha $.
 The 
 propagation vector $\tau=(0,0,0.3962) (\phi=71^o)$ 
 points to a pitch between neighboring Cu moments intermediate between 
 linarite (0.186, $\phi=33.48^o$) and LiVCuO$_4$ ($\phi=83.6^o$)
 and at first glance to a decreased $\mid J_1\mid $ caused by the larger 
 $\Phi$ 
 changing from $\Phi=91.5^o$ to  $\Phi=94.5^o$. Slight deviations
 from a perfect FM interchain ordering point to the presence of
 weak AFM frustrating IC
 in addition to stronger FM ones.
 Ignoring that weak IC and adopting $\alpha=1/4$ for 
 simplicity (close to 0.244 proposed 
 by the author), one arrives from Eq.\ (\ref{satfield})
 with $H_s=0$ generic for a FM at $T=0$ at
 $J_1=104 \pm 2.5$~K, only, in accord 
 with the result of their DFT-calculations slightly above
 a large $U\approx 9$~eV (i.e.\ 8~eV in their notation of $U_{eff}=U-J_{\rm \tiny Cu}$)
 and a $J_2=26~K$~K (2.24~meV) [ 108.7~K and 27.5~K, resp.].
 The effective $J_2$ 
 is somewhat reduced by the twisting of non-ideal Cu$_2$O$_4$ unites 
 and it is also smaller 
 than our value for CYCO of $\approx$ 64~K. 
In view of the discussion given in the MT and the actual $\Phi$ close to $\pi/2$, the small value
of $J_1$ is surprising. It points to a sizable AFM contribution of the $dd$-channel in some
conflict with a large $U_d$ but it is supported by the shorter
Cu-Cu distance as compared to CYCO and Li$_2$CuO$_2$ (see Tab.\ I). 
Using the experimental 
 $H_s\approx$ 14~T, 
 $g$=-2.1 from EPR 
 and $\Theta_{\rm \tiny CW}=26.3$~K for the Sb sister
 CuSb$_2$CuO$_4$,
 ignoring for 
 simplicity possible 
 nematicity, one arrives in the one-magnon picture
 at $-J_1=124.95$~K (10.77~meV) and $J_2=62.5$~K (5.15~meV) adopting their proposed $\alpha=0.5$
 compatible with their DFT result but  now near $U_{eff}=6$~eV [ 117~K and 63.8~K, respectively]
 Within this picture increasing 
 $\Phi$, 
  surprisingly $-J_1$ is slightly {\it increased}
 whereas $J_2$ is twice as large as in the As sister compound.
The inspection of the structural data of CuSb$_2$O$_4$ with two different
Cu-O-bond lengths points to a more general $J_1$,$\prime{J}_1$-$J_2$ model 
with {\it alternating} FM NN couplings $J_1$ and $\prime{J}_1$ at least,
similarly to the spin and nematic liquid system LiCuSbO$_4$ \cite{Dutton2011}
above 125~mK \cite{Bosiocic2017} with a subsequent 3D ordering below and
 with -160~K and -90~K, respectively,
and $J_2=24$~K according to our DFT-results \cite{Grafe2017}.

Another example of interest is 
NaCuMO$_4$(OH) \cite{Nawa2014} [ a
candidate for multipolar
physics]. Its 
$H_s \approx $~26~T. But at variance with the cases considered above, 
 it exhibits a small {\it negative} 
$\Theta_{\rm \tiny CW}=-5$~K. Using again Eq.\ (\ref{satfield}) and the experimental 
$g=2.18$,
one arrives at $\mid J_1 \mid = J_2 +9$~K which 
a bit
differs from
the 
values $J_1=51~$K and $J_2=36$~K derived from their fits of 
$\chi(T)$-data.  Using instead somewhat closer values such as 
$J_1=-48~$K (-4.12~meV) and $J_2= 39$~K (3.36~meV) self-consistency
can be achieved. The latter value is $\sim$
usually obtained
by DFT+$U$ calculations but the former is at first glance anomalously small.
The very different $\Phi$ for the two non-equivalent 
O(1) and O(4) suggests that the small total $J_1=J_{\rm 1,O(1)}+J_{1, O(4)} -J_{\rm O(1),O(4)}$ 
is the sum of two opposite contributions connected with each of this
non-equivalent O  and the AFM higher order
interaction between them discussed in the MT and illustrated in Fig.\ 14 therein.
 Thereby each of them
is expected $\approx$ 
one half of a normal FM contribution like in LICU for the former
from $\Phi_{\rm \tiny O(1)}=91.97^o$ 
  and an AFM $\sim$ 1/2 of that in CuGeO$_3$.
The still larger $\Phi_{\rm \tiny O(4)}=103.65^o$ 
  nearly "compensates" 
  the weaker splitting effect caused by the more moderate crystal field expected in NaCuMoO$_4$(OH).
  Thus, we  estimate $J_{1}\sim$ -0.5$\cdot$22~meV+0.5$\cdot$14~meV=-4~meV
  in accord with Eq.\ (\ref{satfield})
  yielding -4.15~meV.
\vspace{-0.0cm}
\subsection{FM NN-couplings or contributions in some cuprates with edge-sharing elements}
In Tab.\ I we collect available data of the FM NN
exchange coupling in frustrated CuO$_2$ chain and ladder compunds as well as in alternating
chain FM-AFM and FM-FM systems with the aim (i) that the cases already considered in
Ref.\ \onlinecite{mizuno} show a systematic 
underestimation which we ascribed in the MT
to an underestimated direct Cu-O FM exchange $K_{pd}$ and to stress its 
{\it non}universality, (ii) to demonstrate that the sign change of $J_1$ happens often
at larger $\Phi >$ 95$^o$ than suggested in Ref.\ \onlinecite{mizuno},
and (iii) find candidates where the so far ignored AFM {\it dd}-channel might essentially contribute
to small or even tiny $J_1$ in view of moderate $\Phi$. 
The list of materials
is by no means complete.
Anomalous small or even AFM NN $J_1$ values might be
realized only for special reasons not considered in Ref.\ \onlinecite{mizuno},
e.g., in the case of green dioptase\cite{Janson2010} where the NN-coupling
in the dimer Cu$_2$O$_6$ double plaquettes amounts about 3~meV, only, 
or being even AFM as in the case of {\it malachite} Cu$_2$(OH)$_2$ CO$_3$ representing alternating
AFM-AFM chains due to the buckled non-ideal corner-sharing (with 
$\Phi \approx 122^o$) of the connected  double
dimer-plaquettes. 

To illustrate  higher orders of the perturbation
theory, we reconsider here the results for the exchange 
between non-equivalebt Cu-A and B-sites  
in
Ba(Sr)$_2$Cu$_3$O$_4$Cl$_2$ \cite{Rosner1999} up to  6th order 
in an 11-band $pd$ model
and use them 
to estimate 
  the  FM exchange from the recent 
INS \cite{Kim2001,Babkevich2017} and RIXS \cite{Fatale2017} data.
The FM  $J_{\rm \tiny AB}$ results from
the edge-sharing between the 
and Cu$_B$ subsystems. 
We start with the large (small) AFM 
exchanges 
$J_{\rm \tiny AA}$ ($J_{\rm \tiny BB}$), respectively:
\small
\begin{equation} 
J_{\rm \tiny AA}=\frac{4t^2_{pd}}{\left(\Delta_A+V_{pd}\right)^2}
\left(
\frac{2t^2_{pd}}{\Delta_A+U_p}
+\frac{t^2_{pd}}{U_d}-K_{pd}+
\frac{K^2_{pd}}{\Delta_A+V_{pd}}
 \right)\ .
\label{JAA}
\end{equation}
\normalsize
With
$K_{pd}$=0.2~eV, $V_{pd}$=1.2~eV   in accord with the MT , 
and standard
$t_{pd}$=1.3~eV, $U_d$=9~eV, and $U_p$=4~eV as well as $\Delta_{A}$=3.2~eV ones, we 
arrive at $J_{AA}$=163~meV in accord with
$J_{\rm \tiny AA}$=169~meV from INS \cite{Babkevich2017}
and $J_{\rm \tiny AA}$=165~meV (RIXS \cite{Fatale2017}. They are
reproduced exactly for fixed other parameters at $K_{pd}=180$~meV and 192~meV, respectively,
and $U_{\rm \tiny eff}\approx \Delta_A +V_{pd}=4.4$~eV, and Raman data
and that of 130~meV estimated in Ref.\ \onlinecite{Kim2001}.
if $K_{pd}=0.22$~eV and $\Delta_A=3.5$~eV are adopted. 
Also the effective hoppings in the used there 1-band
Hubbard model $\mid t_{\rm \tiny AA}\approx t^2_{pd}/(\Delta_a+V_{pd}
=0.497$~eV
comes out close, as compared with 0.475~eV \cite{Fatale2017}.
Thus, the experimental $J_{\rm \tiny AA}\ \approx 165$~meV 
is reproduced for $K_{pd}\approx 0.2$~eV
like 0.2$\pm$0.02~eV from QCC 
for  La$_2$CuO$_4$ \cite{Hybertsen1990,Hybertsen1992}.
It amounts $\approx$ 400 \% of the $K_{pd}$ from Ref.\ \onlinecite{mizuno}.
For $K_{pd}$=0, we would have 
$J_{\rm \tiny AA}$$\approx $174~meV which is 5\% too large. 
Now, turning to 
$J_{\rm \tiny BB}$, we have
\small
\begin{equation}
J_{\rm \tiny BB}=
\frac{8t^2_{pd,B}t^2_{\pi \pi }}{\Delta^2_B}
\left[ 
\frac{1}{\left( \Delta_B+V_{pd} \right)^2} 
\left(
\frac{t^2_{pd,B}}{U_d} -K_{pd}
\right) + t^2_{pd,B}
\left(\frac{1}{2\Delta_B+U_p} +\frac{1}{\Delta_B} \right)
\left( \frac{1}{\Delta_B +V_{pd}}+\frac{1}{\Delta_B} \right)^2
\right] ,
\end{equation}
\begin{equation}
\mbox{The FM contribution to  $J_{\rm \tiny AB}$ reads:} \quad
J_{\rm \tiny AB}=-\frac{8t^2_{pd}t^2_{pd,B}K_{pd}}
{\left(\Delta_A+\Delta_B +V_{pd}\right)^2}
\left(1-\frac{K_{pd}}{\Delta_{\rm \tiny AB}+V_{pd}}\right)\left(\frac{1}{\Delta_A}+\frac{1}{\Delta_B} \right)^2 \ ,
\end{equation}
\normalsize
where $\Delta_A$=$\varepsilon_p$-$\varepsilon_d$, 
$\Delta_B$ =$\varepsilon_{\pi}$-$\varepsilon_d$, $\Delta_{\rm \tiny AB}$=$\varepsilon_p$-$\varepsilon_{p,B}$ .
Ignoring small O on-site energy differences and using
$t_{pd,B}\approx 0.5 t_{pd}$ 
\cite{Rosner1999}, we have
$J_{\rm \tiny AA}$$ \gg $$J_{\rm \tiny BB}$ in accord with experiment.
The FM $J_{\rm \tiny AB} \approx -20$~meV agrees
with the LSDA from \cite{Yaresko2002}
where a spiral GS was supposed 
for Ba$_2$Cu$_3$O$_4$Cl$_2$ but exceeds the LSWT values twice,
pointing to many-body effects near a QCP
between AFM and spiral orderings with a pitch $\phi_p \approx \pi / 2$
due to the frustrated  Cu$_{\rm \tiny B}$-spins embedded
into a strongly coupled  AFM lattice of the Cu$_{\rm \tiny A}$ spins.
This frustration is independent on sign($J_{\rm \tiny AB}$) 
but here sizably FM 
for plaquettes with $\Phi \approx \pi /2$.
Our estimate with $K_{pd}\sim$~0.1~eV
well exceeds  $\approx -6.6$~meV, for a moderate $J_H$, only,
\cite{Yushankhai1999}.

 \subsection{Enlarged NNN $J_2$-values and related disorder effects in 
 the sister compounds Li$_2$ZrCuO$_4$ and LiCu$_2$O$_2$}
Both systems have attracted attention already 15 years ago.
Nevertheless
 open questions 
 do still exist partly due to missing single crystals 
 and unclear Li (split) positions for the former and Li and Cu$^{2+}$ disorder
 in the latter. 
 Despite
  the large FM $J_1$ problem for both compounds also the 
  enhanced empirical $J_2$ values probably caused by not yet clearly understood
  site substitutional disorder effects, too. 
  Such effects might play some role for the unusally  enhanced $J_2$ values
  for CYCO, 
  La$_6$Ca$_8$Cu$_{24}$O$_{41}$
  and their weakly $h$-doped derivatives.
  \vspace{-0.0cm}
Within the first analysis of $\chi (T)$ (with $g=2$ and $\chi_0=0$ taken
for simplicity to show the generic behavior of the thermodynamic properties
with decreasing $\alpha$-values near  criticality)
and specific heat data analyzed within a  1D $J_1$-$J_2$ model 
point to $\alpha \approx 0.3$ 
and large values: $J_1=-$23.53~meV, $J_2$=7.06 meV \cite{drechsler}
has been fully reproduced in Ref.\ \onlinecite{Sirker2010}. Its author
noticed also, that adopting $g=2.2, \chi_0=0$, and $\alpha=0.4$ instead
yields also a "reasonable" fit of $\chi(T)$ but now with puzzling much
smaller $J_1=-6.46$`meV and $J_2= 2.585$~meV.
Both fits differ slightly in the vicinity of the maximum of $\chi(T)$ (s.\ the inset
in Fig.\ 12 of Ref.\ \onlinecite{Sirker2010}).
Since these values are in conflict with our microscopic analysis in the MT and in
Chapt.\ H, we argue that {\it intermediate} values of  $g \approx 2.1$ and $\alpha \approx0.35$
in accord with $\alpha=0.33$ reported by Y.\ Tarui {\it et al.} \cite{Tarui2008}
from a simulation of an adopted antiphase ordering of spirals in the simulation
of their
might resolve both the fitting problem 
and provide also more realistic intermediate
$J_1$ and $J_2$-values closer to the disordered DFT-description \cite{Schmitt2011}
with 17~meV and 4.5 (5.6~meV for $\alpha=0.33$) (s.\ Tab.\ I).
The latter is close to 5.5~meV extracted in the MT for CYCO. It
might provide a 
disorder scenario for that relatively large
value. 
Then the record value of $J_1$ for FFESC is  
given now by CYCO with 24~meV. But even higher values are expected for
La$_6$Ca$_8$Cu$_{24}$O$_{41}$ with 28-29~meV (s.\ Chapt.\  D) where disorder
and/or damping effects
are larger, leading to $J_2\approx$ 11~meV.

Returning to Li$_2$CuZrO$_4$, we note that
the authors of 
a recent neutron diffraction study doubted these nearly 
critical $\alpha$-values and suggest 
instead
 an {\it in-phase} spiral ordering
\cite{Yasui2015}. They
report a large propagation vector of a spiral structure and a 
pitch of 87.7$^o$ close to $\pi /2$ like in
LiVCuO$_4$ (84.5$^o$). 
Anyhow,
Yasui {\it et al.}  have applied incorrectly 
the classical pitch-Eq.\
{\it NOT} valid here \cite{bursil,Nishimoto2012} since 
 their estimated $\alpha \approx 6$(!) 
 (is at odds with the reported and fitted $\chi(T)$ and specific
 heat data \cite{drechsler,Sirker2010})
corresponds
to a situation of {\it extremely} weakly FM coupled interpenetrating
coupled
Heisenberg chains
where quantum fluctuations at weak IC 
are very important.
Our previous calculations for
LiVCuO$_4$ demonstrated 
a strong reduction of the resulting 
"quantum" value of
$\alpha$ \cite{Nishimoto2012}. Anyhow, we admit
that our previous estimate of $\alpha$ and $J_1$ for Li$_2$ZrCuO$_4$
\cite{drechsler}
should be considered also with some 
caution 
due to the uncertainty caused by fitting polycrystalline data
and ignoring the sensitive IC, the symmetric and antisymmetric
spin anisotropies allowed in a scenario with disorder  as well as a weak 3rd neighbor
coupling $J_3$ which slightly shifts the critical point.

 A phenomenological analysis 
 for LiCu$_2$O$_2$
 \cite{Sirker2010}
 yields 
 rather large 
  $J_2$-values: 6.9 to 10.3~meV
 at 
 reasonably 
 $J_1\sim 17$~meV. We ascribe the former to well-known
 disorder effect for that compound containing very often some Li-ions on Cu(2)-sites
 and Cu(2) on Li-sites \cite{Hibble1990} due to similar ionic radii.
 This provides a realistic scenario for the observed
 multiferroicity. 
 The failure of the simple
 $J_1$-$J_2$ model to describe the propagation vector of the observed spiral
 state is 
 not surpring and further theoretical
 studies are necessary also in view of the very complex magnetically  ordered
 phases. 
 For the isomorphic Na-compound without Cu-Na disorder a comparable value of 
 $J_1=-15$~meV but a much smaller $J_2 \approx 5$~meV has been calculated 
 \cite{Schmitt2011} which points again to the importance of disorder for enlarged
 $J_2$-values.

\small

\normalsize
\begin{table}[b!]
\caption{{\small
Empirical (e) and theoretical (t) values for $J_1$ in various
edge-sharing cuprates: S (single chain - pure edge-sharing), 
AS (alternating FM-AFM single chain) D(L) (single double (zigzag)-chain /
2L(P2L)- two-leg, 3L (three-leg) of  laddered (pseudo-ladder) edge- $ 
and$ single edge-sharing chains) and frustration ratio
$\alpha = J_2/\mid J_1\mid$ (In case of AS Li(Na)$_3$Cu$_2$SbO$_6$ 
the second coupling $J_2$ measures the AFM NN coupling via the
longer Cu-Cu bond bridging  Sb or Te. The mineral notation
 linarite
and {\it malachite}  
stand for 
the chemical notations 
PbCuSO$_4$OH
and Cu$_3$(AsO$_4$)(OH)$_3$,respectively,
PW means "present work" and the rows with "GPa" refer to studies 
under
pressure.}
}
\begin{tabular}{llccccccl}
\hline
\hline
system & type &$ -J_1$   & $J_2$  & $\alpha $ & $\Phi$  & $d_{Cu-Cu}$  &$d_{Cu-O}$ &Ref.\ \\
      & (sharing) & (meV)& (meV) & & $^o$ &(\AA ) & (\AA ) & \\
\hline
CYCO            & S & 2.2(t)      & 4.7 & 2.14& 94.5  & 2.82  &1.92 & \cite{mizuno} \\
CYCO            & S & 24 (e)     & 5.5 & 0.23 &94.1, 92.1   & 2.818, 2.826 &1.953 & PW 
\cite{Gotoh}  \\
Li$_2$CuO$_2$   & S & 19.8 (e) & 0.4 & 0.31 &93.97  & 2.86 & 1.956& \cite{lorenz} \\
Li$_2$CuO$_2$   & S & 8.6 (t)    & 5.5 & 0.62 &94  & 2.86 & 1.956 & \cite{mizuno}\\
Li$_2$CuO$_2$   & S & 14.5 (t) & 7.4 & 0.53 &93.97  & 2.86 & 1.956& \cite{Xiang2007} \\
Li$_2$CuO$_2$   & S & 18.5 (t) & 5.8 & 0.31 &93.97  & 2.86 & 1.956& \cite{Miloslavlevic}, PW \\
Li$_2$CuO$_2$   & S & 19.5 (t) & 4.9 & 0.31 &93.97  & 2.86 & 1.956& \cite{Drechsler2010} \\
Li$_2$CuO$_2$   & S & 19.8 (t) & 0.4 & 0.31 &93.97  & 2.86 & 1.958& PW \\
Li$_2$CuO$_2$   & S & 19.8 (t) & 5.6 & 0.31 &93.97  & 2.86 & 1.956& \cite{Tsirlin} \\
Li$_2$CuO$_2$   & S & --  & -- & -- &94, 90.66  & 2.782, 2.86 & 1.955,1956& \cite{You2009} 0, 32.5 GPa\\
Li$_2$CuO$_2$   & S & -- & -- & -- &92  & 2.785 & 1.938& \cite{Li2011} 28.8 GPa\\
Li$_2$CuO$_2$   & S &20,15,11  &5,7,9&0.34,0.5,0.9 & --  & --  & -- & \cite{Nitzsche2011} 0, 3, 5 GPa\\
Li$_2$ZrCuO$_4$ & S & 23.5 (e)   & 7.05 & 0.3& 94  & 2.88 & 2.00 & \cite{drechsler,Sirker2010} \\
Li$_2$ZrCuO$_4$ & S & 17 (t)   & 4.5 & 0.3& 94.13  & 2.88 & 2.00 & \cite{Schmitt2011} \\
LiCu$_2$O$_2$   & S & 13.9       & 3.7 & 0.33&93   & 2.86 & 1.98& \cite{Gippius2004,Hibble1990}, PW \\
LiCu$_2$O$_2$   & S & $\sim$17.2       & 10.9 & 0.6&93   &  & & \cite{Sirker2010} \\
NaCu$_2$O$_2$   & S &       &  & &92.9   & 2.93 &1.971 & \cite{Capogna2010} \\
NaCu$_2$O$_2$   & S & 15       & 5 & 0.33&92.9   & 2.93 &1.971 & \cite{Schmitt2011} \\
LiCuVO$_4$   & S & 6.4-15.5  & 5.2-7.8 & 0.5-0.85 &96.  & 2.899 & 1.965& \cite{Nishimoto2012,Sirker2010} \\
{\it linarite}        & S & 10          & 5.5 & 0.36 &90.8, 94.2    & 2.823 &1.915, 1.961 & \cite{Schaepers2014,Rule2017} \\
CuGeO$_3$       & S & -10         & 4.99 & 0.241-0.36 &97.654  & 2.935 && \cite{Braden1996} \\
Li$_3$Cu$_2$SbO$_6$ & AS &      24.56  &13.8 & 0.56 &88.95   & 2.92 &2.014& \cite{Koo2016} \\
Na$_3$Cu$_2$SbO$_6$ & AS &      17.8,18.7  &14.7, 14.99 & 1.21 &95.27, 95  
& 2.96 &2.003& \cite{Schmitt2014} \\
Na$_3$Cu$_2$TeO$_6$ & AS &      18.7  &15.0 & 1.25 1&95.27 , 95  & 2.86 &1.9& \cite{Schmitt2014} \\
SrCuO$_2$       &D& 23-48 & 226 &  4-10 &87.5, 92.96   & 2.795  &1.896, 1.958& \cite{Zaliznyak2004}\\
SrCu$_2$O$_3$   &2L& 35.8  & 166 (e); 155 (t) & 4.03 (t) &88x& 2.78  & 1.976(2L)&
 \cite{Sparta2006,Bordas2006} \\
SrCu$_2$O$_3$   &2L& 30.6        & 226 & 7.39& 87.4   & 2.781 & 1.97({\tiny L}),1.87({\tiny R}) &  
 \cite{mizuno}\\
CaCu$_2$O$_3$  & P2L& 28.2 (t)         & 139 (t), 160 (e) &  4.93 (t) &  90
  & 2.8 & 1.908 & \cite{Ruck2001,Bordas2006,Bisogni2013} \\
Sr$_2$Cu$_3$O$_5$   & 3L & 35.8         & 166 (e); 155 (t) & 4.03(t) &&  
2.781 &  1.976 (2L) &  \cite{Sparta2006,Bordas2006} \\ 
Sr$_{14}$Cu$_{24}$O$_{41}$   &2L+S &30.6        & 130 & 2.8&53   & 90 & & \cite{Bordas}\\
Sr$_{14}$Cu$_{24}$O$_{41}$   &2L+S& 26.1        & 226 & &88.6   & 2.75  &1.9({\tiny L})m 1.97({\tiny R})& \cite{mizuno}\\
SrCa$_{13}$Cu$_{24}$O$_{41}$   & 2L+S& -5.4(!?) (S)  
     & ? (S) & ?&88, 95   & 2.746  &1.9(L)m 1.97(r)& \cite{Deng2018,Deng2011}\\
La$_5$Ca$_9$Cu$_{24}$O$_{41}$   & 2L+S & 26        &  112 & &88.6   & 2.761  && \cite{Matsuda2003}\\
La$_6$Ca$_8$Cu$_{24}$O$_{41}$   &2L+S&1.8 (S) & 0 & 0 &91.6   & 2.76  &1.927& \cite{Carter1996}\\
La$_6$Ca$_8$Cu$_{24}$O$_{41}$   &2L+S& $\sim$30(S),$\sim$35(L) &  $\sim12$(S), 139(L)  & $\sim$0.4(S),$\sim$4(L) &91.6 
  & 2.761  &1.927& \cite{Matsuda1996} PW\\
La$_6$Ca$_8$Cu$_{24}$O$_{41}$   &2L+S&26(L) 18.5(S)    
   &6.72& 0.36(S)&91.6   & 2.76  &1.927& \cite{mizuno}\\  
   La$_6$Ca$_8$Cu$_{24}$O$_{41}$   &2L+S& 26(L) 18.5(S)    
   &6.72& 0.36(S)&91.6   &2.76&1.927& \cite{mizuno}\\
   {\it green dioptase } &  AS &  3.2     
   & 6.7  & 0.66 &97.4, 107   & 2.76  &1.96& \cite{Janson2010}\\ 
   CuAs$_2$O$_4$ &S& 8.96       
   &  2.24& 0.25 &91.5   & 2.7865  &1.95& \cite{Caslin2015,Caslin2014}\\ 
      CuSb$_2$O$_4$ & AFFS&  10.78    
   & 5.15 & 0.5-0.55 & 94.5   & 2.7893  &1.877, 2.073& \cite{Caslin2015}\\
  LiCuSbO$_4$ &S& 13.79     
   &  3.24& 0.24 & 89.8, 95.0  & 2.87 &1.95, 2.04 & bond 1
   \cite{Grafe2017,Dutton2011}\\ 
   LiCuSbO$_4$ &AFFS& 7.76      
   &  3.24& 0.42 & 92.0, 96.8   & 2.87 &1.92, 2.0 & bond 2
   \cite{Grafe2017,Dutton2011}\\ 
   {\it malachite} &AS& 12.11      
   &  6.73& 0.56 & 94.75 , 106.41  & 2.787  &1.95& \cite{Lebernegg2013a}\\
   NaCuMoO$_4$(OH) &S& 4.121      
   &  3.36& 0.56 & 91.97, 103.65   & 2.787  &1.899, 2.07& \cite{Nawa2014}\\  
   BaCu$_2$VO$_2$O$_8$ &AS& 11.97      
   &  40.92& 0.295 & 93.7   & 2.86  & 1.94, 1.983, & \cite{Klyushina2018}\\
   BaCu$_2$VO$_2$O$_8$ &AS& 14.80      
   &  46.& 0.0.322 & 93.7   & 2.87  & 1.94, 1.983& \cite{Janson}\\ 
   Ba$_2$Cu$_3$O$_4$Cl$_2$ &2D& 10.3      
   &  8.4& 0.82 & 90.0   & 2.57  &1.899 & \cite{Babkevich2017}\\ 
 Ba$_2$Cu$_3$O$_4$Cl$_2$ &2D& 18     
   &  10.5& 1.05 & 90.0   & 2.787  &1.899 & \cite{Yaresko2002}\\
    Ba$_2$Cu$_3$O$_4$Cl$_2$ &2D& 10.3      
   &  8.4& 0.82 & 90.0   & 2.57  &1.899 & \cite{Babkevich2017}\\ 
 Sr$_2$Cu$_3$O$_4$Cl$_2$ &2D& 10-14      
   &  10.5& 1.05 & 90.0   & 2.7285  &1.929 & \cite{Kim2001}\\
Sr$_2$Cu$_3$O$_4$Cl$_2$ &2D& 10-14      
   &  13.7& 1.05& 90.0   & 2.7285  &1.929 & \cite{Holmlund2009}\\     
\hline
\hline
\end{tabular} 
\end{table}